\documentclass[12pt]{article}
\usepackage{amsmath}

\usepackage{amssymb}
\usepackage{bm}
\usepackage{braket}
\usepackage[dvips]{graphicx}

\begin{document}

\baselineskip=17pt

\begin{titlepage}
%\rightline{\tt arXiv:****.*****}
%\rightline{\texttt{\jobname.tex}}
\rightline\today
\begin{center}
\vskip 1.5cm
\baselineskip=22pt
{\Large \bf {Gauge-invariant operators of open bosonic}}\\
{\Large \bf {string field theory in the low-energy limit}}
\end{center}
\begin{center}
\vskip 1.0cm
{\large Daiki Koyama${}^{\,1}$, Yuji Okawa${}^{\,1}$ and Nanami Suzuki${}^{\,2}$}
\vskip 1.0cm
{\it {${}^1$Institute of Physics, The University of Tokyo}}\\
{\it {3-8-1 Komaba, Meguro-ku, Tokyo 153-8902, Japan}}\\
kouyama@g.ecc.u-tokyo.ac.jp, okawa@g.ecc.u-tokyo.ac.jp

\vskip 0.5cm
{\it {${}^2$Department of Integrated Sciences,}}\\
{\it {College of Arts and Sciences, The University of Tokyo}}\\
{\it {3-8-1 Komaba, Meguro-ku, Tokyo 153-8902, Japan}}\\
nnm.szk@gmail.com

\vskip 1.0cm

{\bf Abstract}
\end{center}

\noindent
In the AdS/CFT correspondence
we consider correlation functions of gauge-invariant operators on the gauge theory side,
which we obtain in the low-energy limit of the open string sector.
To investigate this low-energy limit
we consider the action of open bosonic string field theory
including source terms for gauge-invariant operators
and classically integrate out massive fields
to obtain the effective action for massless fields.
While the gauge-invariant operators depend linearly on the open string field
and do not resemble the corresponding operators such as the energy-momentum tensor
in the low-energy limit,
we find that nonlinear dependence is generated
in the process of integrating out massive fields.
We also find that the gauge transformation is modified
in such a way that the effective action and the modified gauge transformation
can be written in terms of the same set of multi-string products
which satisfy weak $A_\infty$ relations,
and we present explicit expressions for the multi-string products.

\end{titlepage}

\tableofcontents

%\newpage

\section{Introduction and summary}
\label{section-1}
\setcounter{equation}{0}

The AdS/CFT correspondence~\cite{Maldacena:1997re} can be thought of
as providing a nonperturbative definition of closed string theory
in terms of a gauge theory without containing gravity.
In the standard explanation~\cite{Aharony:1999ti} of the AdS/CFT correspondence
we consider the theory on D-branes and take the low-energy limit.
The gauge theory which provides
a nonperturbative definition of closed string theory
is obtained from the low-energy limit of the open string sector.

A key ingredient for proving the AdS/CFT correspondence
is the equivalence of string theory with holes in the world-sheet
and string theory on a different background
without holes in the world-sheet.
While such equivalence has been shown in the topological string~\cite{Ooguri:2002gx},
it is still challenging to prove it in the physical string.\footnote{
See, for example, \cite{Berkovits:2003pq, Berkovits:2007rj, Berkovits:2019ulm}
for developments in this direction.
}
Even if we assume this equivalence,
it is difficult to see gravity directly from the gauge theory,
and one reason for this is that the world-sheet picture is gone
after taking the low-energy limit.
This motivates us to consider
{\it open string field theory}
as a theory on D-branes before taking the low-energy limit.

In the AdS/CFT correspondence we consider correlation functions
of gauge-invariant operators on the gauge theory side.
We are therefore interested in correlation functions
of gauge-invariant operators of open string field theory in this context.
While it is in general difficult to construct gauge-invariant operators
in string field theory,
gauge-invariant operators
for open bosonic string field theory~\cite{Witten:1985cc}
have been constructed in~\cite{Hashimoto:2001sm, Gaiotto:2001ji}.
If we consider the theory on $N$ D-branes,
it was shown that
the $1/N$ expansion of correlation functions of gauge-invariant operators
has an interpretation as the closed-string genus expansion
under the assumption
of the equivalence between string theory with holes in the world-sheet
and string theory without holes~\cite{Okawa}.
Thus to consider open string field theory as a theory
before taking the low-energy limit
can be a promising way for proving the AdS/CFT correspondence
because we can keep track of the world-sheet picture in the limit.

Of course, we need to extend the discussion to open superstring field theory,
as quantization of open bosonic string field theory is formal
because of tachyons in generic backgrounds.
Recently, there have been significant developments in constructing
actions of open superstring field theory
including the Ramond sector~\cite{Kunitomo:2015usa, Sen:2015uaa, Erler:2016ybs, Konopka:2016grr},
and we consider that it is time to study open superstring field theory
in this context.
On the other hand, it would be also useful
to consider gauge-invariant operators
of open bosonic string field theory
in the noncritical string or in the topological string
where tachyons are absent.
For example, it was shown in~\cite{Gaiotto:2003yb}
that the Kontsevich model~\cite{Kontsevich:1992ti}
can be realized as open bosonic string field theory
in the noncritical string,
and open bosonic string field theory
on topological A-branes
is equivalent to Chern-Simons theory~\cite{Witten:1992fb}.
See also~\cite{Costello:2018zrm} for recent discussion
on holography in the B-model topological string.
It would be also interesting
to consider D-instanton contributions in two-dimensional
string theory~\cite{Balthazar:2019rnh, Sen:2019qqg, Balthazar:2019ypi, Sen:2020cef, Sen:2020oqr} from
the viewpoint of open bosonic string field theory with gauge-invariant operators. 
With the extension to open superstring field theory in mind,
let us begin with the discussion
in open bosonic string field theory.

Construction of the low-energy effective action
of string field theory was discussed by Sen~\cite{Sen:2016qap}
for closed superstring field theory.
The string field projected onto the massless sector
is used to describe the low-energy effective action,
and it was shown that the gauge invariance
of the low-energy effective action inherited from that of the original theory.
The same strategy can be applied to open bosonic string field theory,
and in this paper we consider the action
including source terms for the gauge-invariant operators
and use the string field projected onto the massless sector
to study gauge-invariant operators in the low energy.
In the case of open bosonic string field theory, however,
we can integrate out massive fields only classically
because, as we mentioned before, the existence of tachyons
in the open string and in the closed string
renders the quantization formal.
Furthermore, we also integrate out the open-string tachyon
and construct the effective action in terms of massless fields,
although the tachyon plays an important role in the low energy.
Actually, we can also use the string field
projected onto the tachyonic and massless sectors
to construct the effective action
in terms of massless and tachyonic fields,
and, as we will see, the structure of the effective action
does not depend on the detail of the projection.
At any rate, this issue is an artifact of the bosonic theory
and we are interested in stable backgrounds of the superstring
where tachyons are absent.

One puzzling feature regarding gauge-invariant operators
of open bosonic string field theory in our context
is that they depend linearly on the open string field.
For example, the energy-momentum tensor
is a typical example of the gauge-invariant operators
we consider in the AdS/CFT correspondence,
but the gauge-invariant operators
of open bosonic string field theory
do not resemble familiar energy-momentum tensors.
In this paper we show that nonlinear dependence
on the open string field is generated
in the process of integrating out massive fields.
Although our discussion is in the bosonic theory
and classical, the mechanism of generating nonlinear dependence
can be easily understood in terms of Feynman diagrams,
and we expect that the same mechanism will work
in the quantum theory of the superstring.

While the generation of nonlinear dependence
of the gauge-invariant operators on the open string field
is what we expected,
an unexpected feature regarding the effective action
is that the gauge transformation is modified
when we include the source terms,
and gauge invariance requires terms which are nonlinear
with respect to the sources.
There are some similarities between terms in the effective action
and terms in the gauge transformation,
and we find that the effective action
and the modified gauge transformation
can be written in terms of the same set
of multi-string products
which satisfy weak $A_\infty$ relations.\footnote{
The $A_\infty$ algebra consists of a set of multi-string products
satisfying $A_\infty$ relations
and plays an important role in open string field theory~\cite{Stasheff:I, Stasheff:II, Getzler-Jones, Markl, Penkava:1994mu, Gaberdiel:1997ia}.
When zero-string products are incorporated,
it is referred to as the weak $A_\infty$ algebra
or the curved $A_\infty$ algebra.
}
With hindsight, the original action including the source terms
has a weak $A_\infty$ structure in a rather trivial fashion,
and the weak $A_\infty$ structure of the effective action
can be understood as being inherited from that of the original action
in the process of integrating out massive fields
in accord with the general consideration by Sen~\cite{Sen:2016qap}.

An advantage 
of considering open string field theory based on the star product
is that expressions for terms
in the effective action are simpler and more explicit
compared to closed string field theory.
However, expressions for terms in the effective action
become rather lengthy at higher orders
even in open bosonic string field theory based on the star product.
The weak $A_\infty$ structure provides us
with analytic control over terms in the effective action,
and in this paper we present explicit expressions for the multi-string products
to all orders.

The rest of the paper is organized as follows.
We begin with presenting the action and the gauge transformation
of open bosonic string field theory in section~\ref{section-2},
and we explain the gauge-invariant operators of open bosonic string field theory
in section~\ref{section-3}.
We discuss the effective action of open bosonic string field theory
for massless fields in section~\ref{section-4},
and then we incorporate the gauge-invariant operators
into the discussion of the effective action for massless fields in section~\ref{section-5}.
We expand the effective action and the gauge transformation
in powers of the sources and in powers of the fields,
and we explicitly construct terms at lower orders.
In section~\ref{section-6} we present the basics of the weak $A_\infty$ structure,
and we show that the terms of the effective action and the gauge transformation
constructed in section~\ref{section-5} can be expressed
using multi-string products satisfying weak $A_\infty$ relations.
In section~\ref{section-7}
we explain the coalgebra representation of the weak $A_\infty$ structure,
and we make use of this coalgebra representation
to construct the multi-string products to all orders in section~\ref{section-8}.
Section~\ref{section-9} is devoted to conclusions and discussion.

\bigskip

When the draft of this paper was almost complete,
we were informed of independent work
addressing the same problem
by Erbin, Maccaferri, Schnabl and Vosmera~\cite{EMSV},
which is arranged to appear simultaneously with ours.

\section{Open bosonic string field theory}
\label{section-2}
\setcounter{equation}{0}

In this section we present the action and the gauge transformation
of open bosonic string field theory~\cite{Witten:1985cc}.
The open string field $\Psi$ is a Grassmann-odd state of ghost number $1$
in the Hilbert space
of the boundary conformal field theory describing the open string background we consider,
which consists of the matter sector and the $bc$ ghost sector.
The action of the free theory is given by
\begin{equation}
S = {}-\frac{1}{2} \, \langle \, \Psi, Q \Psi \, \rangle \,,
\end{equation}
where $Q$ is the BRST operator which is Grassmann odd
and $\langle \, A_1, A_2 \, \rangle$ is the BPZ inner product
of a pair of states $A_1$ and $A_2$.
Three basic properties of the BPZ inner product
and the BRST operator $Q$ are
\begin{equation}
\langle \, A_1, A_2 \, \rangle
= (-1)^{A_1 A_2} \langle \, A_2, A_1 \, \rangle \,, \quad
Q^2 = 0 \,, \quad
\langle \, Q A_1, A_2 \, \rangle
= {}-(-1)^{A_1} \langle \, A_1, QA_2 \, \rangle \,.
\label{BPZ-Q-properties}
\end{equation}
Here and in what follows,
a state in the exponent of $-1$ represents
its Grassmann parity: it is 0 mod 2 for a Grassmann-even state
and 1 mod 2 for a Grassmann-odd state.
The action of the free theory is invariant
under the gauge transformation given by
\begin{equation}
\delta_\Lambda \Psi = Q \Lambda \,,
\end{equation}
where the gauge parameter $\Lambda$ is a Grassmann-even state of ghost number $0$.

The interacting theory by Witten~\cite{Witten:1985cc} was constructed
by introducing the star product $A_1 \ast A_2$
defined for a pair of states $A_1$ and $A_2$.
The Grassmann parity of $A_1 \ast A_2$ is $\epsilon( A_1 ) +\epsilon( A_2 )$ mod $2$,
where $\epsilon ( A_i )$ is the Grassmann parity of $A_i$ mod $2$ for $i = 1, 2$.
Three important properties involving the star product are
\begin{equation}
\begin{split}
& \langle \, A_1, A_2 \ast A_3 \, \rangle
= \langle \, A_1 \ast A_2, A_3 \, \rangle \,, \qquad
Q ( A_1 \ast A_2 ) = Q A_1 \ast A_2 +(-1)^{A_1} A_1 \ast Q A_2 \,, \\
& ( A_1 \ast A_2 ) \ast A_3 = A_1 \ast ( A_2 \ast A_3 ) \,.
\end{split}
\label{star-product-properties}
\end{equation}
The action of the interacting theory is given by
\begin{equation}
S = {}-\frac{1}{2} \, \langle \, \Psi, Q \Psi \, \rangle
-\frac{g}{3} \, \langle \, \Psi, \Psi \ast \Psi \, \rangle \,,
\label{open-bosonic-string-field-theory-action}
\end{equation}
where $g$ is the open string coupling constant.
Using~\eqref{BPZ-Q-properties} and~\eqref{star-product-properties},
we can show that this action
is invariant under the gauge transformation given by
\begin{equation}
\delta_\Lambda \Psi = Q \Lambda
+g \, ( \, \Psi \ast \Lambda -\Lambda \ast \Psi\, ) \,.
\label{gauge-transformation}
\end{equation}

\section{Gauge-invariant operators}
\label{section-3}
\setcounter{equation}{0}

It is in general difficult to construct gauge-invariant operators
in string field theory,
but a class of gauge-invariant operators have been constructed
in open bosonic string field theory
and we can define a gauge-invariant operator
for each on-shell closed string vertex operator~\cite{Hashimoto:2001sm, Gaiotto:2001ji}.

We label on-shell closed string vertex operators of ghost number $2$
as $\mathcal{V}_\alpha$,
where the collective label $\alpha$ generically contains
both continuous and discrete variables.
The gauge-invariant operator $\mathcal{A}_{\mathcal{V}_\alpha} [ \, \Psi \, ]$ 
for $\mathcal{V}_\alpha$ is defined
by the following correlation function on the upper half-plane:
\begin{equation}
\mathcal{A}_{\mathcal{V}_\alpha} [ \, \Psi \, ]
= \langle \, f_{\rm mid} \circ \mathcal{V}_\alpha (0) \,
f_I \circ \Psi (0) \, \rangle_{\rm UHP} \,,
\end{equation}
where $\Psi (0)$ is the operator corresponding
to the state $\Psi$ in the state-operator correspondence.
Here and in what follows we denote the operator mapped from $\phi (0)$ 
at $\xi=0$ in the local coordinate $\xi$
under a conformal transformation $f (\xi)$ by $f \circ \phi (0)$
both for boundary and bulk operators.
The conformal transformation $f_I (\xi)$
associated with the identity string field is given by
\begin{equation}
f_ I (\xi) = \tan \Bigl( \, 2 \, \arctan \xi \, \Bigr)
= \frac{2 \, \xi}{1-\xi^2}
\end{equation}
and $f_{\rm mid} (\xi)$ is the translation to the open-string midpoint:
\begin{equation}
f_{\rm mid} (\xi) = \xi +i \,.
\end{equation}

We denote the closed string state corresponding to $\mathcal{V}_\alpha$
in the state-operator correspondence by $\Phi_\alpha$,
which is annihilated by the BRST operator:
\begin{equation}
Q \Phi_\alpha = 0 \,.
\end{equation}
Then the gauge-invariant operator $\mathcal{A}_{\mathcal{V}_\alpha} [ \, \Psi \, ]$
can be expressed as a BPZ inner product with $\Phi_\alpha$:
\begin{equation}
\mathcal{A}_{\mathcal{V}_\alpha} [ \, \Psi \, ]
= \langle \, \Phi_\alpha, D ( \Psi ) \, \rangle \,,
\end{equation}
where the closed string field $D (A)$ constructed from an open string field $A$
is defined in terms of the BPZ inner product with an arbitrary closed string field $B$ as
\begin{equation}
\langle \, B, D (A) \, \rangle
= \langle \, f_{\rm mid} \circ B (0) \, f_I \circ A (0) \, \rangle_{\rm UHP} \,,
\end{equation}
where $A(0)$ is the open string field corresponding to $A$
and $B(0)$ is the closed string field corresponding to $B$
in the state-operator correspondence.
Since the gauge-invariant operator $\mathcal{A}_{\mathcal{V}_\alpha} [ \, \Psi \, ]$
is linear in $\Psi$, it can also be expressed
in terms of a BPZ inner product with $\Psi$ as
\begin{equation}
\mathcal{A}_{\mathcal{V}_\alpha} [ \, \Psi \, ]
= \langle \, J (\Phi_\alpha), \Psi \, \rangle \,,
\end{equation}
where the open string field $J (B)$ constructed from a closed string field $B$
is defined in terms of the BPZ inner product with an arbitrary open string field $A$ as
\begin{equation}
\langle \, J(B), A \, \rangle
= \langle \, f_{\rm mid} \circ B (0) \, f_I \circ A (0) \, \rangle_{\rm UHP} \,.
\end{equation}
The Grassmann parity of $J(B)$ is the same as that of $B$ mod $2$.
Two important properties associated with $J(B)$ are as follows:
\begin{equation}
\begin{split}
Q J(B) & = J(QB) \,, \\
J(B) \ast A & = (-1)^{AB} A \ast J(B) \,,
\end{split}
\end{equation}
where $A$ is an arbitrary open string field.
We can use these properties to show the gauge invariance
of $\mathcal{A}_{\mathcal{V}_\alpha} [ \, \Psi \, ]$.
The gauge variation of $\mathcal{A}_{\mathcal{V}_\alpha} [ \, \Psi \, ]$ vanishes because
\begin{equation}
\begin{split}
\delta_\Lambda \mathcal{A}_{\mathcal{V}_\alpha} [ \, \Psi \, ]
& = \mathcal{A}_{\mathcal{V}_\alpha} [ \, Q \Lambda +g \, ( \, \Psi \ast \Lambda -\Lambda \ast \Psi \, ) \, ] \\
& = \langle \, J (\Phi_\alpha), Q \Lambda \, \rangle
+g \, \langle \, J (\Phi_\alpha), \Psi \ast \Lambda \, \rangle
-g \, \langle \, J (\Phi_\alpha), \Lambda \ast \Psi \, \rangle \\
& = {}-\langle \, Q J (\Phi_\alpha), \Lambda \, \rangle
+g \, \langle \, J (\Phi_\alpha) \ast \Psi, \Lambda \, \rangle
-g \, \langle \, \Psi \ast J (\Phi_\alpha), \Lambda \, \rangle \\
& = {}-\langle \, J ( Q \Phi_\alpha), \Lambda \, \rangle
+g \, \langle \, J (\Phi_\alpha) \ast \Psi, \Lambda \, \rangle
-g \, \langle \, J (\Phi_\alpha) \ast \Psi, \Lambda \, \rangle = 0 \,,
\end{split}
\end{equation}
where in the last step we used $Q \Phi_\alpha = 0$.

We introduce the source $\mathcal{G}_\alpha$ for $\mathcal{A}_{\mathcal{V}_\alpha} [ \, \Psi \, ]$.
The source term $S_{\rm source}$ added to the action is given by
\begin{equation}
S_{\rm source} = \sum_{\alpha} \mathcal{G}_\alpha \, \mathcal{A}_{\mathcal{V}_\alpha} [ \, \Psi \, ] \,,
\end{equation}
where the summation over $\alpha$ should be understood
to include integrals for continuous variables.
The source term can be written in terms of $D (\Psi)$ as
\begin{equation}
S_{\rm source} = \sum_{\alpha} \mathcal{G}_\alpha \,
\langle \, \Phi_\alpha, D (\Psi) \, \rangle
\end{equation}
or in terms of $J (\Phi_\alpha)$ as
\begin{equation}
S_{\rm source} = \sum_{\alpha} \mathcal{G}_\alpha \,
\langle \, J (\Phi_\alpha), \Psi \, \rangle \,.
\end{equation}
On-shell closed string states can be incorporated into a single closed string field $\Phi$ as
\begin{equation}
\Phi = \sum_{\alpha} \mathcal{G}_\alpha \, \Phi_\alpha \,.
\end{equation}
The on-shell closed string field $\Phi$
is a Grassmann-even state of ghost number $2$
and is annihilated by the BRST operator:
\begin{equation}
Q \Phi = 0 \,.
\end{equation}
Then the source term $S_{\rm source}$ can be expressed as
\begin{equation}
S_{\rm source} = \langle \, \Phi, D ( \Psi ) \, \rangle
\end{equation}
or as
\begin{equation}
S_{\rm source} = \langle \, J (\Phi), \Psi \, \rangle \,.
\end{equation}
In the rest of the paper we will use this form of the source term.

The action with the source term is given by
\begin{equation}
S = {}-\frac{1}{2} \, \langle \, \Psi, Q \Psi \, \rangle
-\frac{g}{3} \, \langle \, \Psi, \Psi \ast \Psi \, \rangle
+\frac{\kappa}{g} \, \langle \, J(\Phi), \Psi \, \rangle \,,
\label{action+source}
\end{equation}
where we introduced the parameter $\kappa$ to count
the power of the sources.
We can show that the action with the source term~\eqref{action+source}
is invariant under the gauge transformation~\eqref{gauge-transformation}
using the following properties of $J(\Phi)$:
\begin{equation}
\begin{split}
Q J(\Phi) & = 0 \,, \\
J(\Phi) \ast A & = A \ast J(\Phi)
\end{split}
\end{equation}
for any open string field $A$.

\section{Open string field theory in the low-energy limit}
\label{section-4}
\setcounter{equation}{0}

We are interested in the low-energy limit of correlation functions
of gauge-invariant operators in open string field theory.
Let us first discuss the low-energy limit
of open bosonic string field theory
following the approach developed in~\cite{Sen:2016qap}
without introducing gauge-invariant operators.

The low-energy limit of closed superstring field theory
discussed in~\cite{Sen:2016qap}
is described by the string field projected onto the massless sector.
The action for the massless fields
corresponds to the effective action
obtained by integrating out massive fields.
The idea can be applied to open string field theory
at least in the classical theory,
which corresponds to classically integrating out massive fields.
In the case of open bosonic string field theory
the low-energy limit is formal because of the existence
of the tachyon in the spectrum.
Since this is an artifact of the bosonic string
and is absent in the superstring,
we ignore this issue
and we classically integrate out the tachyon field
in addition to massive fields.
Alternatively, we can also use the string field
projected onto the tachyonic and massless sectors
to construct the effective action
in terms of massless and tachyonic fields,
and, as we will see, the structure of the effective action
does not depend on the detail of the projection.

An open string field $\Psi$ for massless fields is annihilated
by $L_0 -\alpha' p^2$,
\begin{equation}
( \, L_0 -\alpha' p^2 \, ) \, \Psi = 0 \,,
\end{equation}
where $L_0$ is the zero mode of the energy-momentum tensor,
$p_\mu$ is the spacetime momentum operator,
and $p^2 = p_\mu \, p^\mu$.
We denote the projection operator
onto the massless sector by $P$.
The string field $\Psi$ for massless fields satisfies
\begin{equation}
P \, \Psi = \Psi \,.
\end{equation}
The projection operator $P$ has the following properties:
\begin{equation}
P^2 = P \,, \qquad
P \, Q = Q \, P \,, \qquad
\langle \, A_1, P A_2 \, \rangle
= \langle \, P A_1, A_2 \, \rangle
\end{equation}
for any pair of states $A_1$ and $A_2$.

\begin{figure}[t]
\begin{align*}
\raisebox{-0.46\height}{\includegraphics[width=2.5cm]{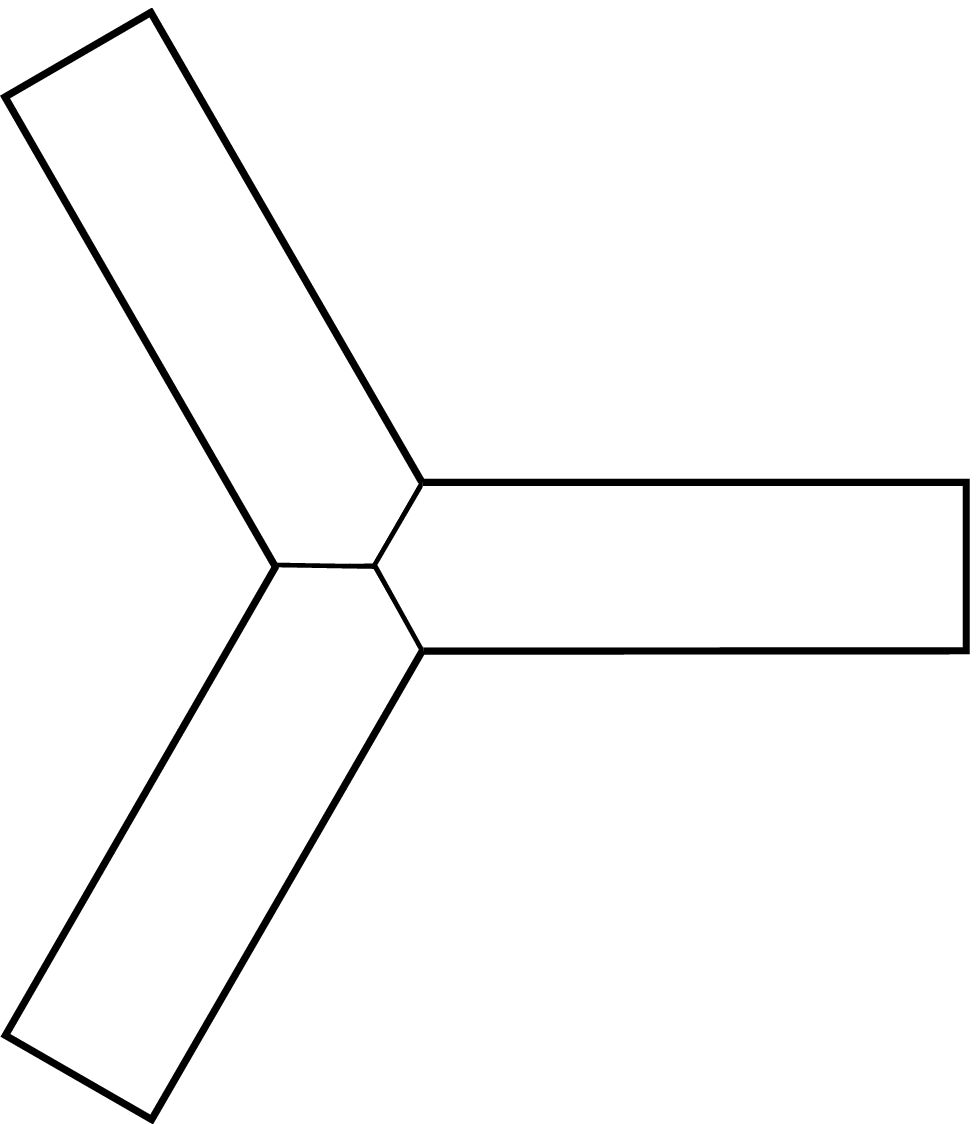}}
\quad + \quad
\raisebox{-0.46\height}{\includegraphics[width=2.5cm]{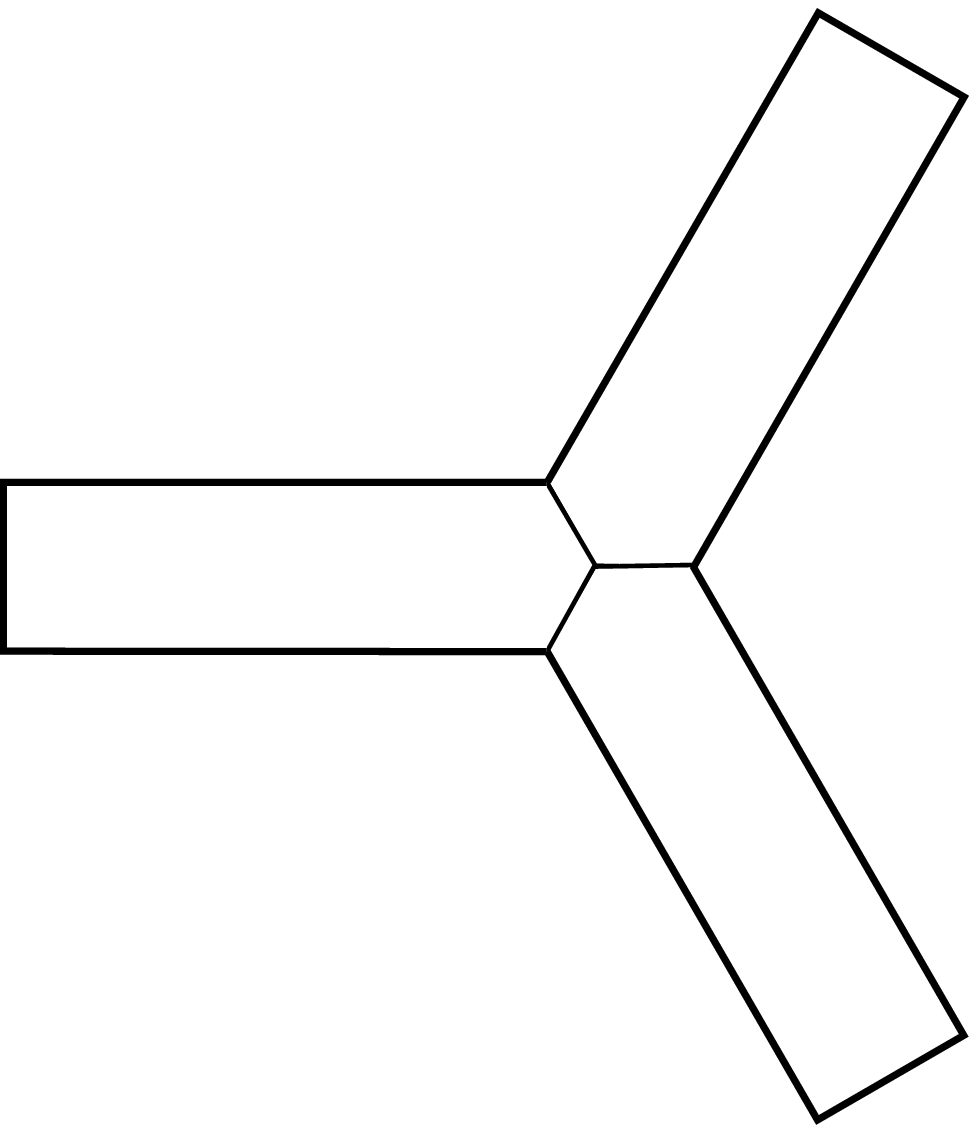}}
\quad \longrightarrow \quad
\raisebox{-0.46\height}{\includegraphics[width=3.7cm]{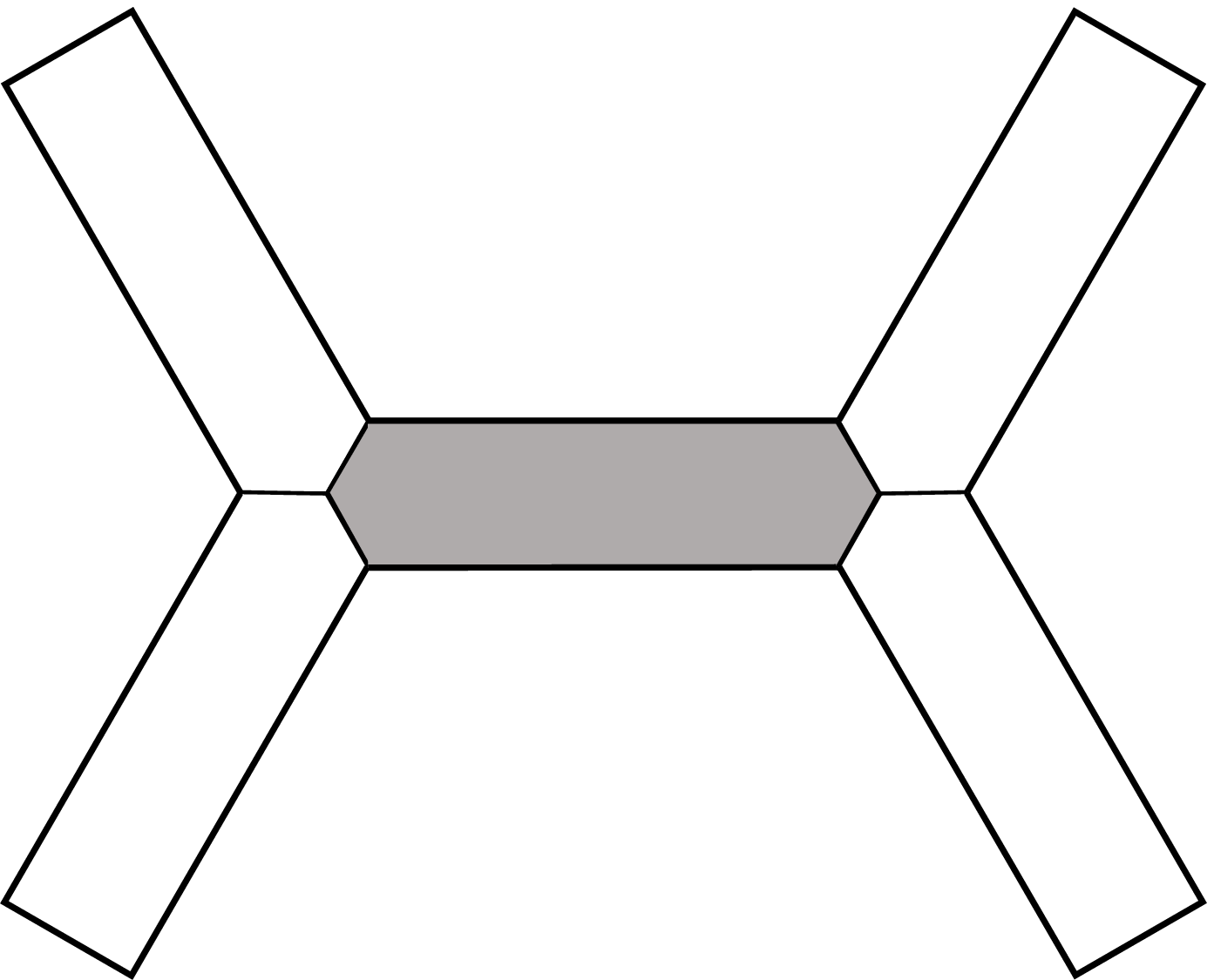}}
\end{align*}
\caption{The quartic term in the effective action is generated from two cubic vertices
by integrating out massive and tachyonic fields.
The propagator for massive and tachyonic fields is shaded in the figure.
}
\label{quartic-vertex-figure}
\end{figure}

Let us consider open bosonic string field theory
in terms of this projected string field.
The kinetic term of the theory is given by
\begin{equation}
S = {}-\frac{1}{2} \, \langle \, \Psi, Q \Psi \, \rangle \,,
\end{equation}
where the string field $\Psi$ satisfies the condition
\begin{equation}
P \, \Psi = \Psi \,.
\end{equation}
This kinetic term is invariant under the gauge transformation
given by
\begin{equation}
\delta_\Lambda \Psi = Q \Lambda \,,
\end{equation}
where the gauge parameter $\Lambda$ satisfies the condition
\begin{equation}
P \, \Lambda = \Lambda \,.
\end{equation}
Let us add the cubic interaction to this kinetic term:
\begin{equation}
S = {}-\frac{1}{2} \, \langle \, \Psi, Q \Psi \, \rangle
-\frac{g}{3} \, \langle \, \Psi, \Psi \ast \Psi \, \rangle \,.
\end{equation}
The variation of the action
under the gauge transformation
\begin{equation}
\delta_\Lambda \Psi = Q \Lambda
+g P \, ( \, \Psi \ast \Lambda -\Lambda \ast \Psi\, )
\end{equation}
vanishes at $O(g)$,
but it is nonvanishing at $O(g^2)$.
Following the approach developed in~\cite{Sen:2016qap},
let us add the quartic term obtained
by classically integrating out massive and tachyonic fields:
\begin{equation}
S = {}-\frac{1}{2} \, \langle \, \Psi, Q \Psi \, \rangle
-\frac{g}{3} \, \langle \, \Psi, \Psi \ast \Psi \, \rangle
+\frac{g^2}{2} \, \langle \, \Psi \ast \Psi,
\frac{b_0}{L_0} \, (1-P) \, ( \Psi \ast \Psi ) \, \rangle
+O(g^3) \,,
\end{equation}
where $b_0$ is the zero mode of the $b$ ghost.
See figure~\ref{quartic-vertex-figure}.
%%%
%%%
The propagator for massive and tachyonic fields appears frequently in what follows,
and we denote it by $h$:
\begin{equation}
h \equiv \frac{b_0}{L_0} \,(1-P) \,.
\end{equation}
It is Grassmann odd and is BPZ even:
\begin{equation}
\langle \, A_1, h \, A_2 \, \rangle
= (-1)^{A_1} \langle \, h \, A_1, A_2 \, \rangle
\end{equation}
for any pair of states $A_1$ and $A_2$.
The important property of $h$ is
\begin{equation}
Q \, h +h \, Q = 1-P \,.
\end{equation}
We can show that the action is invariant up to~$O(g^2)$
under the gauge transformation given by
\begin{equation}
\begin{split}
\delta_\Lambda \Psi & = Q \Lambda
+g P \, ( \, \Psi \ast \Lambda -\Lambda \ast \Psi\, ) \\
& \quad~
+g^2 P \, [ \,
-h \, ( \Psi \ast \Psi ) \ast \Lambda
+h \, ( \Psi \ast \Lambda ) \ast \Psi
-h \, ( \Lambda \ast \Psi ) \ast \Psi \\
& \qquad \qquad \quad
-\Psi \ast h \, ( \Psi \ast \Lambda )
+\Psi \ast h \, ( \Lambda \ast \Psi )
+\Lambda \ast h \, ( \Psi \ast \Psi ) \, ]
+O(g^3) \,.
\end{split}
\end{equation}

\section{Open string field theory with the source term in the low-energy limit}
\label{section-5}
\setcounter{equation}{0}

\subsection{Modification of the gauge transformation}

\begin{figure}[t]
\begin{align*}
\raisebox{-0.33\height}{\includegraphics[width=1.8cm]{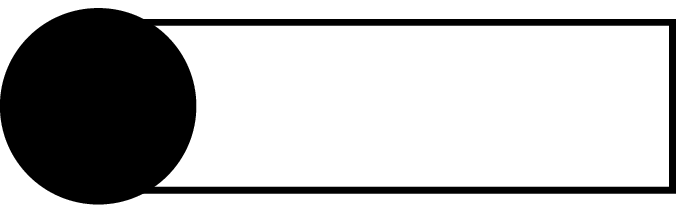}}
\quad + \quad
\raisebox{-0.46\height}{\includegraphics[width=2.5cm]{massless.diagram.vertex.string2.2.eps}}
\quad \longrightarrow \quad
\raisebox{-0.46\height}{\includegraphics[width=3.0cm]{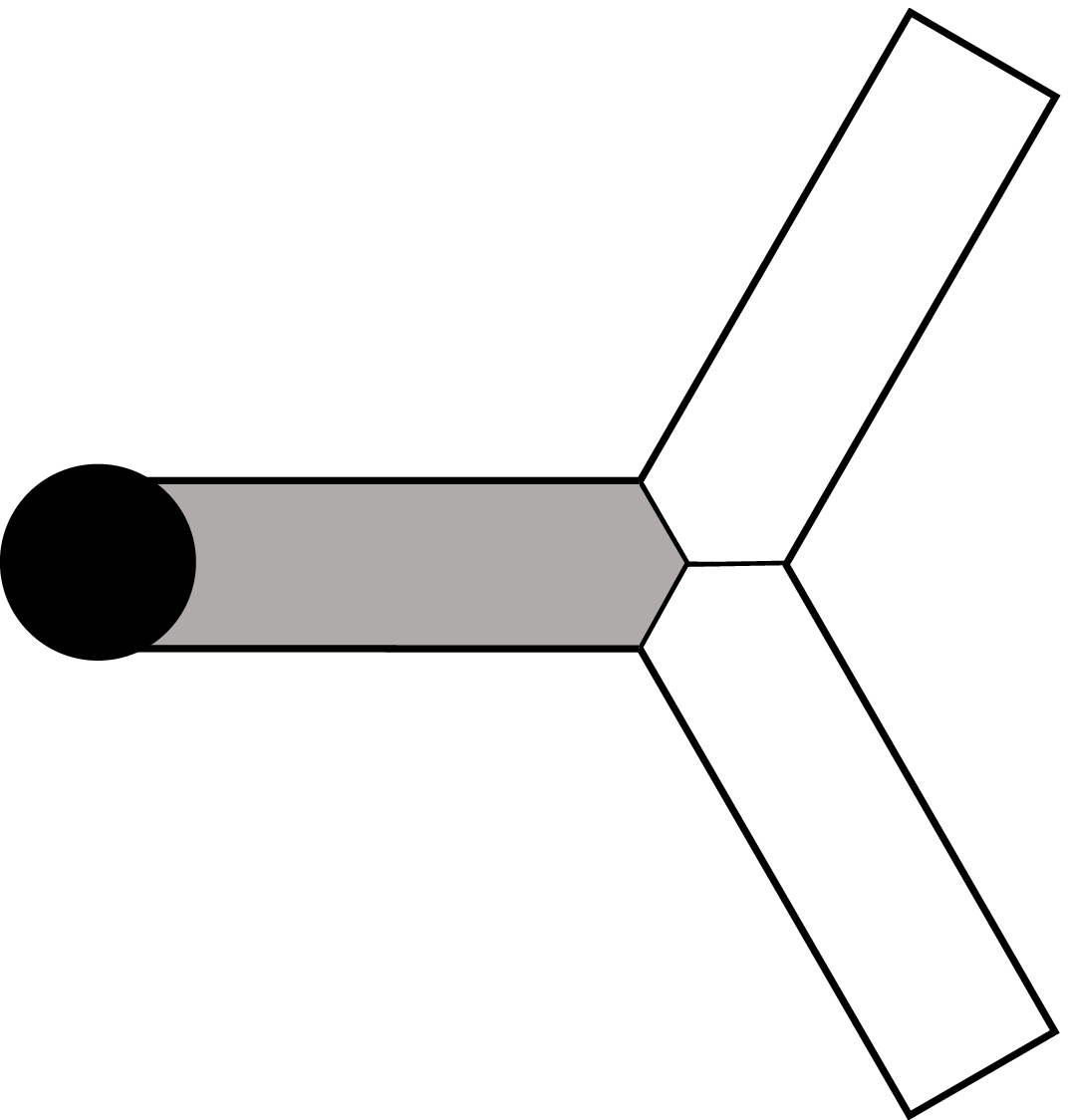}}
\end{align*}
\caption{Nonlinear coupling to the gauge-invariant operators is generated
in the effective action from the source term and the cubic vertex
by integrating out massive and tachyonic fields.
The black dot denotes the source,
and the propagator for massive and tachyonic fields is shaded in the figure.}
\label{J-Psi-Psi-figure}
\end{figure}

Let us now incorporate gauge-invariant operators
by adding the source term in the low-energy limit.
We denote the action in the low-energy limit
before introducing the source term by $S^{(0)}$
and we expand it in $g$ as follows:
\begin{equation}
S^{(0)} = S^{(0)}_2 +g \, S^{(0)}_3 +g^2 \, S^{(0)}_4 +O(g^3) \,,
\label{low-energy-action}
\end{equation}
where
\begin{equation}
\begin{split}
S^{(0)}_2 & = {}-\frac{1}{2} \, \langle \, \Psi, Q \Psi \, \rangle \,, \quad
S^{(0)}_3 = {}-\frac{1}{3} \, \langle \, \Psi, \Psi \ast \Psi \, \rangle \,, \quad
S^{(0)}_4 = \frac{1}{2} \, \langle \, \Psi \ast \Psi,
h \, ( \Psi \ast \Psi ) \, \rangle \,.
\end{split}
\end{equation}
We denote the gauge transformation by $\delta^{(0)} \Psi$,
and we also expand it in $g$ as follows:
\begin{equation}
\delta^{(0)} \Psi
= \delta^{(0)}_0 \Psi +g \, \delta^{(0)}_1 \Psi
+g^2 \, \delta^{(0)}_2 \Psi +O(g^3) \,,
\end{equation}
where
\begin{equation}
\begin{split}
\delta^{(0)}_0 \Psi & = Q \Lambda \,, \qquad
\delta^{(0)}_1 \Psi = P \, ( \, \Psi \ast \Lambda -\Lambda \ast \Psi\, ) \\
\delta^{(0)}_2 \Psi
& = P \, [ \,
-h \, ( \Psi \ast \Psi ) \ast \Lambda
+h \, ( \Psi \ast \Lambda ) \ast \Psi
-h \, ( \Lambda \ast \Psi ) \ast \Psi \\
& \qquad \quad
-\Psi \ast h \, ( \Psi \ast \Lambda )
+\Psi \ast h \, ( \Lambda \ast \Psi )
+\Lambda \ast h \, ( \Psi \ast \Psi ) \, ] \,.
\end{split}
\end{equation}
We add the coupling
\begin{equation}
\frac{\kappa}{g} \, \langle \, J(\Phi), \Psi \, \rangle
\end{equation}
to the action $S^{(0)}$,
where $\Psi$ satisfies the condition $P \, \Psi = \Psi$.
It is easy to see that this coupling is invariant
under $\delta^{(0)}_0 \Psi$,
\begin{equation}
\delta^{(0)}_0 \biggl[ \, \frac{\kappa}{g} \, \langle \, J(\Phi), \Psi \, \rangle \, \biggr]
= 0 \,,
\end{equation}
but it is not invariant when we include the correction $\delta^{(0)}_1 \Psi$:
\begin{equation}
g \, \delta^{(0)}_1 \biggl[ \, \frac{\kappa}{g} \, \langle \, J(\Phi), \Psi \, \rangle \, \biggr] \ne 0 \,.
\end{equation}
Based on the insight we learned from the approach developed in~\cite{Sen:2016qap},
we expect that the gauge invariance can be recovered
if we include terms
which are obtained from the source term and the cubic vertex 
by classically integrating out massive and tachyonic fields.
To leading order in $g$, the following term is generated:
\begin{equation}
{}-\kappa \, \langle \, J(\Phi), h \, ( \Psi \ast \Psi ) \, \rangle \,.
\label{J-Psi-Psi}
\end{equation}
Note that this term depends nonlinearly on $\Psi$.
See figure~\ref{J-Psi-Psi-figure}.
%%%
%%%
Let us calculate the variation under the gauge transformation
\begin{equation}
\delta^{(0)} \biggl[ \,
\frac{\kappa}{g} \, \langle \, J(\Phi), \Psi \, \rangle \,
{}-\kappa \, \langle \, J(\Phi), h \, ( \Psi \ast \Psi ) \, \rangle
+O(g) \, \biggr] \,,
\end{equation}
which can be expanded in $g$ as
\begin{equation}
\frac{1}{g} \, \biggl[ \,
\kappa \, \delta^{(0)}_0 \langle \, J(\Phi), \Psi \, \rangle \,
\biggr] \,
+\biggl[ \,
\kappa \, \delta^{(0)}_1 \langle \, J(\Phi), \Psi \, \rangle \,
{}-\kappa \, \delta^{(0)}_0 \langle \, J(\Phi), h \, ( \Psi \ast \Psi ) \, \rangle
\biggr] +O(g) \,.
\end{equation}
We find that the variation at~$O(g^0)$ is still nonvanishing,
\begin{equation}
\kappa \, \delta^{(0)}_1 \langle \, J(\Phi), \Psi \, \rangle \,
{}-\kappa \, \delta^{(0)}_0 \langle \, J(\Phi),
h \, ( \Psi \ast \Psi ) \, \rangle \ne 0 \,,
\end{equation}
but this can be written as
\begin{equation}
\kappa \, \delta^{(0)}_1 \langle \, J(\Phi), \Psi \, \rangle \,
{}-\kappa \, \delta^{(0)}_0 \langle \, J(\Phi),
h \, ( \Psi \ast \Psi ) \, \rangle
= \kappa \, \delta^{(1)}_0 \biggl[ \,
\frac{1}{2} \, \langle \, \Psi, Q \Psi \, \rangle \, \biggr]
\end{equation}
with $\delta^{(1)}_0 \Psi$ given by
\begin{equation}
\delta^{(1)}_0 \Psi = P \, [ \,
h  \, J(\Phi) \ast \Lambda -\Lambda \ast h \, J(\Phi) \, ] \,.
\end{equation}
This means that the gauge invariance can be restored at this order
if we modify the gauge transformation at~$O(\kappa)$.

As we expected, the gauge invariance was recovered
by adding the term~\eqref{J-Psi-Psi}.
This term is nonlinear in $\Psi$,
and we have learned that such nonlinear terms
can be generated by integrating out massive fields.
This has resolved the puzzle we mentioned in section~\ref{section-1}.
On the other hand, what we did not expect was
that the gauge transformation is modified
when we include the source term.
Let us explore more about this structure
by investigating higher-order terms in the action.

\subsection{Expansion in $\kappa$ and $g$}

Since we have learned that the gauge transformation is modified
when we include the coupling to the gauge-invariant operators,
it is convenient to expand the action and the gauge transformation in $\kappa$
to discuss the gauge invariance systematically.
We expand the action $S$ and the gauge transformation $\delta_\Lambda \Psi$ as
\begin{align}
S & = S^{(0)}
+\kappa \, S^{(1)} +\kappa^2 \, S^{(2)} +\kappa^3 \, S^{(3)} +O(\kappa^4) \,, \\
\delta_\Lambda \Psi & = \delta^{(0)} \Psi
+\kappa \, \delta^{(1)} \Psi +\kappa^2 \, \delta^{(2)} \Psi +O(\kappa^3) \,.
\end{align}
Then the condition $\delta_\Lambda S = 0$ for the gauge invariance
can be expanded as
\begin{equation}
\begin{split}
\delta_\Lambda S
& = \delta^{(0)} S^{(0)}
+\kappa \, \Bigl[ \, \delta^{(0)} S^{(1)} +\delta^{(1)} S^{(0)} \, \Bigr]
+\kappa^2 \, \Bigl[ \, \delta^{(0)} S^{(2)} +\delta^{(1)} S^{(1)}
+\delta^{(2)} S^{(0)} \, \Bigr] \\
& \quad~ {}+\kappa^3 \, \Bigl[ \, \delta^{(0)} S^{(3)} +\delta^{(1)} S^{(2)}
+\delta^{(2)} S^{(1)} +\delta^{(3)} S^{(0)} \, \Bigr]
+O(\kappa^4) = 0 \,.
\end{split}
\end{equation}

\subsubsection{Construction at $O(\kappa)$}

\begin{figure}[h]
\begin{align*}
\raisebox{-0.47\height}{\includegraphics[width=2.7cm]{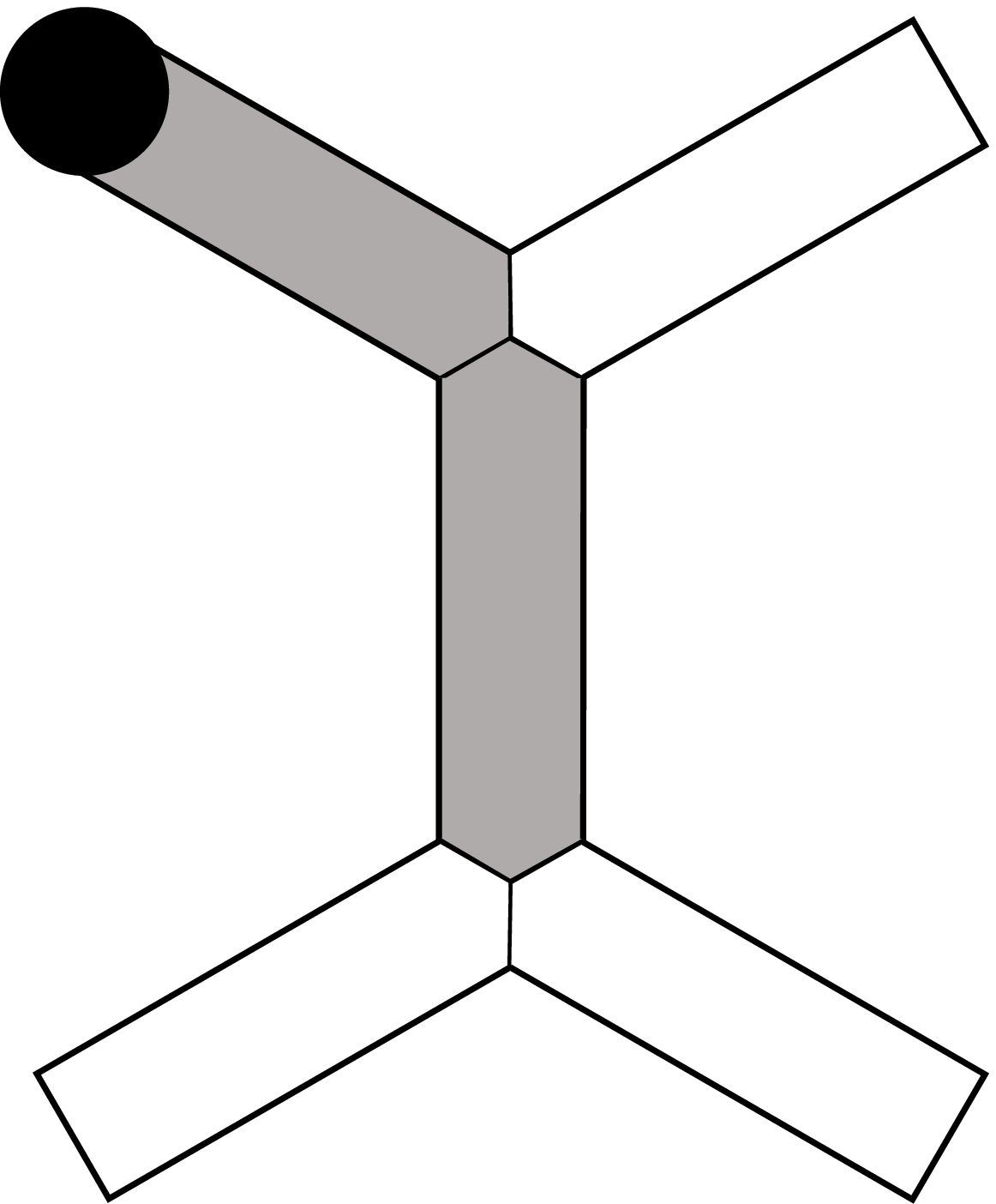}}
\quad + \quad
\raisebox{-0.45\height}{\includegraphics[width=3.5cm]{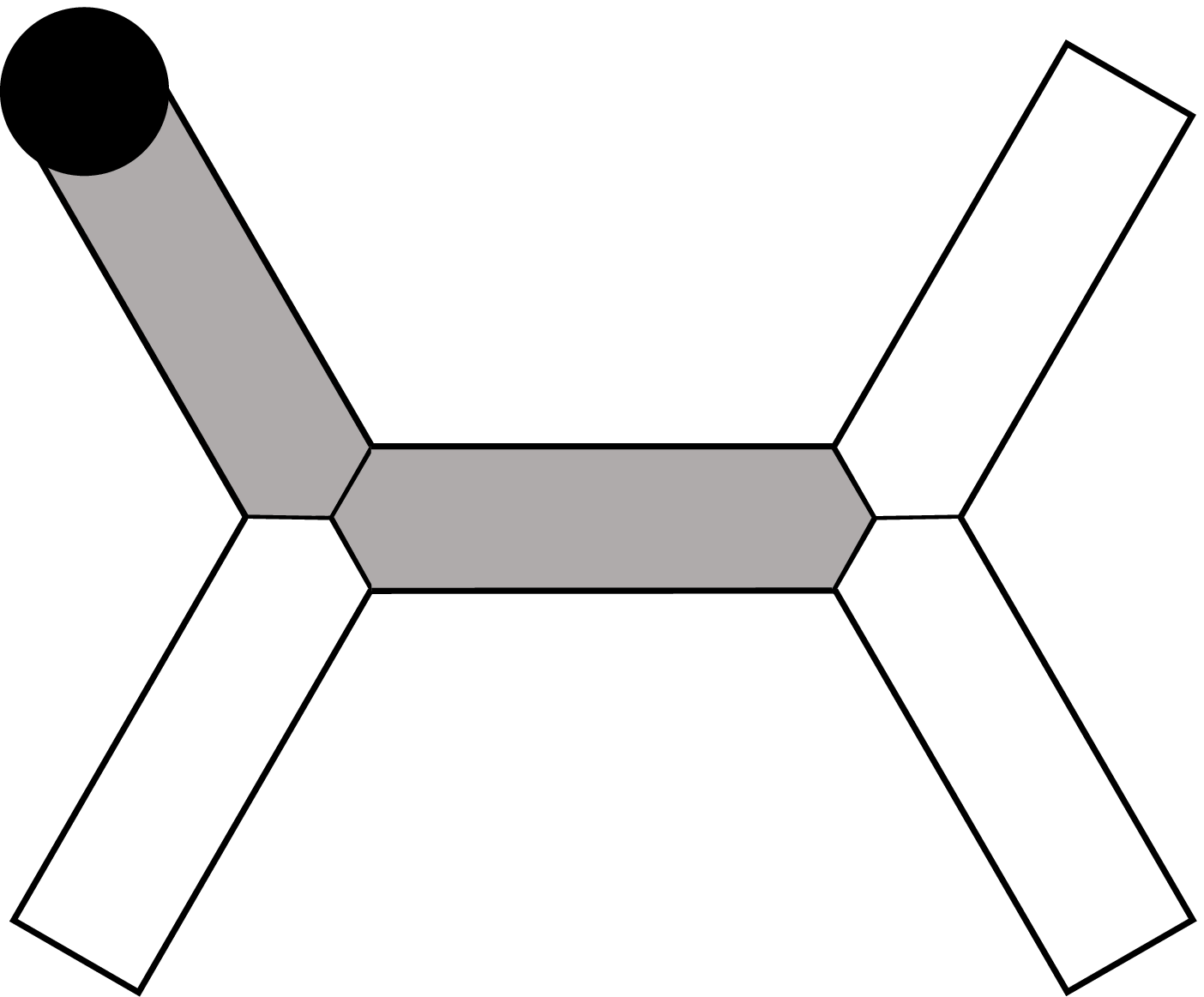}}
\end{align*}
\caption{Illustration of the two terms in $S^{(1)}_3$.
Propagators for massive and tachyonic fields are shaded in the figure.}
\label{corrected-cubic-vertex-figure}
\end{figure}

We further expand $S^{(1)}$ and $\delta^{(1)} \Psi$ in~$g$ as
\begin{align}
S^{(1)} & = \frac{1}{g} \, S^{(1)}_1
+S^{(1)}_2 +g \, S^{(1)}_3 +O(g^2) \,, \\
\delta^{(1)} \Psi & = \delta^{(1)}_0 \Psi +g \, \delta^{(1)}_1 \Psi +O(g^2) \,,
\end{align}
where $S^{(1)}_1$ and $S^{(1)}_2$ are given by
\begin{equation}
S^{(1)}_1 = \langle \, J(\Phi), \Psi \, \rangle \,, \qquad
S^{(1)}_2 = {}-\langle \, J(\Phi),
h \, ( \Psi \ast \Psi ) \, \rangle \,.
\end{equation}
Then the condition
\begin{equation}
\delta^{(0)} S^{(1)} +\delta^{(1)} S^{(0)} = 0
\end{equation}
for the gauge invariance at~$O(\kappa)$ can be expanded in~$g$ as
\begin{equation}
\begin{split}
& \frac{1}{g} \, \delta^{(0)}_0 S^{(1)}_1
+\biggl[ \, \delta^{(0)}_1 S^{(1)}_1 +\delta^{(0)}_0 S^{(1)}_2
+\delta^{(1)}_0 S^{(0)}_2 \, \biggr] \\
& {}+g \, \biggl[ \,
\delta^{(0)}_2 S^{(1)}_1 +\delta^{(0)}_1 S^{(1)}_2 +\delta^{(0)}_0 S^{(1)}_3
+\delta^{(1)}_1 S^{(0)}_2 +\delta^{(1)}_0 S^{(0)}_3 \, \biggr] 
+O(g^2) = 0 \,.
\end{split}
\end{equation}
We have so far shown that
\begin{equation}
\delta^{(0)}_0 S^{(1)}_1 = 0 \,, \qquad
\delta^{(0)}_1 S^{(1)}_1 +\delta^{(0)}_0 S^{(1)}_2
+\delta^{(1)}_0 S^{(0)}_2 = 0 \,,
\end{equation}
and the condition at the next order in $g$ is
\begin{equation}
\delta^{(0)}_2 S^{(1)}_1 +\delta^{(0)}_1 S^{(1)}_2 +\delta^{(0)}_0 S^{(1)}_3
+\delta^{(1)}_1 S^{(0)}_2 +\delta^{(1)}_0 S^{(0)}_3 = 0 \,.
\label{next-condition}
\end{equation}
As before, we infer $S^{(1)}_3$ based on Feynman diagrams as
\begin{equation}
\begin{split}
S^{(1)}_3 & = \langle \, J(\Phi) \,,
h \, ( \, h \, ( \, \Psi \ast \Psi \, ) \ast \Psi \, ) \, \rangle
+\langle \, J(\Phi) \,,
h \, ( \, \Psi \ast h \, ( \, \Psi \ast \Psi \, ) \, ) \, \rangle \,,
\end{split}
\end{equation}
and we indeed find that the condition~\eqref{next-condition}
can be satisfied by choosing $\delta^{(1)}_1 \Psi$ as
\begin{equation}
\begin{split}
\delta^{(1)}_1 \Psi & = P \, [ \, 
h \, ( \, \Psi \ast \Lambda \, ) \ast h \, J(\Phi)
-h \, ( \, \Lambda \ast \Psi \, ) \ast h \, J(\Phi)
-h \, J(\Phi) \ast h \, ( \, \Psi \ast \Lambda \, ) \\
& \qquad \quad
+h \, J(\Phi) \ast h \, ( \, \Lambda \ast \Psi  \, )
-\Psi \ast h \, ( \, h \, J(\Phi) \ast \Lambda \, )
+\Psi \ast h \, ( \, \Lambda \ast h \, J(\Phi) \, ) \\
& \qquad \quad
+h \, ( \, h \, J(\Phi) \ast \Lambda \, ) \ast \Psi
-h \, ( \, \Lambda \ast h \, J(\Phi) \, ) \ast \Psi
+\Lambda \ast h \, ( \, h \, J(\Phi) \ast \Psi \, ) \\
& \qquad \quad
+\Lambda \ast h \, ( \, \Psi \ast h \, J(\Phi) \, )
-h \, ( \, h \, J(\Phi) \ast \Psi \, ) \ast \Lambda
-h \, ( \, \Psi \ast h \, J(\Phi) \, ) \ast \Lambda \, ] \,.
\end{split}
\end{equation}
See figure~\ref{corrected-cubic-vertex-figure}
for the two terms in~$S^{(1)}_3$.
%%%
%%%

\subsubsection{Construction at $O(\kappa^2)$}

Let us next consider the gauge invariance at~$O(\kappa^2)$.
The condition at this order is
\begin{equation}
\delta^{(0)} S^{(2)} +\delta^{(1)} S^{(1)} +\delta^{(2)} S^{(0)} = 0 \,.
\label{condition-at-O(kappa^2)}
\end{equation}
We expand $S^{(2)}$ and $\delta^{(2)} \Psi$ in~$g$ as
\begin{align}
S^{(2)} & = \frac{1}{g} \, S^{(2)}_1 +S^{(2)}_2 +O(g) \,, \\
\delta^{(2)} \Psi & = \delta^{(2)}_0 \Psi +O(g) \,,
\end{align}
and the condition~\eqref{condition-at-O(kappa^2)} can be expanded as
\begin{align}
& \delta^{(0)} S^{(2)} +\delta^{(1)} S^{(1)} +\delta^{(2)} S^{(0)} \\
& = \frac{1}{g} \, \biggl[ \,
\delta^{(0)}_0 S^{(2)}_1 +\delta^{(1)}_0 S^{(1)}_1 \, \biggr]
+\biggl[ \, \delta^{(0)}_1 S^{(2)}_1 +\delta^{(0)}_0 S^{(2)}_2
+\delta^{(1)}_1 S^{(1)}_1 +\delta^{(1)}_0 S^{(1)}_2
+\delta^{(2)}_0 S^{(0)}_2 \, \biggr] +O(g) = 0 \,. \nonumber
\end{align}
We find that $\delta^{(1)}_0 S^{(1)}_1$ is nonvanishing,
\begin{equation}
\delta^{(1)}_0 S^{(1)}_1 \ne 0 \,,
\end{equation}
but the condition
\begin{equation}
\delta^{(0)}_0 S^{(2)}_1 +\delta^{(1)}_0 S^{(1)}_1 = 0
\end{equation}
can be satisfied with $S^{(2)}_1$ given by
\begin{equation}
S^{(2)}_1 = {}-\langle \, J(\Phi) \,,\,
h \, ( \, h \, J(\Phi) \ast \Psi \, ) \, \rangle
\end{equation}
in accord with the consideration based on Feynman diagrams.

At the next order in $g$, we learn from known terms
in the effective action and the gauge transformation that
\begin{equation}
\delta^{(0)}_1 S^{(2)}_1 +\delta^{(1)}_1 S^{(1)}_1 +\delta^{(1)}_0 S^{(1)}_2 \ne 0 \,.
\end{equation}
By explicit calculations we find that the condition
\begin{equation}
\delta^{(0)}_1 S^{(2)}_1 +\delta^{(0)}_0 S^{(2)}_2
+\delta^{(1)}_1 S^{(1)}_1 +\delta^{(1)}_0 S^{(1)}_2
+\delta^{(2)}_0 S^{(0)}_2 = 0
\end{equation}
is satisfied with $S^{(2)}_2$ and $\delta^{(2)}_0 \Psi$ given by
\begin{equation}
\begin{split}
S^{(2)}_2 & =
\langle \, h \, J(\Phi)
\ast h \, J(\Phi) \,,\,
h \, ( \, \Psi \ast \Psi \, ) \, \rangle
{}+\langle \, h \, J(\Phi)
\ast \Psi  \,,\,
h \, ( \, \Psi \ast h \, J(\Phi) \, ) \, \rangle \\
& \quad~
{}+\frac{1}{2} \, \langle \, h \, J(\Phi)
\ast \Psi  \,,\,
h \, ( \, h \, J(\Phi) \ast \Psi \, ) \, \rangle
{}+\frac{1}{2} \, \langle \, \Psi \ast h \, J(\Phi) \,,\,
h \, ( \, \Psi \ast h \, J(\Phi) \, ) \, \rangle
\end{split}
\end{equation}
and
\begin{equation}
\begin{split}
\delta^{(2)}_0 \Psi & = P \, [ \,
\Lambda \ast h \, ( \, h \, J(\Phi) \ast
h \, J(\Phi) \, )
{}-h \, ( \, h \, J(\Phi) \ast
h \, J(\Phi) \, ) \ast \Lambda \\
& \qquad \quad
{}-h \, ( \, \Lambda \ast
h \, J(\Phi) \, ) \ast h \, J(\Phi)
+h \, J(\Phi) \ast h \, ( \, \Lambda \ast h \, J(\Phi) \, ) \\
& \qquad \quad
{}-h \, J(\Phi) \ast h \, ( \, h \, J(\Phi) \ast \Lambda \, )
{}+h \, ( \, h \, J(\Phi) \ast \Lambda \, ) \ast h \, J(\Phi) \, ] \,.
\end{split}
\end{equation}

\subsubsection{Construction at $O(\kappa^3)$}

The condition for the gauge invariance at~$O(\kappa^3)$ is
\begin{equation}
\delta^{(0)} S^{(3)} +\delta^{(1)} S^{(2)}
+\delta^{(2)} S^{(1)} +\delta^{(3)} S^{(0)} = 0 \,.
\label{condition-at-O(kappa^3)}
\end{equation}
We expand $S^{(3)}$ in $g$ as
\begin{equation}
S^{(3)} = \frac{1}{g} \, S^{(3)}_1 +O(g^0) \,.
\end{equation}
Since $\delta^{(3)} \Psi$ is of $O(g^0)$,
\begin{equation}
\delta^{(3)} \Psi = O(g^0) \,,
\end{equation}
the condition~\eqref{condition-at-O(kappa^3)} can be expanded as
\begin{equation}
\begin{split}
& \delta^{(0)} S^{(3)} +\delta^{(1)} S^{(2)}
+\delta^{(2)} S^{(1)} +\delta^{(3)} S^{(0)}
= \frac{1}{g} \, \biggl[ \,
\delta^{(0)}_0 S^{(3)}_1 +\delta^{(1)}_0 S^{(2)}_1
+\delta^{(2)}_0 S^{(1)}_1 \, \biggr]
+O(g^0) = 0 \,.
\end{split}
\end{equation}
From known terms in the effective action and the gauge transformation
we find that
\begin{equation}
\delta^{(1)}_0 S^{(2)}_1
+\delta^{(2)}_0 S^{(1)}_1 \ne 0 \,.
\end{equation}
We determine $S^{(3)}_1$ such that
the condition
\begin{equation}
\delta^{(0)}_0 S^{(3)}_1
+\delta^{(1)}_0 S^{(2)}_1
+\delta^{(2)}_0 S^{(1)}_1 = 0
\end{equation}
is satisfied. Our result is as follows:
\begin{equation}
S^{(3)}_1 = \langle \, 
h \, J(\Phi) \ast h \, J(\Phi) \,,\, h \, ( \, \Psi \ast h \, J(\Phi) \, ) \, \rangle
+\langle \, 
h \, J(\Phi) \ast h \, J(\Phi) \,,\, h \, ( \, h \, J(\Phi) \ast \Psi \, ) \, \rangle \,.
\end{equation}

\subsection{Structure}

To summarize, we can organize the terms in the action we have constructed as follows:
\begin{equation}
\begin{split}
S & = \qquad \qquad \qquad~~
S^{(0)}_2 +g \, S^{(0)}_3 +g^2 \, S^{(0)}_4 +O(g^3) \\
& \quad~ {}+\kappa^{\phantom{1}} \, \biggl[ \,
\frac{1}{g} \, S^{(1)}_1 +S^{(1)}_2 +g \, S^{(1)}_3 +O(g^2) \, \biggr] \\
& \quad~ {}+\kappa^2 \, \biggl[ \,
\frac{1}{g} \, S^{(2)}_1 +S^{(2)}_2 +O(g) \, \biggr] \\
& \quad~ {}+\kappa^3 \, \biggl[ \,
\frac{1}{g} \, S^{(3)}_1 +O(g^0) \, \biggr] \\
& \quad~ +O(\kappa^4) \,.
\end{split}
\end{equation}
We can also organize the terms in the gauge transformation as follows:
\begin{equation}
\begin{split}
\delta_\Lambda \Psi & = \qquad \quad
\delta^{(0)}_0 \Psi +g \, \delta^{(0)}_1 \Psi +g^2 \delta^{(0)}_2 \Psi +O(g^3) \\
& \quad~ +\kappa^{\phantom{1}} \, \biggl[ \,
\delta^{(1)}_0 \Psi +g \, \delta^{(1)}_1 \Psi +O(g^2) \, \biggr] \\
& \quad~ {}+\kappa^2 \, \biggl[ \,
\delta^{(2)}_0 \Psi +O(g) \, \biggr] \\
& \quad~ +O(\kappa^3) \,.
\end{split}
\end{equation}
When we compare $S^{(n)}_m$ and $\delta^{(n)}_{m-2} \Psi$ with $m \ge 2$,
we recognize some similarities.
In fact, we translate the result in~\cite{Sen:2016qap}
into our setup to learn that
the action $S^{(0)}$
before introducing the source term
and the gauge transformation $\delta^{(0)} \Psi$
can be written in terms of the same set of multi-string products
and this is reflected, for example, in a similarity
between $S^{(0)}_4$ and $\delta^{(0)}_2 \Psi$.
We will find in the following sections that
the action $S$
after introducing the source term
and the gauge transformation $\delta_\Lambda \Psi$
are also written in terms of the same set of multi-string products.
This explains, for example, the similarity between $S^{(1)}_2$ and $\delta^{(1)}_0 \Psi$.

The multi-string products before introducing the source term
satisfy a set of relations called $A_\infty$ relations.
We say that an action has an $A_\infty$ structure
when it is written in terms of multi-string products
which satisfy the $A_\infty$ relations.
While the expressions for the action and the gauge transformation
become very complicated at higher orders,
there is an efficient way to describe the $A_\infty$ structure 
which provides us with control over all-order expressions.

The multi-string products after introducing the source term
contain a zero-string product,
and we will show that they satisfy a set of relations called weak $A_\infty$ relations.
We say that an action has a weak $A_\infty$ structure
when it is written in terms of multi-string products
which satisfy the weak $A_\infty$ relations.
There is also an efficient way to describe the weak $A_\infty$ structure,
and we will use it to construct the action and the gauge transformation
after introducing the source term to all orders.

\section{Weak $A_\infty$ structure}
\label{section-6}
\setcounter{equation}{0}

In this section we will show that the action and the gauge transformation
constructed in the preceding section can be expressed
in terms of the same set of multi-string products
which satisfy the weak $A_\infty$ relations.
While the $A_\infty$ structure plays an important role
in recent research on open string field theory,
the weak $A_\infty$ structure is less familiar.
We therefore begin with the basics of the $A_\infty$ structure
and then motivate the generalization to the weak $A_\infty$ structure.

\subsection{$A_\infty$ structure up to quartic interactions}

Let us consider an action of the form
\begin{equation}
S = {}-\frac{1}{2} \, \langle \, \Psi, Q \Psi \, \rangle
-\frac{g}{3} \, \langle \, \Psi, V_2 ( \Psi, \Psi ) \, \rangle
-\frac{g^2}{4} \, \langle \, \Psi, V_3 ( \Psi, \Psi, \Psi ) \, \rangle +O(g^3) \,,
\label{action}
\end{equation}
where $V_2 ( A_1, A_2 )$ defined for a pair of string fields $A_1$ and $A_2$
is a two-string product
and $V_3 ( A_1, A_2, A_3 )$ defined for three string fields $A_1$, $A_2$, and $A_3$
is a three-string product.
The Grassmann parity of $V_2 ( A_1, A_2 )$ is $\epsilon( A_1 ) +\epsilon( A_2 )$ mod $2$
and the Grassmann parity of $V_3 ( A_1, A_2, A_3 )$
is $\epsilon( A_1 ) +\epsilon( A_2 ) +\epsilon( A_3 ) +1$ mod $2$,
where $\epsilon ( A_i )$ is the Grassmann parity of $A_i$ mod $2$ for $i = 1, 2, 3$.
These string products are assumed to have the following cyclic properties:
\begin{align}
\langle \, A_1, V_2 ( A_2, A_3 ) \, \rangle
& = (-1)^{A_1 (A_2+A_3)} \langle \, A_2, V_2 ( A_3, A_1 ) \, \rangle \,, \\
\langle \, A_1, V_3 ( A_2, A_3, A_4 ) \, \rangle
& = {}-(-1)^{A_1+A_2+A_1 (A_2+A_3+A_4)} \langle \, A_2, V_3 ( A_3, A_4, A_1 ) \, \rangle \,.
\end{align}
These are equivalently described as
\begin{align}
\langle \, A_1, V_2 ( A_2, A_3 ) \, \rangle
& = \langle \, V_2 ( A_1, A_2 ), A_3 \, \rangle \,,
\label{two-string-cyclicity}
\\
\langle \, A_1, V_3 ( A_2, A_3, A_4 ) \, \rangle
& = {}-(-1)^{A_1} \langle \, V_3 ( A_1, A_2, A_3 ), A_4 \, \rangle \,.
\label{three-string-cyclicity}
\end{align}
The variation of the action~\eqref{action} is then given by
\begin{equation}
\delta S = {}-\, \langle \, \delta \Psi, Q \Psi \, \rangle
-g \, \langle \, \delta \Psi, V_2 ( \Psi, \Psi ) \, \rangle
-g^2 \, \langle \, \delta \Psi, V_3 ( \Psi, \Psi, \Psi ) \, \rangle +O(g^3) \,.
\end{equation}
Under the gauge transformation given by
\begin{equation}
\begin{split}
\delta_\Lambda \Psi & = Q \Lambda
+g \, ( \, V_2 ( \Psi, \Lambda ) -V_2 ( \Lambda, \Psi ) \, ) \\
& \quad~
+g^2 \, ( \, V_3 ( \Psi, \Psi, \Lambda )
-V_3 ( \Psi, \Lambda, \Psi )
+V_3 ( \Lambda, \Psi, \Psi ) \, ) +O(g^3) \,,
\end{split}
\end{equation}
the action~\eqref{action} is invariant up to $O(g^3)$,
\begin{equation}
\delta_\Lambda S = O(g^3) \,,
\end{equation}
if the following relations hold:
\begin{align}
& Q^2 = 0 \,, \\
& Q \, V_2 ( A_1, A_2 ) -V_2 ( Q A_1, A_2 ) -(-1)^{A_1} V_2 ( A_1, Q A_2 ) = 0 \,,
\label{A_infinity-2} \\
& Q \, V_3 ( A_1, A_2, A_3 )
+V_3 ( Q A_1, A_2, A_3 )
+(-1)^{A_1} V_3 ( A_1, Q A_2, A_3 )
+(-1)^{A_1+A_2} V_3 ( A_1, A_2, Q A_3 ) \nonumber \\
& -V_2 ( V_2 ( A_1, A_2 ), A_3 )
+V_2 ( A_1, V_2 ( A_2, A_3 ) ) = 0 \,.
\label{A_infinity-3}
\end{align}
These relations can be extended to all orders in $g$,
and they are called $A_\infty$ relations.
As we mentioned before,
we say that an action has an $A_\infty$ structure
when it is written in terms of multi-string products
which satisfy the $A_\infty$ relations.

\subsection{$A_\infty$ structure in the low energy}

The action~\eqref{open-bosonic-string-field-theory-action}
of open bosonic string field theory has an $A_\infty$ structure
up to $O(g^3)$ under the identification
\begin{equation}
V_2 ( A_1, A_2 ) = A_1 \ast A_2 \,, \qquad
V_3 ( A_1, A_2, A_3 ) = 0 \,.
\end{equation}
This two-string product has the cyclic property~\eqref{two-string-cyclicity},
and the $A_\infty$ relations~\eqref{A_infinity-2} and~\eqref{A_infinity-3}
are satisfied with the vanishing three-string product.
In fact, the action~\eqref{open-bosonic-string-field-theory-action}
has an $A_\infty$ structure to all orders, as we will see later.

The observation in~\cite{Sen:2016qap}
for the gauge invariance of the low-energy effective action 
in the context of closed string field theory
can be translated into our setup,
and it implies that
the action~$S^{(0)}$ in the low energy~\eqref{low-energy-action}
also has an $A_\infty$ structure.
In fact, the action~$S^{(0)}$ up to $O(g^3)$
has an $A_\infty$ structure under the identification
\begin{align}
V_2 \, ( A_1, A_2 )
& = P \, ( A_1 \ast A_2 ) \,,
\label{V_2} \\
V_3 \, ( A_1, A_2, A_3 )
& = (-1)^{A_1} P \, ( \, A_1 \ast h \, ( A_2 \ast A_3 ) \, )
-P \, ( \, h \, ( A_1 \ast A_2 ) \ast A_3 \, ) \,.
\label{V_3}
\end{align}
These string products have the cyclic properties~\eqref{two-string-cyclicity}
and~\eqref{three-string-cyclicity}
for string fields projected onto the massless sector satisfying
\begin{equation}
P A_i = A_i
\end{equation}
for $i=1, 2, 3$.
We can also confirm that the $A_\infty$ relations~\eqref{A_infinity-2} and~\eqref{A_infinity-3} are satisfied.

\subsection{The BRST operator as a one-string product}

As a step to the generalization to the weak $A_\infty$ structure,
let us slightly change the notation for the $A_\infty$ structure.
The action of the BRST operator can be regarded as a one-string product.
We define the one-string product $V_1 ( A_1 )$ for a string field $A_1$ by
\begin{equation}
V_1 ( A_1 ) = Q A_1 \,.
\end{equation}
The Grassmann parity of $V_1 ( A_1 )$ is then $\epsilon( A_1 ) +1$ mod $2$.
The BRST operator is BPZ odd,
\begin{equation}
\langle \, Q A_1, A_2 \, \rangle
= {}-(-1)^{A_1} \langle \, A_1, QA_2 \, \rangle \,,
\end{equation}
and this is translated into the following cyclic property
of the one-string product:
\begin{equation}
\langle \, A_1, V_1 ( A_2 ) \, \rangle
= {}-(-1)^{A_1+A_2+A_1 A_2} \langle \, A_2, V_1 ( A_1 ) \, \rangle \,,
\end{equation}
which is equivalently described as
\begin{equation}
\langle \, A_1, V_1 ( A_2 ) \, \rangle
= {}-(-1)^{A_1} \langle \, V_1 ( A_1 ), A_2 \, \rangle \,.
\end{equation}
The action~\eqref{action} is then written as
\begin{equation}
S = {}-\frac{1}{2} \, \langle \, \Psi, V_1 ( \Psi ) \, \rangle
-\frac{g}{3} \, \langle \, \Psi, V_2 ( \Psi, \Psi ) \, \rangle
-\frac{g^2}{4} \, \langle \, \Psi, V_3 ( \Psi, \Psi, \Psi ) \, \rangle +O(g^3) \,,
\end{equation}
and the variation of the action~\eqref{action} is
\begin{equation}
\delta S = {}-\, \langle \, \delta \Psi, V_1 ( \Psi ) \, \rangle
-g \, \langle \, \delta \Psi, V_2 ( \Psi, \Psi ) \, \rangle
-g^2 \, \langle \, \delta \Psi, V_3 ( \Psi, \Psi, \Psi ) \, \rangle +O(g^3) \,.
\end{equation}
The $A_\infty$ relations up to $O(g^3)$ can also be expressed as
\begin{align}
& V_1 ( V_1 ( A_1 ) ) = 0 \,, \\
& V_1 ( V_2 ( A_1, A_2 ) )
-V_2 ( V_1 ( A_1 ), A_2 ) -(-1)^{A_1} V_2 ( A_1, V_1 ( A_2 ) ) = 0 \,, \\
& V_1 ( V_3 ( A_1, A_2, A_3 ) )
+V_3 ( V_1 ( A_1 ), A_2, A_3 )
+(-1)^{A_1} V_3 ( A_1, V_1 ( A_2 ), A_3 ) \nonumber \\
& +(-1)^{A_1+A_2} V_3 ( A_1, A_2, V_1 ( A_3 ) )
-V_2 ( V_2 ( A_1, A_2 ), A_3 )
+V_2 ( A_1, V_2 ( A_2, A_3 ) ) = 0 \,.
\end{align}

\subsection{Action with linear terms}

Let us now consider an action including a term linear in $\Psi$.
We write it in the following form:
\begin{equation}
\begin{split}
S & = {}-\frac{1}{g} \, \langle \, \Psi, V_0 \, \rangle
-\frac{1}{2} \, \langle \, \Psi, V_1 ( \Psi ) \, \rangle
-\frac{g}{3} \, \langle \, \Psi, V_2 ( \Psi, \Psi ) \, \rangle
-\frac{g^2}{4} \, \langle \, \Psi, V_3 ( \Psi, \Psi, \Psi ) \, \rangle \\
& \quad~
{}-\frac{g^3}{5} \, \langle \, \Psi, V_4 ( \Psi, \Psi, \Psi, \Psi ) \, \rangle
+O(g^4) \,,
\end{split}
\end{equation}
where $V_0$ is a Grassmann-even string field of ghost number $2$
and $V_4 ( A_1, A_2, A_3, A_4 )$ is a four-string product
defined for four string fields $A_1$, $A_2$, $A_3$, and $A_4$.
The Grassmann parity of $V_4 ( A_1, A_2, A_3, A_4 )$
is $\epsilon( A_1 ) +\epsilon( A_2 ) +\epsilon( A_3 ) +\epsilon( A_4 )$ mod $2$,
where $\epsilon ( A_i )$ is the Grassmann parity of $A_i$ mod $2$ for $i = 1, 2, 3, 4$.
The four-string product is assumed to have the following cyclic property:
\begin{equation}
\langle \, A_1, V_4 ( A_2, A_3, A_4, A_5 ) \, \rangle
= (-1)^{A_1 (A_2+A_3+A_4+A_5)}
\langle \, A_2, V_4 ( A_3, A_4, A_5, A_1 ) \, \rangle \,,
\end{equation}
which is equivalently described as
\begin{equation}
\langle \, A_1, V_4 ( A_2, A_3, A_4, A_5 ) \, \rangle
= \langle \, V_4 ( A_1, A_2, A_3, A_4 ), A_5 \, \rangle \,.
\label{four-string-cyclicity}
\end{equation}
We can think of $V_0$ as a zero-string product.
Since it is a Grassmann-even string field, we have
\begin{equation}
\langle \, A_1, V_0 \, \rangle = \langle \, V_0, A_1 \, \rangle
\end{equation}
for any string field $A_1$,
and this can be regarded as the cyclic property of the zero string product.
The variation of the action is given by
\begin{equation}
\begin{split}
\delta S & = {}-\frac{1}{g} \, \langle \, \delta \Psi, V_0 \, \rangle
-\langle \, \delta \Psi, V_1 ( \Psi ) \, \rangle
-g \, \langle \, \delta \Psi, V_2 ( \Psi, \Psi ) \, \rangle
-g^2 \, \langle \, \delta \Psi, V_3 ( \Psi, \Psi, \Psi ) \, \rangle \\
& \quad~
{}-g^3 \, \langle \, \delta \Psi, V_4 ( \Psi, \Psi, \Psi, \Psi ) \, \rangle
+O(g^4) \,.
\end{split}
\end{equation}
We can show that this action is invariant up to $O(g^3)$
under the gauge transformation given by
\begin{align}
\delta_\Lambda \Psi & = V_1 ( \Lambda )
+g \, ( \, V_2 ( \Psi, \Lambda ) -V_2 ( \Lambda, \Psi ) \, ) \nonumber \\
& \quad~
+g^2 \, ( \, V_3 ( \Psi, \Psi, \Lambda )
-V_3 ( \Psi, \Lambda, \Psi )
+V_3 ( \Lambda, \Psi, \Psi ) \, ) \\
& \quad~
+g^3 \, ( \, V_4 ( \Psi, \Psi, \Psi, \Lambda )
-V_4 ( \Psi, \Psi, \Lambda, \Psi )
+V_4 ( \Psi, \Lambda, \Psi, \Psi )
-V_4 ( \Lambda, \Psi, \Psi, \Psi ) \, ) +O(g^4) \nonumber
\end{align}
if the multi-string products $V_0$, $V_1$, $V_2$, $V_3$, and $V_4$
satisfy the following relations:
\begin{align}
& V_1 ( V_0 ) = 0 \,,
\label{weak-A_infinity-1} \\
& V_1 ( V_1 ( A_1 ) )
-V_2 ( V_0, A_1 ) +V_2 ( A_1, V_0 ) = 0 \,,
\label{weak-A_infinity-2} \\
& V_1 ( V_2 ( A_1, A_2 ) )
-V_2 ( V_1 ( A_1 ), A_2 ) -(-1)^{A_1} V_2 ( A_1, V_1 ( A_2 ) ) \nonumber \\
& +V_3 ( V_0, A_1, A_2 ) -V_3 ( A_1, V_0, A_2 ) +V_3 ( A_1, A_2, V_0 ) = 0 \,,
\label{weak-A_infinity-3} \\
& V_1 ( V_3 ( A_1, A_2, A_3 ) )
+V_3 ( V_1 ( A_1 ), A_2, A_3 )
+(-1)^{A_1} V_3 ( A_1, V_1 ( A_2 ), A_3 ) \nonumber \\
& +(-1)^{A_1+A_2} V_3 ( A_1, A_2, V_1 ( A_3 ) )
-V_2 ( V_2 ( A_1, A_2 ), A_3 )
+V_2 ( A_1, V_2 ( A_2, A_3 ) ) \nonumber \\
& {}-V_4 ( V_0, A_1, A_2, A_3 ) +V_4 ( A_1, V_0, A_2, A_3 )
-V_4 ( A_1, A_2, V_0, A_3 ) +V_4 ( A_1, A_2, A_3, V_0 ) = 0 \,.
\label{weak-A_infinity-4}
\end{align}
These relations can be extended to all orders in $g$,
and they are called weak $A_\infty$ relations.
As we mentioned before,
we say that an action has a weak $A_\infty$ structure
when the action is written in terms of the multi-string products
which satisfy the weak $A_\infty$ relations.

Note that the relation~\eqref{weak-A_infinity-2} implies
that the one-string product does not square to zero
when $V_2 ( V_0, A_1 ) \ne V_2 ( A_1, V_0 ) $.
This indicates that the one-string product is in general
different from the BRST operator.

\subsection{Weak $A_\infty$ structure in the low energy}

The action~\eqref{action+source} has a weak $A_\infty$ structure
up to $O(g^2)$ under the identification
\begin{equation}
\begin{split}
& V_0 = {}-\kappa \, J(\Phi) \,, \quad
V_1 ( A_1 ) = Q A_1 \,, \quad
V_2 ( A_1, A_2 ) = A_1 \ast A_2 \,, \\
& V_3 ( A_1, A_2, A_3 ) = 0 \,, \quad
V_4 ( A_1, A_2, A_3, A_4 ) = 0 \,.
\end{split}
\end{equation}
In fact, the action~\eqref{action+source} has a weak $A_\infty$ structure
to all orders, as we will see later.

One of the main observations in this paper
is that the action of open bosonic string field theory with the source term
in the low-energy limit also has a weak $A_\infty$ structure.
The action and the gauge transformation
we constructed in section~\ref{section-5}
can be expressed in terms of the same set of multi-string products.
For example, the term $S^{(1)}_2$ in the action
and $\delta^{(1)}_0 \Psi$ in the gauge transformation given by
\begin{equation}
S^{(1)}_2 = {}-\langle \, J(\Phi),
h \, ( \Psi \ast \Psi ) \, \rangle \,, \qquad
\delta^{(1)}_0 \Psi = P \, [ \,
h  \, J(\Phi) \ast \Lambda -\Lambda \ast h \, J(\Phi) \, ]
\end{equation}
can be written as
\begin{equation}
S^{(1)}_2 = {}-\frac{1}{2} \, \langle \, \Psi, V_1^{(1)} ( \Psi ) \, \rangle \,, \qquad
\delta^{(1)}_0 \Psi = V_1^{(1)} ( \Lambda )
\end{equation}
in terms of the one-string product $V_1^{(1)} ( A_1 )$ given by
\begin{equation}
V_1^{(1)} ( A_1 ) = P \, [ \,
h \, J(\Phi) \ast A_1
-(-1)^{A_1} A_1 \ast h \, J(\Phi) \, ] \,.
\end{equation}
All the terms in the action and the gauge transformation
we constructed in section~\ref{section-5} can be written
using the following multi-string products $V_0$, $V_1$,
$V_2$, and $V_3$.
The zero-string product $V_0$ is given by
\begin{equation}
V_0 = \kappa \, V_0^{(1)} +\kappa^2 \, V_0^{(2)} +\kappa^3 \, V_0^{(3)} +O(\kappa^4) \,,
\end{equation}
where
\begin{align}
V_0^{(1)} & = {}-P J(\Phi) \,, \\
V_0^{(2)} & = P \, [ \,
h \, J(\Phi) \ast h \, J(\Phi) \, ] \,, \\
V_0^{(3)} & = {}-P \, [ \,
h \, J(\Phi)
\ast h \, ( \,
h \, J(\Phi) \ast h \, J(\Phi) \, )
+h \, ( \, h \, J(\Phi) \ast h \, J(\Phi) \, )
\ast h \, J(\Phi) \, ] \,.
\end{align}
The one-string product $V_1$ is given by
\begin{equation}
V_1 ( A_1 ) = V_1^{(0)} ( A_1 ) +\kappa \, V_1^{(1)} ( A_1 )
+\kappa^2 \, V_2^{(2)} ( A_1 ) +O(\kappa^3) \,,
\end{equation}
where
\begin{align}
V_1^{(0)} ( A_1 ) & = Q A_1 \,, \\
V_1^{(1)} ( A_1 ) & = P \, [ \,
h \, J(\Phi) \ast A_1
-(-1)^{A_1} A_1 \ast h \, J(\Phi) \, ] \,, \\
V_1^{(2)} ( A_1 ) & = P \, [ \,
(-1)^{A_1} A_1 \ast h \, ( \, h \, J(\Phi) \ast
h \, J(\Phi) \, )
{}-h \, ( \, h \, J(\Phi) \ast
h \, J(\Phi) \, ) \ast A_1 \nonumber \\
& \qquad \quad
{}-h \, ( \, A_1 \ast
h \, J(\Phi) \, ) \ast h \, J(\Phi)
{}+(-1)^{A_1} h \, J(\Phi) \ast h \, ( \, A_1 \ast h \, J(\Phi) \, ) \nonumber \\
& \qquad \quad
{}-h \, J(\Phi) \ast h \, ( \, h \, J(\Phi) \ast A_1 \, )
{}+(-1)^{A_1} h \, ( \, h \, J(\Phi) \ast A_1 \, ) \ast h \, J(\Phi) \, ] \,.
\end{align}
The two-string product $V_2$ is given by
\begin{equation}
V_2 ( A_1, A_2 ) = V_2^{(0)} ( A_1, A_2 ) +\kappa \, V_2^{(1)} ( A_1, A_2 )
+O(\kappa^2) \,,
\end{equation}
where
\begin{align}
V_2^{(0)} ( A_1, A_2 ) & = P \, [ \, A_1 \ast A_2 \, ] \,, \\
V_2^{(1)} ( A_1, A_2 ) & = P \, [ \,
{}-h \, ( \, h \, J(\Phi) \ast A_1 \, ) \ast A_2
+(-1)^{A_1} h \, ( \, A_1 \ast h \, J(\Phi) \, ) \ast A_2 \nonumber \\
& \qquad \quad
{}-(-1)^{A_1+A_2} h \, ( \, A_1 \ast A_2 \, ) \ast h \, J(\Phi)
-h \, J(\Phi) \ast h \, ( \, A_1 \ast A_2 \, ) \nonumber \\
& \qquad \quad
{}-A_1 \ast h \, ( \, h \, J(\Phi) \ast A_2 \, )
+(-1)^{A_2} A_1 \ast h \, ( \, A_2 \ast h \, J(\Phi) \, ) \, ] \,.
\end{align}
The three-string product $V_3$ is given by
\begin{equation}
V_3 ( A_1, A_2, A_3 ) = V_3^{(0)} ( A_1, A_2, A_3 ) +O(\kappa) \,,
\end{equation}
where
\begin{equation}
V_3^{(0)} ( A_1, A_2, A_3 )
= P \, [ \, (-1)^{A_1} A_1 \ast h \, ( A_2 \ast A_3 )
-h \, ( A_1 \ast A_2 ) \ast A_3 \, ] \,.
\end{equation}

For the four-string product $V_4$,
we do not need its explicit form,
but obviously it of $O(\kappa^0)$:
\begin{equation}
V_4 ( A_1, A_2, A_3, A_4 ) = O(\kappa^0) \,.
\end{equation}
Thus the four-string product is of $O(\kappa)$
when one of the four string fields $A_1$, $A_2$, $A_3$, and $A_4$ is $V_0$.
We can show that the weak $A_\infty$ relations~\eqref{weak-A_infinity-1},
\eqref{weak-A_infinity-2}, \eqref{weak-A_infinity-3}
and~\eqref{weak-A_infinity-4} are satisfied
with these multi-string products up to the following orders:
\begin{align}
& V_1 ( V_0 ) = O(\kappa^4) \,, \\
& V_1 ( V_1 ( A_1 ) )
-V_2 ( V_0, A_1 ) +V_2 ( A_1, V_0 ) = O(\kappa^3) \,, \\
& V_1 ( V_2 ( A_1, A_2 ) )
-V_2 ( V_1 ( A_1 ), A_2 ) -(-1)^{A_1} V_2 ( A_1, V_1 ( A_2 ) ) \nonumber \\
& +V_3 ( V_0, A_1, A_2 ) -V_3 ( A_1, V_0, A_2 ) +V_3 ( A_1, A_2, V_0 ) = O(\kappa^2) \,, \\
& V_1 ( V_3 ( A_1, A_2, A_3 ) )
+V_3 ( V_1 ( A_1 ), A_2, A_3 )
+(-1)^{A_1} V_3 ( A_1, V_1 ( A_2 ), A_3 ) \nonumber \\
& +(-1)^{A_1+A_2} V_3 ( A_1, A_2, V_1 ( A_3 ) )
-V_2 ( V_2 ( A_1, A_2 ), A_3 )
+V_2 ( A_1, V_2 ( A_2, A_3 ) ) \\
& {}-V_4 ( V_0, A_1, A_2, A_3 ) +V_4 ( A_1, V_0, A_2, A_3 )
-V_4 ( A_1, A_2, V_0, A_3 ) +V_4 ( A_1, A_2, A_3, V_0 ) = O(\kappa) \,. \nonumber
\end{align}
We can also confirm that the cyclic properties
\begin{equation}
\begin{split}
\langle \, A_1, V_0^{(1)} \, \rangle & = \langle \, V_0^{(1)}, A_1 \, \rangle \,, \\
\langle \, A_1, V_0^{(2)} \, \rangle & = \langle \, V_0^{(2)}, A_1 \, \rangle \,, \\
\langle \, A_1, V_0^{(3)} \, \rangle & = \langle \, V_0^{(3)}, A_1 \, \rangle \,, \\
\langle \, A_1, V_1^{(1)} ( A_2 ) \, \rangle
& = {}-(-1)^{A_1} \langle \, V_1^{(1)} ( A_1 ), A_2 \, \rangle \,, \\
\langle \, A_1, V_1^{(2)} ( A_2 ) \, \rangle
& = {}-(-1)^{A_1} \langle \, V_1^{(2)} ( A_1 ), A_2 \, \rangle \,, \\
\langle \, A_1, V_2^{(1)} ( A_2, A_3 ) \, \rangle
& = \langle \, V_2^{(1)} ( A_1, A_2 ), A_3 \, \rangle
\end{split}
\end{equation}
hold for string fields projected onto the massless sector satisfying
\begin{equation}
P A_i = A_i
\end{equation}
with $i=1, 2, 3$.

\section{Coalgebra representation}
\label{section-7}
\setcounter{equation}{0}

Explicit expressions for the weak $A_\infty$ relations
are complicated at higher orders.
The fourth relation~\eqref{weak-A_infinity-4} is already lengthy,
and we may hesitate to write down the fifth relation explicitly.
Fortunately, there is an efficient way called coalgebra representation
to describe the weak $A_\infty$ structure.
While it is a straightforward generalization
of the description for the $A_\infty$ structure
which has been used frequently in recent research
on open string field theory,\footnote{
The coalgebra representation of the $A_\infty$ structure
is explained in detail, for example, in appendix A of~\cite{Erler:2015uba}.
We mostly follow the conventions used in this appendix.
}
we seldom find
an explicit description for the weak $A_\infty$ structure
in the literature
so we describe it in this section.
We will simplify the description of the weak $A_\infty$ structure
in three steps.
Each of the first three subsections corresponds to one of the three steps.

Another ingredient which is necessary in our context
is the projection onto a subspace of the Hilbert space.
While it is intuitively understood in~\cite{Sen:2016qap}
using Feynman diagrams,
one advantage of considering open string field theory
based on the star product
compared to closed string field theory
is that various expressions can be represented
very explicitly.
In \S\ref{projection-subsection}
we present an algebraic framework
to incorporate the projection.

\subsection{Degree}

The first step to simplify the description of the weak $A_\infty$ structure
is to  introduce {\it degree}
for a string field $A$ denoted by ${\rm deg} (A)$. It is defined by
\begin{equation}
{\rm deg} (A) = \epsilon (A) +1 \mod 2 \,,
\end{equation}
where $\epsilon (A)$ is the Grassmann parity of $A$.
We define $\omega ( A_1, A_2 )$, $M_0$, $M_1 ( A_1 )$,
$M_2 ( A_1, A_2 )$,
$M_3 ( A_1, A_2, A_3 )$,
and $M_4 ( A_1, A_2, A_3, A_4 )$ by
\begin{equation}
\begin{split}
\omega ( A_1, A_2 ) & = (-1)^{{\rm deg} (A_1)} \langle \, A_1, A_2 \, \rangle \,, \\
M_0 & = V_0 \,, \\
M_1 ( A_1 ) & = V_1 ( A_1 ) \,, \\
M_2 ( A_1, A_2 ) & = (-1)^{{\rm deg} (A_1)} \, V_2 ( A_1, A_2 ) \,, \\
M_3 ( A_1, A_2, A_3 ) & = (-1)^{{\rm deg} (A_2)} \, V_3 ( A_1, A_2, A_3 ) \,, \\
M_4 ( A_1, A_2, A_3, A_4 ) &
= (-1)^{{\rm deg} (A_1)+{\rm deg} (A_3)} \, V_4 ( A_1, A_2, A_3, A_4 ) \,.
\end{split}
\label{V-to-M}
\end{equation}
The inner product $\omega ( A_1, A_2 )$ is graded antisymmetric
with respect to degree:
\begin{equation}
\omega ( A_1, A_2 )
= {}-(-1)^{{\rm deg} (A_1) \, {\rm deg} (A_2)} \, \omega ( A_2, A_1 ) \,.
\end{equation}
We have
\begin{equation}
\begin{split}
{\rm deg} ( M_0 )
& = 1 \,, \\
{\rm deg} ( M_1 (A_1) )
& = {\rm deg} (A_1) +1 \,, \\
{\rm deg} ( M_2 ( A_1, A_2 ) )
& = {\rm deg} (A_1)+{\rm deg} (A_2) +1 \,, \\
{\rm deg} ( M_3 ( A_1, A_2, A_3 ) )
& = {\rm deg} (A_1)+{\rm deg} (A_2)+{\rm deg} (A_3) +1 \,, \\
{\rm deg} ( M_4 ( A_1, A_2, A_3, A_4 ) )
& = {\rm deg} (A_1)+{\rm deg} (A_2)+{\rm deg} (A_3) +{\rm deg} (A_4) +1 \,,
\end{split}
\end{equation}
and we say that $M_0$, $M_1$, $M_2$, $M_3$, and $M_4$ are degree odd.
The cyclic properties of the multi-string products are translated into
\begin{equation}
\begin{split}
\omega ( A_1, M_0 )
& = {}-(-1)^{{\rm deg} (A_1)} \, \omega ( M_0, A_1 ) \,, \\
\omega ( A_1, M_1 (A_2) )
& = {}-(-1)^{{\rm deg} (A_1)} \, \omega ( M_1 (A_1), A_2 ) \,, \\
\omega ( A_1, M_2 ( A_2, A_3 ) )
& = {}-(-1)^{{\rm deg} (A_1)} \, \omega ( M_2 ( A_1, A_2 ), A_3 ) \,, \\
\omega ( A_1, M_3 ( A_2, A_3, A_4 ) )
& = {}-(-1)^{{\rm deg} (A_1)} \, \omega ( M_3 ( A_1, A_2, A_3 ), A_4 ) \,, \\
\omega ( A_1, M_4 ( A_2, A_3, A_4, A_5 ) )
& = {}-(-1)^{{\rm deg} (A_1)} \, \omega ( M_4 ( A_1, A_2, A_3, A_4 ), A_5 ) \,,
\end{split}
\end{equation}
and the weak $A_\infty$ relations~\eqref{weak-A_infinity-1},
\eqref{weak-A_infinity-2}, \eqref{weak-A_infinity-3},
and \eqref{weak-A_infinity-4} are written as
\begin{align}
& M_1 ( M_0 ) = 0 \,,
\label{weak-M-1} \\
& M_1 ( M_1 ( A_1 ) ) +M_2 ( M_0, A_1 ) +(-1)^{{\rm deg} (A_1)} M_2 ( A_1, M_0 ) = 0 \,,
\label{weak-M-2} \\
& M_1 ( M_2 ( A_1, A_2 ) )
+M_2 ( M_1 ( A_1 ), A_2 ) +(-1)^{{\rm deg} (A_1)} M_2 ( A_1, M_1 ( A_2 ) )
\label{weak-M-3} \\
& +M_3 ( M_0, A_1, A_2 ) +(-1)^{{\rm deg} (A_1)} M_3 ( A_1, M_0, A_2 )
+(-1)^{{\rm deg} (A_1) +{\rm deg} (A_2)} M_3 ( A_1, A_2, M_0 ) = 0 \,, \nonumber \\
& M_1 ( M_3 ( A_1, A_2, A_3 ) )
+M_3 ( M_1 ( A_1 ), A_2, A_3 ) \nonumber \\
& +(-1)^{{\rm deg} (A_1)} M_3 ( A_1, M_1 ( A_2 ), A_3 )
+(-1)^{{\rm deg} (A_1) +{\rm deg} (A_2)} M_3 ( A_1, A_2, M_1 ( A_3 ) ) \nonumber \\
& +M_2 ( M_2 ( A_1, A_2 ), A_3 )
+(-1)^{{\rm deg} (A_1)} M_2 ( A_1, M_2 ( A_2, A_3 ) )
+M_4 ( M_0, A_1, A_2, A_3 ) \nonumber \\
& +(-1)^{{\rm deg} (A_1)} M_4 ( A_1, M_0, A_2, A_3 )
+(-1)^{{\rm deg} (A_1) +{\rm deg} (A_2)} M_4 ( A_1, A_2, M_0, A_3 ) \nonumber \\
& +(-1)^{{\rm deg} (A_1) +{\rm deg} (A_2) +{\rm deg} (A_3)}
M_4 ( A_1, A_2, A_3, M_0 ) = 0 \,.
\label{weak-M-4}
\end{align}

The three important string products in this paper
are the BRST operator as a one-string product,
the star product as a two-string product,
and the string field $J(\Phi)$ as a zero-string product.
The BRST operator is a one-string product of degree odd because
\begin{equation}
{\rm deg} ( Q A ) = {\rm deg} (A) +1
\end{equation}
for any string field $A$.
The cyclic property of the BRST operator is expressed as
\begin{equation}
\omega ( A_1, Q A_2 )
= {}-(-1)^{{\rm deg} (A_1)} \, \omega ( Q A_1, A_2 ) \,.
\end{equation}
Associated with the star product, we define $m_2 ( A_1, A_2 )$ by
\begin{equation}
m_2 ( A_1, A_2 ) = (-1)^{{\rm deg} (A_1)} A_1 \ast A_2 \,.
\end{equation}
The two-string product $m_2$ is degree odd,
\begin{equation}
{\rm deg} ( m_2 ( A_1, A_2 ) )
= {\rm deg} (A_1)+{\rm deg} (A_2) +1 \,,
\end{equation}
and it has the following cyclic property:
\begin{equation}
\omega ( A_1, m_2 ( A_2, A_3 ) )
= {}-(-1)^{{\rm deg} (A_1)} \, \omega ( m_2 ( A_1, A_2 ), A_3 ) \,.
\label{m_2-cyclic-property}
\end{equation}
We define the zero-string product $w_0$ by
\begin{equation}
w_0 = {}-J (\Phi)
\end{equation}
with the closed string field $\Phi$ annihilated by the BRST operator.
It is degree odd,
\begin{equation}
{\rm deg} (w_0) = 1 \,,
\end{equation}
and the cyclic property of $w_0$ can be represented as
\begin{equation}
\omega ( A_1, w_0 )
= {}-(-1)^{{\rm deg} (A_1)} \, \omega ( w_0, A_1 )
\label{w_0-cyclic-property}
\end{equation}
for any string field $A_1$.
The basic relations involving the BRST operator,
the star product, and $J (\Phi)$ given by
\begin{equation}
\begin{split}
Q^2 & = 0 \,, \\
Q \, ( A_1 \ast A_2 ) & = Q A_1 \ast A_2 +(-1)^{A_1} A_1 \ast Q A_2 \,, \\
( A_1 \ast A_2 ) \ast A_3 & = A_1 \ast ( A_2 \ast A_3 ) \,, \\
Q J(\Phi) & = 0 \,, \\
J(\Phi) \ast A_1 & = A_1 \ast J (\Phi)
\end{split}
\end{equation}
for any string fields $A_1$, $A_2$, and $A_3$
are expressed in terms of $Q$, $m_2$, and $w_0$ as
\begin{equation}
\begin{split}
Q^2 & = 0 \,, \\
Q \, m_2 ( A_1, A_2 ) +m_2 ( Q A_1, A_2 ) +(-1)^{{\rm deg} (A_1)} \, m_2 ( A_1, Q A_2 )
& = 0 \,, \\
m_2 ( m_2 ( A_1, A_2 ), A_3 ) +(-1)^{{\rm deg} (A_1)} \, m_2 ( A_1, m_2 ( A_2, A_3 ) ) & = 0 \,, \\
Q \, w_0 & = 0 \,, \\
m_2 ( w_0, A_1 ) +(-1)^{{\rm deg} (A_1)} m_2 ( A_1, w_0 ) & = 0 \,.
\end{split}
\label{relations-multi-string-products}
\end{equation}

\subsection{Tensor products of the Hilbert space}

The second step to simplify the description of the $A_\infty$ structure
is to use operators
acting on tensor products of the Hilbert space $\mathcal{H}$.
We denote the tensor product of $n$ copies of the Hilbert space $\mathcal{H}$
by $\mathcal{H}^{\otimes n}$.
For an $n$-string product $c_n ( A_1, A_2, \ldots, A_n)$
we define a corresponding operator $c_n$
which maps $\mathcal{H}^{\otimes n}$ into $\mathcal{H}$ by
\begin{equation}
c_n \, ( A_1 \otimes A_2 \otimes \ldots \otimes A_n )
\equiv c_n ( A_1, A_2, \ldots, A_n) \,.
\end{equation}
We use the same symbol $c_n$ to denote this operator,
and to distinguish the operator from the $n$-string product
we call it {\it $n$-string operator}.

Let us use examples to demonstrate how the introduction of $n$-string operators
helps simplify the description of the weak $A_\infty$ structure.
The first example is the relation
\begin{equation}
Q \, m_2 ( A_1, A_2 ) +m_2 ( Q A_1, A_2 ) +(-1)^{{\rm deg} (A_1)} \, m_2 ( A_1, Q A_2 )
= 0 \,.
\label{Q-m_2}
\end{equation}
Using $Q$ as a one-string operator and $m_2$ as a two-string operator,
this relation can be expressed as
\begin{equation}
Q \, m_2 \, ( A_1 \otimes A_2 ) +m_2 \, ( Q A_1 \otimes A_2 )
+(-1)^{{\rm deg} (A_1)} \, m_2 \, ( A_1 \otimes Q A_2 ) = 0 \,.
\end{equation}
We denote the identity map as a one-string operator from $\mathcal{H}$ to $\mathcal{H}$
by ${\mathbb I} \,$,
\begin{equation}
\mathbb I \, ( A ) = A
\end{equation}
for any string field $A$,
and we write $Q A_1 \otimes A_2$ and $(-1)^{{\rm deg} (A_1)} A_1 \otimes Q A_2$ as
\begin{equation}
\begin{split}
Q A_1 \otimes A_2 & = ( \, Q \otimes {\mathbb I} \, ) \, ( A_1 \otimes A_2 ) \,, \\ 
(-1)^{{\rm deg} (A_1)} A_1 \otimes Q A_2
& = ( \, {\mathbb I} \otimes Q \, ) \, ( A_1 \otimes A_2 ) \,,
\end{split}
\end{equation}
where in the second equation the sign factor $(-1)^{{\rm deg} (A_1)}$ is canceled
on the right-hand side
when the degree-odd operator $Q$ passes through $A_1$.
We then have
\begin{equation}
\begin{split}
& Q \, m_2 \, ( A_1 \otimes A_2 ) +m_2 \, ( Q A_1 \otimes A_2 )
+(-1)^{{\rm deg} (A_1)} \, m_2 \, ( A_1 \otimes Q A_2 ) \\
& = Q \, m_2 \, ( A_1 \otimes A_2 )
+m_2 \, ( \, Q \otimes {\mathbb I} \, ) \, ( A_1 \otimes A_2 )
+m_2 \, ( \, {\mathbb I} \otimes Q \, ) \, ( A_1 \otimes A_2 ) \\
& = ( \, Q \, m_2 +m_2 \, ( \, Q \otimes {\mathbb I} \, )
+m_2 \, ( \, {\mathbb I} \otimes Q \, ) \, ) \, ( A_1 \otimes A_2 ) \,,
\end{split}
\end{equation}
and the relation~\eqref{Q-m_2} can be expressed
without using $A_1$ and $A_2$ as
\begin{equation}
Q \, m_2 +m_2 \, ( \, Q \otimes {\mathbb I} \, )
+m_2 \, ( \, {\mathbb I} \otimes Q \, ) = 0 \,.
\end{equation}
Similarly, the left-hand side of the relation
\begin{equation}
m_2 ( m_2 ( A_1, A_2 ), A_3 ) +(-1)^{{\rm deg} (A_1)} \, m_2 ( A_1, m_2 ( A_2, A_3 ) ) = 0
\label{m_2-m_2}
\end{equation}
can be written as
\begin{equation}
\begin{split}
& m_2 ( m_2 ( A_1, A_2 ), A_3 ) +(-1)^{{\rm deg} (A_1)} \, m_2 ( A_1, m_2 ( A_2, A_3 ) ) \\
& = m_2 ( m_2 ( A_1 \otimes A_2 ) \otimes A_3 )
+(-1)^{{\rm deg} (A_1)} \, m_2 ( A_1 \otimes m_2 ( A_2 \otimes A_3 ) ) \\
& = m_2 \, ( \, m_2 \otimes \mathbb{I} +\mathbb{I} \otimes m_2 \, ) \,
( A_1 \otimes A_2 \otimes A_3 ) \,,
\end{split}
\end{equation}
so the relation~\eqref{m_2-m_2} can be expressed
without using $A_1$, $A_2$, and $A_3$ as
\begin{equation}
m_2 \, ( \, m_2 \otimes \mathbb{I} +\mathbb{I} \otimes m_2 \, ) = 0 \,.
\end{equation}

We also introduce the vector space for the zero-string space
denoted by $\mathcal{H}^{\otimes 0}$.
It is a one-dimensional vector space
given by multiplying a single basis vector {\bf 1} by complex numbers.
The vector {\bf 1} satisfies
\begin{equation}
{\bf 1} \otimes A = A \,, \qquad A \otimes {\bf 1} = A
\end{equation}
for any string field $A$.
For a zero-string product $c_0$
we define a corresponding operator $c_0$
which maps $\mathcal{H}^{\otimes 0}$ into $\mathcal{H}$ by
\begin{equation}
c_0 \, {\bf 1} \equiv c_0 \,,
\end{equation}
where $c_0$ on the right-hand side is the zero-string product
and $c_0$ on the left-hand side is the zero-string operator
which maps $\mathcal{H}^{\otimes 0}$ into $\mathcal{H}$.
The relation $Q \, w_0 = 0$ for the zero-string product $w_0$
is translated into
$Q \, w_0 \, {\bf 1} = 0$
for the zero-string operator $w_0$,
and thus we can express it without using ${\bf 1}$ as the relation
\begin{equation}
Q \, w_0 = 0
\end{equation}
for the zero-string operator $w_0$.
The left-hand side of the relation
\begin{equation}
m_2 ( w_0, A_1 ) +(-1)^{{\rm deg} (A_1)} m_2 ( A_1, w_0 ) = 0
\label{m_2-w_0}
\end{equation}
can be written as
\begin{equation}
\begin{split}
& m_2 ( w_0, A_1 ) +(-1)^{{\rm deg} (A_1)} m_2 ( A_1, w_0 ) \\
& = m_2 \, ( \, w_0 \otimes A_1 \, )
+(-1)^{{\rm deg} (A_1)} \, m_2 \, ( \, A_1 \otimes w_0 \, ) \\
& = m_2 \, ( \, w_0 \otimes {\mathbb I} +{\mathbb I} \otimes w_0 \, ) \, ( \, A_1 \, ) \,,
\end{split}
\end{equation}
so the relation~\eqref{m_2-w_0} can be expressed
without using $A_1$ as
\begin{equation}
m_2 \, ( \, w_0 \otimes {\mathbb I} +{\mathbb I} \otimes w_0 \, ) = 0 \,.
\end{equation}

To summarize, the relations~\eqref{relations-multi-string-products}
in terms of multi-string products
are written as the relations
\begin{align}
Q^2 & = 0 \,, \\
Q \, m_2 +m_2 \, ( \, Q \otimes {\mathbb I} +{\mathbb I} \otimes Q \, ) & = 0 \,,
\label{Q-m_2-operators} \\
m_2 \, ( \, m_2 \otimes \mathbb{I} +\mathbb{I} \otimes m_2 \, ) & = 0 \,,
\label{m_2-m_2-operators} \\
Q \, w_0 & = 0 \,,
\label{Q-w_0-operators} \\
m_2 \, ( \, w_0 \otimes {\mathbb I} +{\mathbb I} \otimes w_0 \, ) & = 0
\label{m_2-w_0-operators}
\end{align}
in terms of multi-string operators.
We no longer need string fields to express these relations,
and all the sign factors in~\eqref{relations-multi-string-products} are now gone.

For the BPZ inner product, we define $\langle \omega |$
which maps $\mathcal{H}^{\otimes 2}$ into a complex number by
\begin{equation}
\langle \omega | \, A_1 \otimes A_2 \equiv \omega \, ( A_1, A_2 )
\end{equation}
for a pair of string fields $A_1$ and $A_2$.
The left-hand side of the relation
\begin{equation}
\begin{split}
\omega ( A_1, Q A_2 )
+(-1)^{{\rm deg} (A_1)} \, \omega ( Q A_1, A_2 ) = 0
\end{split}
\end{equation}
can be written as
\begin{equation}
\begin{split}
& \omega ( A_1, Q A_2 )
+(-1)^{{\rm deg} (A_1)} \, \omega ( Q A_1, A_2 ) \\
& = \langle \omega | \, A_1 \otimes Q A_2
+(-1)^{{\rm deg} (A_1)} \langle \omega | \, Q A_1 \otimes A_2 \\
& = \langle \omega | \, ( \, {\mathbb I} \otimes Q +Q \otimes {\mathbb I} \, ) \,
( \, A_1 \otimes A_2 \, ) \,,
\end{split}
\end{equation}
so the cyclic property of the BRST operator can be expressed
without using $A_1$ and $A_2$ as
\begin{equation}
\langle \omega | \, ( \, {\mathbb I} \otimes Q +Q \otimes {\mathbb I} \, ) = 0 \,.
\end{equation}
Similarly, the cyclic properties of $m_2$ and $w_0$
in~\eqref{m_2-cyclic-property} and~\eqref{w_0-cyclic-property}, respectively,
can be written as
\begin{equation}
\begin{split}
\langle \omega | \, ( \, {\mathbb I} \otimes m_2 +m_2 \otimes {\mathbb I} \, ) & = 0 \,, \\
\langle \omega | \, ( \, {\mathbb I} \otimes w_0 +w_0 \otimes {\mathbb I} \, ) & = 0 \,.
\label{m_2-w_0-cyclic-properties}
\end{split}
\end{equation}
The operators $P$ and $h$ are BPZ even:
\begin{equation}
\langle \, A_1, P A_2 \, \rangle
= \langle \, P A_1, A_2 \, \rangle \,, \qquad
\langle \, A_1, h \, A_2 \, \rangle
= (-1)^{A_1} \langle \, h \, A_1, A_2 \, \rangle
\end{equation}
for any pair of states $A_1$ and $A_2$.
These relations can be expressed using $\langle \omega |$ as
\begin{equation}
\langle \omega | \, {\mathbb I} \otimes P
= \langle \omega | \, P \otimes {\mathbb I} \,, \qquad
\langle \omega | \, {\mathbb I} \otimes h
= \langle \omega | \, h \otimes {\mathbb I} \,.
\end{equation}

Finally, let us use the $n$-string operator $M_n$
to simplify the description of the weak $A_\infty$ structure.
The weak $A_\infty$ relations are written as
\begin{equation}
\begin{split}
& M_1 \, M_0 = 0 \,, \\
& M_1 \, M_1 +M_2 \, ( \, M_0 \otimes {\mathbb I} +{\mathbb I} \otimes M_0 \, ) = 0 \,, \\
& M_1 \, M_2 +M_2 \, ( \, M_1 \otimes {\mathbb I} +{\mathbb I} \otimes M_1 \, )
+M_3 \, ( \, M_0 \otimes {\mathbb I} \otimes {\mathbb I}
+{\mathbb I} \otimes M_0 \otimes {\mathbb I}
+{\mathbb I} \otimes {\mathbb I} \otimes M_0 \, ) = 0 \,, \\
& M_1 \, M_3
+M_3 \, ( \, M_1 \otimes {\mathbb I} \otimes {\mathbb I}
+{\mathbb I} \otimes M_1 \otimes {\mathbb I}
+{\mathbb I} \otimes {\mathbb I} \otimes M_1\, )
+M_2 \, ( \, M_2 \otimes {\mathbb I} +{\mathbb I} \otimes M_2 \, ) \\
& +M_4 \, ( \, M_0 \otimes {\mathbb I} \otimes {\mathbb I} \otimes {\mathbb I}
+{\mathbb I} \otimes M_0 \otimes {\mathbb I} \otimes {\mathbb I}
+{\mathbb I} \otimes {\mathbb I} \otimes M_0 \otimes {\mathbb I}
+{\mathbb I} \otimes {\mathbb I} \otimes {\mathbb I} \otimes M_0 \, )
= 0 \,,
\end{split}
\label{4-weak-A_infinity}
\end{equation}
and the cyclic properties are
\begin{equation}
\begin{split}
\langle \omega | \, ( \, {\mathbb I} \otimes M_0 +M_0 \otimes {\mathbb I} \, )
& = 0 \,, \\
\langle \omega | \, ( \, {\mathbb I} \otimes M_1 +M_1 \otimes {\mathbb I} \, )
& = 0 \,, \\
\langle \omega | \, ( \, {\mathbb I} \otimes M_2 +M_2 \otimes {\mathbb I} \, )
& = 0 \,, \\
\langle \omega | \, ( \, {\mathbb I} \otimes M_3 +M_3 \otimes {\mathbb I} \, )
& = 0 \,, \\
\langle \omega | \, ( \, {\mathbb I} \otimes M_4 +M_4 \otimes {\mathbb I} \, )
& = 0 \,.
\end{split}
\label{5-cyclic-properties}
\end{equation}

\subsection{Coderivations}

To describe the weak $A_\infty$ structure to all orders,
it is convenient to consider linear operators
acting on the vector space $T \mathcal{H}$ defined by
\begin{equation}
T \mathcal{H}
= \mathcal{H}^{\otimes 0} \, \oplus \mathcal{H}
\oplus \mathcal{H}^{\otimes 2} \oplus \mathcal{H}^{\otimes 3} \oplus  \ldots \,.
\end{equation}
We denote the projection operator onto $\mathcal{H}^{\otimes n}$ by $\pi_n$.
For an $n$-string operator $c_n$,
we define an associated operator ${\bm c}_n$ acting on $T \mathcal{H}$ as follows.
The action on the sector $\mathcal{H}^{\otimes m}$ vanishes when $m < n$:
\begin{equation}
{\bm c}_n \, \pi_m = 0 \quad \text{for} \quad m < n \,.
\end{equation}
The action on the sector $\mathcal{H}^{\otimes n}$
is given by the $n$-string operator $c_n$:
\begin{equation}
{\bm c}_n \, \pi_n = c_n \, \pi_n \,.
\end{equation}
The action on the sector $\mathcal{H}^{\otimes n+1}$ is given by
\begin{equation}
{\bm c}_n \, \pi_{n+1}
= ( \, c_n \otimes {\mathbb I} +{\mathbb I} \otimes c_n \, ) \, \pi_{n+1} \,.
\end{equation}
The action on the sector $\mathcal{H}^{\otimes m}$
for $m > n+1$ is given by
\begin{equation}
{\bm c}_n \, \pi_m = \Bigl( \, c_n \otimes {\mathbb I}^{\otimes (m-n)}
+\sum_{k=1}^{m-n-1} {\mathbb I}^{\otimes k} \otimes c_n \otimes {\mathbb I}^{\otimes (m-n-k)}
+{\mathbb I}^{\otimes (m-n)} \otimes c_n \, \Bigr) \, \pi_m
\quad \text{for} \quad m > n+1 \,,
\end{equation}
where
\begin{equation}
{\mathbb I}^{\otimes k} = \underbrace{ \, {\mathbb I} \otimes {\mathbb I} \otimes \ldots
\otimes {\mathbb I} \, }_{k} \,.
\end{equation}
The degree of ${\bm c}_n$ is defined to be the same as that of $c_n$.
An operator acting on $T \mathcal{H}$ of this form is called a {\it coderivation}.

Coderivations can be characterized using a linear operation called {\it coproduct}
which maps $T \mathcal{H}$
to a tensor product of two copies of $T \mathcal{H}$
denoted as $T \mathcal{H} \otimes' T \mathcal{H}$.
We use the symbol~$\otimes'$ to distinguish
the tensor product of $T \mathcal{H}$ from
the tensor product~$\otimes$ within $T \mathcal{H}$.
We denote the coproduct by $\triangle$,
and its actions on ${\bf 1}$, $A_1$, $A_1 \otimes A_2$,
and $A_1 \otimes A_2 \otimes A_3$ are given by
\begin{equation}
\begin{split}
\triangle \, {\bf 1} & = {\bf 1} \otimes' {\bf 1} \,, \\
\triangle \, A_1 & = {\bf 1} \otimes' A_1 +A_1 \otimes' {\bf 1} \,, \\ 
\triangle \, ( \, A_1 \otimes A_2 \, )
& = {\bf 1} \otimes' ( \, A_1 \otimes A_2 \, )
+A_1 \otimes' A_2
+( \, A_1 \otimes A_2 \, ) \otimes' {\bf 1} \,, \\ 
\triangle \, ( \, A_1 \otimes A_2 \otimes A_3 \, )
& = {\bf 1} \otimes' ( \, A_1 \otimes A_2 \otimes A_3 \, )
+A_1 \otimes' ( \, A_2 \otimes A_3 \, ) \\
& \quad~
+( \, A_1 \otimes A_2 \, ) \otimes' A_3
+( \, A_1 \otimes A_2 \otimes A_3 \, ) \otimes' {\bf 1} \,.
\end{split}
\end{equation}
The action of the coproduct
on $A_1 \otimes A_2 \otimes \ldots \otimes A_n$ for $n > 1$
is as follows:
\begin{equation}
\begin{split}
\triangle \, ( \, A_1 \otimes A_2 \otimes \ldots \otimes A_n \, )
& = {\bf 1} \otimes' ( \, A_1 \otimes A_2 \otimes \ldots \otimes A_n \, ) \\
& \quad~
+\sum_{k=1}^{n-1} ( \, A_1 \otimes A_2 \otimes \ldots \otimes A_k \, )
\otimes' ( \, A_{k+1} \otimes A_{k+2} \otimes \ldots \otimes A_n \, ) \\
& \quad~
+( \, A_1 \otimes A_2 \otimes \ldots \otimes A_n \, ) \otimes' {\bf 1} \,.
\end{split}
\end{equation}
Using the coproduct, a coderivation is defined in the following way.
A linear operator ${\bm a}$ on $T \mathcal{H}$
is a coderivation when it satisfies
\begin{equation}
\triangle \, {\bm a}
= ( \, {\bm a} \otimes' {\bf I} +{\bf I} \otimes' {\bm a} \, ) \, \triangle \,,
\end{equation}
where ${\bf I}$ is the identity operator on $T \mathcal{H}$.
It is clear from this definition that the sum of two coderivations is a coderivation.
In general, a coderivation is associated
with a linear combination of multi-string operators,
and the multi-string operators can be extracted from the coderivation ${\bm a}$
by decomposing $\pi_1 \, {\bm a}$ as
\begin{equation}
\pi_1 \, {\bm a} = \sum_{n=0}^\infty a_n \, \pi_n
\end{equation}
where $a_n$ is an $n$-string operator.
When ${\bm a}$ is a coderivation and $\pi_1 \, {\bm a}$ vanishes,
we can show that ${\bm a}$ itself vanishes.

Let us show that the commutator of two coderivations graded with respect to degree
is a coderivation.
We denote the commutator of two operators ${\bf A}$ and ${\bf B}$
on $T \mathcal{H}$ graded with respect to degree by
\begin{equation}
[ \, {\bf A}, {\bf B} \, ]
= {\bf A} \, {\bf B}
-(-1)^{{\rm deg} ({\bf A}) \, {\rm deg} ({\bf B})} \, {\bf B} \, {\bf A} \,,
\end{equation}
where ${\rm deg} ({\bf O})$ is the degree of the operator ${\bf O}$.
When ${\bm a}$ and ${\bm b}$ are coderivations, we find
\begin{equation}
\begin{split}
\triangle \, [ \, {\bm a}, {\bm b} \, ]
& = ( \, {\bm a} \otimes' {\bf I} +{\bf I} \otimes' {\bm a} \, ) \,
( \, {\bm b} \otimes' {\bf I} +{\bf I} \otimes' {\bm b} \, ) \, \triangle \\
& \quad~ {}-(-1)^{{\rm deg} ({\bm a}) \, {\rm deg} ({\bm b})}
( \, {\bm b} \otimes' {\bf I} +{\bf I} \otimes' {\bm b} \, ) \,
( \, {\bm a} \otimes' {\bf I} +{\bf I} \otimes' {\bm a} \, ) \, \triangle \\
& = ( \, [ \, {\bm a}, {\bm b} \, ] \otimes' {\bf I}
+{\bf I} \otimes' [ \, {\bm a}, {\bm b} \, ] \, ) \, \triangle \,.
\end{split}
\end{equation}
The graded commutator $[ \, {\bm a}, {\bm b} \, ]$ is therefore a coderivation.

Let us present a few examples of coderivations.
The action of the coderivation ${\bf Q}$ associated
with the BRST operator $Q$ as a one-string operator of degree odd is given by
\begin{equation}
\begin{split}
{\bf Q} \, {\bf 1} & = 0 \,, \\
{\bf Q} \, A_1 & = Q \, A_1 \,, \\
{\bf Q} \, ( A_1 \otimes A_2 )
& = Q \, A_1  \otimes A_2 +(-1)^{{\rm deg} (A_1)} A_1 \otimes Q \, A_2 \,, \\
{\bf Q} \, ( A_1 \otimes A_2 \otimes A_3 )
& = Q \, A_1 \otimes A_2 \otimes A_3
+(-1)^{{\rm deg} (A_1)} A_1 \otimes Q \, A_2 \otimes A_3 \\
& \quad~ +(-1)^{{\rm deg} (A_1) +{\rm deg} (A_2)} A_1 \otimes A_2 \otimes Q \, A_3 \,, \\
& \quad \vdots
\end{split}
\end{equation}
The graded commutator $[ \, {\bf Q}, {\bf Q} \, ]$ is a coderivation,
and it follows from $Q^2=0$ that $\pi_1 \, [ \, {\bf Q}, {\bf Q} \, ]$ vanishes:
\begin{equation}
\pi_1 \, [ \, {\bf Q}, {\bf Q} \, ]
= 2 \, \pi_1 \, {\bf Q}^2
= 2 \, Q \, \pi_1 \, {\bf Q}
= 2 \, Q^2 \, \pi_1 = 0 \,.
\end{equation}
Therefore, the coderivation $[ \, {\bf Q}, {\bf Q} \, ]$ vanishes:
\begin{equation}
[ \, {\bf Q}, {\bf Q} \, ] = 0 \,.
\end{equation}

The action of the coderivation ${\bm m}_2$ associated
with the two-string operator $m_2$ of degree odd is given by
\begin{equation}
\begin{split}
{\bm m}_2 \, {\bf 1} & = 0 \,, \\
{\bm m}_2 \, A_1 & = 0 \,, \\
{\bm m}_2 \, ( A_1 \otimes A_2 )
& = m_2 ( A_1, A_2 ) \,, \\
{\bm m}_2 \, ( A_1 \otimes A_2 \otimes A_3 )
& = m_2 ( A_1, A_2 ) \otimes A_3
+(-1)^{{\rm deg} (A_1)} A_1 \otimes m_2 ( A_2, A_3 ) \,, \\
{\bm m}_2 \, ( A_1 \otimes A_2 \otimes A_3 \otimes A_4 )
& = m_2 ( A_1, A_2 ) \otimes A_3 \otimes A_4
+(-1)^{{\rm deg} (A_1)} A_1 \otimes m_2 ( A_2, A_3 ) \otimes A_4 \\
& \quad~ +(-1)^{{\rm deg} (A_1) +{\rm deg} (A_2)} A_1 \otimes A_2 \otimes m_2 ( A_3, A_4 ) \,, \\
& \quad \vdots
\end{split}
\end{equation}
Let us calculate $\pi_1 \, [ \, {\bf Q}, {\bm m}_2 \, ]$. We find
\begin{equation}
\begin{split}
& \pi_1 \, [ \, {\bf Q}, {\bm m}_2 \, ]
= \pi_1 \, {\bf Q} \, {\bm m}_2 +\pi_1 \, {\bm m}_2 \, {\bf Q}
= Q \, \pi_1 \, {\bm m}_2 +m_2 \, \pi_2 \, {\bf Q} \\
& = Q \, m_2 \, \pi_2
+m_2 \, ( \, Q \otimes {\mathbb I} +{\mathbb I} \otimes Q \, ) \, \pi_2
= ( \, Q \, m_2
+m_2 \, ( \, Q \otimes {\mathbb I} +{\mathbb I} \otimes Q \, ) \, ) \, \pi_2 = 0 \,,
\end{split}
\end{equation}
where we used~\eqref{Q-m_2-operators}.
Therefore, the graded commutator $[ \, {\bf Q}, {\bm m}_2 \, ]$ vanishes:
\begin{equation}
[ \, {\bf Q}, {\bm m}_2 \, ] = 0 \,.
\end{equation}
Let us next calculate $\pi_1 \, [ \, {\bm m}_2, {\bm m}_2 \, ]$. We find
\begin{equation}
\pi_1 [ \, {\bm m}_2, {\bm m}_2 \, ]
= 2 \, \pi_1 \, {\bm m}_2^2
= 2 \, m_2 \, \pi_2 \, {\bm m}_2
= 2 \, m_2 \,
( \, m_2 \otimes {\mathbb I} +{\mathbb I} \otimes m_2 \, ) \, \pi_3 = 0 \,,
\end{equation}
where we used~\eqref{m_2-m_2-operators}.
Therefore, the graded commutator $[ \, {\bm m}_2, {\bm m}_2 \, ]$ vanishes:
\begin{equation}
[ \, {\bm m}_2, {\bm m}_2 \, ] = 0 \,.
\end{equation}

The action of the coderivation ${\bm w}_0$ associated
with the zero-string operator $w_0$ of degree odd is given by
\begin{equation}
\begin{split}
{\bm w}_0 \, {\bf 1} & = w_0 \,, \\
{\bm w}_0 \, A_1 & = w_0 \otimes A_1 +(-1)^{{\rm deg} (A_1)} A_1 \otimes w_0 \,, \\
{\bm w}_0 \, ( A_1 \otimes A_2 )
& = w_0 \otimes A_1 \otimes A_2
+(-1)^{{\rm deg} (A_1)} A_1 \otimes w_0 \otimes A_2 \\
& \quad~ +(-1)^{{\rm deg} (A_1) +{\rm deg} (A_2)} A_1 \otimes A_2 \otimes w_0 \,, \\
{\bm w}_0 \, ( A_1 \otimes A_2 \otimes A_3 )
& = w_0 \otimes A_1 \otimes A_2 \otimes A_3
+(-1)^{{\rm deg} (A_1)} A_1 \otimes w_0 \otimes A_2 \otimes A_3 \\
& \quad~ +(-1)^{{\rm deg} (A_1) +{\rm deg} (A_2)}
A_1 \otimes A_2 \otimes w_0 \otimes A_3 \\
& \quad~ +(-1)^{{\rm deg} (A_1) +{\rm deg} (A_2) +{\rm deg} (A_3)}
A_1 \otimes A_2 \otimes A_3 \otimes w_0 \,, \\
& \quad \vdots
\end{split}
\end{equation}
Let us calculate $\pi_1 \, [ \, {\bm w}_0, {\bm w}_0 \, ]$. We find
\begin{equation}
\pi_1 \, [ \, {\bm w}_0, {\bm w}_0 \, ]
= 2 \, \pi_1 \, {\bm w}_0^2
= 2 \, w_0 \, \pi_0 \, {\bm w}_0 = 0 \,.
\end{equation}
Therefore, the graded commutator $[ \, {\bm w}_0, {\bm w}_0 \, ]$ vanishes:
\begin{equation}
[ \, {\bm w}_0, {\bm w}_0 \, ] = 0 \,.
\end{equation}
Let us next calculate $\pi_1 \, [ \, {\bf Q}, {\bm w}_0 \, ]$. We find
\begin{equation}
\pi_1 \, [ \, {\bf Q}, {\bm w}_0 \, ]
= \pi_1 \, {\bf Q} \, {\bm w}_0 +\pi_1 \, {\bm w}_0 \, {\bf Q}
= Q \, \pi_1 \, {\bm w}_0 +w_0 \, \pi_0 \, {\bf Q}
= Q \, w_0 \, \pi_0 = 0 \,,
\end{equation}
where we used~\eqref{Q-w_0-operators}.
Therefore, the graded commutator $[ \, {\bf Q}, {\bm w}_0 \, ]$ vanishes:
\begin{equation}
[ \, {\bf Q}, {\bm w}_0 \, ] = 0 \,.
\end{equation}
Let us also calculate $\pi_1 \, [ \, {\bm m}_2, {\bm w}_0 \, ]$. We find
\begin{equation}
\begin{split}
\pi_1 \, [ \, {\bm m}_2, {\bm w}_0 \, ]
& = \pi_1 \, {\bm m}_2 \, {\bm w}_0 +\pi_1 \, {\bm w}_0 \, {\bm m}_2
= m_2 \, \pi_2 \, {\bm w}_0 +w_0 \, \pi_0 \, {\bm m}_2 \\
& = m_2 \,
( \, w_0 \otimes {\mathbb I} +{\mathbb I} \otimes w_0 \, ) \, \pi_1 = 0 \,,
\end{split}
\end{equation}
where we used~\eqref{m_2-w_0-operators}.
Therefore, the graded commutator $[ \, {\bm m}_2, {\bm w}_0 \, ]$ vanishes:
\begin{equation}
[ \, {\bm m}_2, {\bm w}_0 \, ] = 0 \,.
\end{equation}

The cyclic properties of $Q$, $m_2$, and $w_0$ can be described
in terms of the corresponding coderivations ${\bf Q}$, ${\bm m}_2$, and ${\bm w}_0$
as follows:
\begin{equation}
\langle \omega | \, \pi_2 \, {\bf Q} = 0 \,, \qquad
\langle \omega | \, \pi_2 \, {\bm m}_2 = 0 \,, \qquad
\langle \omega | \, \pi_2 \, {\bm w}_0 = 0 \,.
\end{equation}
In general, we say that a coderivation ${\bm a}$ is cyclic
when it satisfies
\begin{equation}
\langle \omega | \, \pi_2 \, {\bm a} = 0 \,.
\end{equation}
All of the coderivations ${\bf Q}$, ${\bm m}_2$, and ${\bm w}_0$ are cyclic. 

Let us now suppose that a linear operator ${\bf M}$ on $T \mathcal{H}$
is a coderivation of degree odd and satisfies
\begin{equation}
[ \, {\bf M}, {\bf M} \, ] = 0 \,.
\label{M^2=0}
\end{equation}
We decompose $\pi_1 \, {\bf M}$ as
\begin{equation}
\pi_1 \, {\bf M} = \sum_{n=0}^\infty M_n \, \pi_n
\label{M_n-definition}
\end{equation}
where $M_n$ is an $n$-string operator.
It follows from~\eqref{M^2=0} that
\begin{equation}
\pi_1 \, [ \, {\bf M}, {\bf M} \, ] \, \pi_n = 0
\end{equation}
for any non-negative integer $n$.
Using the decomposition~\eqref{M_n-definition}, we find
\begin{equation}
\pi_1 \, [ \, {\bf M}, {\bf M} \, ] \, \pi_n
= 2 \, \pi_1 \, {\bf M}^2 \, \pi_n
= 2 \sum_{k=0}^\infty M_k \, \pi_k \, {\bf M} \, \pi_n \,.
\end{equation}
Since ${\bf M}$ is a coderivation, $\pi_k \, {\bf M} \, \pi_n$ vanishes
when $k=0$ and $k > n+1$.
We therefore have
\begin{equation}
\sum_{k=1}^{n+1} M_k \, \pi_k \, {\bf M} \, \pi_n = 0
\end{equation}
for any non-negative integer $n$.
We find that the four relations in~\eqref{4-weak-A_infinity}
precisely correspond to this equation with $n=0, 1, 2, 3$.
In fact, this enables us to describe the weak $A_\infty$ relations to all orders.
When a degree-odd coderivation ${\bf M}$ satisfies~\eqref{M^2=0},
the set of multi-string operators $\{ \, M_0, M_1, M_2 \ldots \, \}$
satisfy the weak $A_\infty$ relations.

Let us further suppose that the coderivation ${\bf M}$ is cyclic:
\begin{equation}
\langle \omega | \, \pi_2 \, {\bf M} = 0 \,.
\end{equation}
We then have
\begin{equation}
\langle \omega | \, \pi_2 \, {\bf M} \, \pi_n = 0
\label{M_n-cyclic}
\end{equation}
for any non-negative integer $n$.
Since ${\bf M}$ is a coderivation, $\pi_2 \, {\bf M} \, \pi_n$ vanishes
when $n=0$.
For $n > 0$, we have
\begin{equation}
\pi_2 \, {\bf M} \, \pi_n
= ( \, {\mathbb I} \otimes M_{n-1} +M_{n-1} \otimes {\mathbb I} \, ) \, \pi_n \,.
\end{equation}
Therefore, the five relations in~\eqref{5-cyclic-properties}
precisely correspond to~\eqref{M_n-cyclic} with $n=1, 2, 3, 4, 5$.
In fact, this enables us to describe
the cyclic properties of
the multi-string operators $\{ \, M_0, M_1, M_2 \ldots \, \}$ to all orders.
When ${\bf M}$ is cyclic,
the multi-string operators $\{ \, M_0, M_1, M_2 \ldots \, \}$
have appropriate cyclic properties.

To summarize, when a linear operator ${\bf M}$ on $T \mathcal{H}$
is degree odd and satisfies
\begin{align}
& \triangle \, {\bf M}
= ( \, {\bf M} \otimes' {\bf I} +{\bf I} \otimes' {\bf M} \, ) \, \triangle \,, \\
& [ \, {\bf M}, {\bf M} \, ] = 0 \,,
\label{A_infinity-M} \\
& \langle \omega | \, \pi_2 \, {\bf M} = 0 \,,
\end{align}
the corresponding multi-string operators $\{ \, M_0, M_1, M_2 \ldots \, \}$
satisfy the weak $A_\infty$ relations,
and they have the cyclic property given by
\begin{equation}
\langle \omega | \, ( \, M_n \otimes {\mathbb I} +{\mathbb I} \otimes M_n \, ) = 0 \,.
\end{equation}

For the action of open bosonic string field theory
without introducing the source term for the gauge-invariant operators,
the $A_\infty$ structure can be described in terms of ${\bf M}$ given by
\begin{equation}
{\bf M} = {\bf Q} +{\bm m}_2 \,.
\end{equation} 
Since both of ${\bf Q}$ and ${\bm m}_2$ are cyclic coderivations of degree odd,
it is clear that ${\bf M}$ is a cyclic coderivation of degree odd.
It follows from
\begin{equation}
\begin{split}
[ \, {\bf Q}, {\bf Q} \, ] = 0 \,, \qquad
[ \, {\bf Q}, {\bm m}_2 \, ] = 0 \,, \qquad
[ \, {\bm m}_2, {\bm m}_2 \, ] = 0
\end{split}
\label{(Q+m_2)^2=0}
\end{equation}
that the condition~\eqref{A_infinity-M} is satisfied.
The multi-string operators from ${\bf M}$ are
\begin{equation}
M_0 = 0 \,, \qquad
M_1 = Q \,, \qquad
M_2 = m_2 \,, \qquad
M_n = 0 \quad \text{for} \quad n > 2 \,.
\end{equation}

For the action of open bosonic string field theory
including the source term for the gauge-invariant operators,
the weak $A_\infty$ structure can be described in terms of ${\bf M}$ given by
\begin{equation}
{\bf M} = {\bf Q} +{\bm m}_2 +\kappa \, {\bm w}_0 \,.
\end{equation} 
Since ${\bf Q}$, ${\bm m}_2$, and ${\bm w}_0$ are cyclic coderivations of degree odd,
it is clear that ${\bf M}$ is a cyclic coderivation of degree odd.
It follows from~\eqref{(Q+m_2)^2=0} and
\begin{equation}
\begin{split}
[ \, {\bm w}_0, {\bm w}_0 \, ] = 0 \,, \qquad
[ \, {\bf Q}, {\bm w}_0 \, ] = 0 \,, \qquad
[ \, {\bm m}_2, {\bm w}_0 \, ] = 0
\end{split}
\end{equation}
that the condition~\eqref{A_infinity-M} is satisfied.
The multi-string operators from ${\bf M}$ are
\begin{equation}
M_0 = \kappa \, w_0 \,, \qquad
M_1 = Q \,, \qquad
M_2 = m_2 \,, \qquad
M_n = 0 \quad \text{for} \quad n > 2 \,.
\end{equation}

\subsection{Projection to the massless sector}
\label{projection-subsection}

We describe the low-energy limit of open bosonic string field theory
in terms of string fields projected onto the massless sector.
We denote the subspace of $\mathcal{H}$
for the massless sector by $\widetilde{\mathcal{H}}$.
A string field $A$ belongs to $\widetilde{\mathcal{H}}$ if
\begin{equation}
P A = A
\end{equation}
is satisfied.

We are interested in an $n$-string product $c_n ( A_1, A_2, \ldots , A_n )$
satisfying the condition
\begin{equation}
P c_n ( A_1, A_2, \ldots , A_n )  = c_n ( A_1, A_2, \ldots , A_n ) \,,
\end{equation}
where all of the string fields $A_1$, $A_2$, \ldots , $A_{n-1}$, and $A_n$
are in $\widetilde{\mathcal{H}}$.
When we write such an $n$-string product $c_n ( A_1, A_2, \ldots , A_n )$ as
\begin{equation}
c_n ( A_1, A_2, \ldots, A_n)
= c_n \, ( A_1 \otimes A_2 \otimes \ldots \otimes A_n ) \,,
\end{equation}
we require that the $n$-string operator $c_n$
be a map from $\widetilde{\mathcal{H}}^{\otimes n}$
to $\widetilde{\mathcal{H}}$,
where $\widetilde{\mathcal{H}}^{\otimes n}$
is the tensor product of $n$ copies of $\widetilde{\mathcal{H}}$.
Then the $n$-string operator $c_n$ satisfies
\begin{equation}
P \, c_n \, P^{\otimes n} = c_n \,,
\label{c_n-P}
\end{equation}
where
\begin{equation}
P^{\otimes n} = \underbrace{ \, P \otimes P \otimes \ldots
\otimes P \, }_{n} \,.
\end{equation}
We also define the vector space $T \widetilde{\mathcal{H}}$ by
\begin{equation}
T \widetilde{\mathcal{H}}
= \widetilde{\mathcal{H}}^{\otimes 0} \, \oplus \widetilde{\mathcal{H}}
\oplus \widetilde{\mathcal{H}}^{\otimes 2}
\oplus \widetilde{\mathcal{H}}^{\otimes 3} \oplus  \ldots \,,
\end{equation}
where
\begin{equation}
\widetilde{\mathcal{H}}^{\otimes 0} = \mathcal{H}^{\otimes 0} \,.
\end{equation}
We denote the projection operator onto $T \widetilde{\mathcal{H}}$ by ${\bf P}$.
The action of ${\bf P}$ is given by
\begin{equation}
\begin{split}
{\bf P} \, {\bf 1} & = {\bf 1} \,, \\
{\bf P} \, A_1 & = P A_1 \,, \\
{\bf P} \, ( A_1 \otimes A_2 )
& = P A_1 \otimes P A_2 \,, \\
{\bf P} \, ( A_1 \otimes A_2 \otimes A_3 )
& = P A_1 \otimes P A_2 \otimes P A_3 \,, \\
& \quad \vdots
\end{split}
\end{equation}
It is degree even, and it has the following property:
\begin{equation}
{\bf P}^2 = {\bf P} \,, \qquad
\triangle \, {\bf P} = ( \, {\bf P} \otimes' {\bf P} \, ) \, \triangle \,.
\end{equation}
An operator ${\bf A}$ which maps from $T \widetilde{\mathcal{H}}$
into $T \widetilde{\mathcal{H}}$ satisfies
\begin{equation}
{\bf P} \, {\bf A} \, {\bf P} = {\bf A} \,.
\end{equation}
For an $n$-string operator $c_n$ which satisfies~\eqref{c_n-P},
we define an associated operator ${\bm c}_n$
which is a coderivation on $T \widetilde{\mathcal{H}}$ as follows.
The action on the sector $\widetilde{\mathcal{H}}^{\otimes m}$ vanishes when $m < n$:
\begin{equation}
{\bm c}_n \, \pi_m = 0 \quad \text{for} \quad m < n \,.
\end{equation}
The action on the sector $\widetilde{\mathcal{H}}^{\otimes n}$
is given by the $n$-string operator $c_n$:
\begin{equation}
{\bm c}_n \, \pi_n = c_n \, \pi_n \,.
\end{equation}
The action on the sector $\widetilde{\mathcal{H}}^{\otimes n+1}$ is given by
\begin{equation}
{\bm c}_n \, \pi_{n+1}
= ( \, c_n \otimes P +P \otimes c_n \, ) \, \pi_{n+1} \,.
\end{equation}
The action on the sector $\widetilde{\mathcal{H}}^{\otimes m}$
for $m > n+1$ is given by
\begin{equation}
{\bm c}_n \, \pi_m = \Bigl( \, c_n \otimes P^{\otimes (m-n)}
+\sum_{k=1}^{m-n-1} P^{\otimes k} \otimes c_n \otimes P^{\otimes (m-n-k)}
+P^{\otimes (m-n)} \otimes c_n \, \Bigr) \, \pi_m
\quad \text{for} \quad m > n+1 \,.
\end{equation}
The degree of ${\bm c}_n$ is defined to be the same as that of $c_n$.
The operator ${\bm c}_n$ has the following properties:
\begin{equation}
{\bf P} \, {\bm c}_n \, {\bf P} = {\bm c}_n \,, \qquad
\triangle \, {\bm c}_n
= ( \, {\bm c}_n \otimes' {\bf P} +{\bf P} \otimes' {\bm c}_n \, ) \, \triangle \,.
\end{equation}

We are interested in an $n$-string product $c_n ( A_1, A_2, \ldots , A_n )$
of degree odd which has the following cyclic property:
\begin{equation}
\omega ( A_1, c_n ( A_2, A_3, \ldots , A_{n+1} ) )
= {}-(-1)^{{\rm deg} (A_1)} \,
\omega ( c_n ( A_1, A_2, \ldots , A_n ), A_{n+1} )
\end{equation}
for $A_1$, $A_2$, \ldots $A_{n+1}$ in $\widetilde{\mathcal{H}}$.
This cyclic property is described in terms of the corresponding $n$-string operator $c_n$ as
\begin{equation}
\langle \omega | \, ( \, c_n \otimes P +P \otimes c_n \, ) = 0
\end{equation}
and in terms of ${\bm c}_n$ as
\begin{equation}
\langle \omega | \, \pi_2 \, {\bm c}_n = 0 \,.
\end{equation}

Suppose that a linear operator ${\bf M}$ on $T \mathcal{H}$
is degree odd and satisfies
\begin{align}
& {\bf P} \, {\bf M} \, {\bf P} = {\bf M} \,, \\
& \triangle \, {\bf M}
= ( \, {\bf M} \otimes' {\bf P} +{\bf P} \otimes' {\bf M} \, ) \, \triangle \,, \\
& [ \, {\bf M}, {\bf M} \, ] = 0 \,, \\
& \langle \omega | \, \pi_2 \, {\bf M} = 0 \,.
\end{align}
We decompose $\pi_1 \, {\bf M}$ as
\begin{equation}
\pi_1 \, {\bf M} = \sum_{n=0}^\infty M_n \, \pi_n \,,
\end{equation}
where $M_n$ is an $n$-string operator.
Then the multi-string operators $\{ \, M_0, M_1, M_2 \ldots \, \}$
satisfy the weak $A_\infty$ relations,
and they have the cyclic property given by
\begin{equation}
\langle \omega | \, ( \, M_n \otimes P +P \otimes M_n \, ) = 0 \,.
\end{equation}

\section{Construction to all orders}
\label{section-8}
\setcounter{equation}{0}

In this section we use the coalgebra representation of the weak $A_\infty$ structure
explained in the preceding section
to construct multi-string operators satisfying the weak $A_\infty$ relations to all orders
in $g$ and $\kappa$.
We first explain the construction
for open bosonic string field theory without the source term
in~\S\ref{without-source-term-subsection}.
This is not a new result,
and the construction was used in~\cite{Matsunaga:2019fnc}, for example, in a different context.
Once we fully understand the construction in this case,
the generalization to include the source term
is rather straightforward.
We therefore explain the construction without the source term in detail.
We then present the construction
including the source term in~\S\ref{with-source-term-subsection}.

\subsection{Open bosonic string field theory without the source term
in the low-energy limit}
\label{without-source-term-subsection}

\subsubsection{Multi-string products}

For the effective action of open bosonic string field theory
without the source term,
we have presented $V_2 ( A_1, A_2 )$ in~\eqref{V_2}
and $V_3 ( A_1, A_2, A_3 )$ in~\eqref{V_3}.
The two-string product $M^{(0)}_2 ( A_1, A_2 )$ from $V_2 ( A_1, A_2 )$
and the three-string product $M^{(0)}_3 ( A_1, A_2, A_3 )$ from $V_3 ( A_1, A_2, A_3 )$
under the identification in~\eqref{V-to-M}
together with the one-string product $M^{(0)}_1 ( A_1 )$ are
\begin{align}
M^{(0)}_1 ( A_1 ) & = Q A_1 \,, \\
M^{(0)}_2 ( A_1, A_2 ) & = P \, m_2 ( A_1, A_2 ) \,,
\label{M_2} \\
M^{(0)}_3 ( A_1, A_2, A_3 )
& = {}-P \, m_2 ( \, h \, m_2 ( A_1, A_2 ),\, A_3 )
{}-P \, m_2 ( A_1,\, h \, m_2 ( A_2, A_3 ) ) \,.
\label{M_3}
\end{align}
We require the corresponding $n$-string operator $M^{(0)}_n$ to satisfy
\begin{equation}
P \, M^{(0)}_n \, P^{\otimes n} = M^{(0)}_n \,.
\label{M^(0)_n-projection}
\end{equation}
Since $P A_1 = A_1$, $P^2 = P$, and $Q \, P = P \, Q$,
we write $M^{(0)}_1 ( A_1 )$ as
\begin{equation}
M^{(0)}_1 ( A_1 ) = Q \, A_1 = Q \, P \, A_1
= Q \, P^2 \, A_1 = P \, Q \, P \, A_1
\end{equation}
and the one-string operator $M^{(0)}_1$ is given by
\begin{equation}
M^{(0)}_1 = P \, Q \, P \,.
\end{equation}
While $P$ on either side of $Q$ is sufficient because $P$ and $Q$ commute,
we write $P$ on each side of $Q$ to make it manifest
that the condition~\eqref{M^(0)_n-projection} is satisfied.
For $M^{(0)}_2 ( A_1, A_2 )$, we write it as
\begin{equation}
\begin{split}
M^{(0)}_2 ( A_1, A_2 ) & = P \, m_2 ( A_1, A_2 )
= P \, m_2 ( P A_1, P A_2 ) \\
& = P \, m_2 \, ( \, P A_1 \otimes P A_2 \, )
= P \, m_2 \, ( \, P \otimes P \, ) \, ( \, A_1 \otimes A_2 \, ) \,,
\end{split}
\end{equation}
and the two-string operator $M^{(0)}_2$ is given by
\begin{equation}
M^{(0)}_2 = P \, m_2 \, ( \, P \otimes P \, ) \,.
\end{equation}
For $M^{(0)}_3 ( A_1, A_2, A_3 )$, we write it as
\begin{equation}
\begin{split}
M^{(0)}_3 ( A_1, A_2, A_3 )
& = {}-P \, m_2 ( \, h \, m_2 ( A_1, A_2 ),\, A_3 )
{}-P \, m_2 ( A_1,\, h \, m_2 ( A_2, A_3 ) ) \\
& = {}-P \, m_2 ( \, h \, m_2 ( P A_1, P A_2 ),\, P A_3 )
{}-P \, m_2 ( P A_1,\, h \, m_2 ( P A_2, P A_3 ) ) \\
& = {}-P \, m_2 \, ( \, h \, m_2 \otimes P +P \otimes h \, m_2 \, ) \,
( \, P \otimes P \otimes P \, ) \, ( \, A_1 \otimes A_2 \otimes A_3 \, ) \,,
\end{split}
\end{equation}
and the three-string operator $M^{(0)}_3$ is given by
\begin{equation}
M^{(0)}_3 = {}-P \, m_2 \, ( \, h \, m_2 \otimes P +P \otimes h \, m_2 \, ) \,
( \, P \otimes P \otimes P \, ) \,.
\end{equation}
Our goal is thus to construct
a linear operator ${\bf M}^{(0)}$ on $T \mathcal{H}$
of degree odd which satisfies
\begin{align}
& {\bf P} \, {\bf M}^{(0)} \, {\bf P} = {\bf M}^{(0)} \,,
\label{M^(0)-P} \\
& \triangle \, {\bf M}^{(0)}
= ( \, {\bf M}^{(0)} \otimes' {\bf P} +{\bf P} \otimes' {\bf M}^{(0)} \, ) \,
\triangle \,,
\label{M^(0)-coderivation} \\
& [ \, {\bf M}^{(0)}, {\bf M}^{(0)} \, ] = 0 \,,
\label{M^(0)^2=0} \\
& \langle \omega | \, \pi_2 \, {\bf M}^{(0)} = 0
\label{M^(0)-cyclic}
\end{align}
with $M^{(0)}_0$, $M^{(0)}_1$, $M^{(0)}_2$, and $M^{(0)}_3$ given by
\begin{align}
M^{(0)}_0 & = 0 \,, \\
M^{(0)}_1 & = P \, Q \, P \,, \\
M^{(0)}_2 & = P \, m_2 \, ( \, P \otimes P \, ) \,, \\
M^{(0)}_3 & = {}-P \, m_2 \, ( \, h \, m_2 \otimes P +P \otimes h \, m_2 \, ) \,
( \, P \otimes P \otimes P \, ) \,,
\end{align}
where the $n$-string operator $M^{(0)}_n$ is defined in terms of the decomposition
\begin{equation}
\pi_1 \, {\bf M}^{(0)} = \sum_{n=0}^\infty M^{(0)}_n \, \pi_n \,.
\end{equation} 

\subsubsection{Expression to all orders}

In constructing ${\bf M}^{(0)}$, 
we decompose it as
\begin{equation}
{\bf M}^{(0)} = \sum_{n=1}^\infty \, {\bf M}^{(0)}_n \,,
\end{equation}
where ${\bf M}^{(0)}_n$ satisfies
\begin{align}
& {\bf P} \, {\bf M}^{(0)}_n \, {\bf P} = {\bf M}^{(0)}_n \,,
\label{M^(0)_n-P} \\
& \triangle \, {\bf M}^{(0)}_n
= ( \, {\bf M}^{(0)}_n \otimes' {\bf P} +{\bf P} \otimes' {\bf M}^{(0)}_n \, ) \,
\triangle \,,
\label{M^(0)_n-coderivation} \\
& \langle \omega | \, \pi_2 \, {\bf M}^{(0)}_n = 0
\label{M^(0)_n-cyclic}
\end{align}
with
\begin{align}
\pi_1 \, {\bf M}^{(0)}_n = M^{(0)}_n \, \pi_n \,.
\label{pi_1-M^(0)_n}
\end{align}
The condition~\eqref{M^(0)^2=0} is also decomposed as
\begin{equation}
\sum_{m=1}^{n} \, [ \, {\bf M}^{(0)}_m, {\bf M}^{(0)}_{n-m+1} \, ] = 0
\label{M^(0)^2-decomposed}
\end{equation}
for $n = 1, 2, 3, \ldots  \,$.

Let us begin with ${\bf M}^{(0)}_1$.
Since $M^{(0)}_1 = P \, Q \, P$,
we simply replace $P$ and $Q$ with ${\bf P}$ and ${\bf Q}$, respectively,
to obtain
\begin{equation}
{\bf M}^{(0)}_1 = {\bf P} \, {\bf Q} \, {\bf P} \,.
\end{equation}
This satisfies the condition~\eqref{pi_1-M^(0)_n} for $n=1$:
\begin{equation}
\pi_1 \, {\bf M}^{(0)}_1
= \pi_1 \, {\bf P} \, {\bf Q} \, {\bf P}
= P \, Q \, P \, \pi_1 = M^{(0)}_1 \, \pi_1 \,.
\end{equation}
The condition~\eqref{M^(0)_n-P} for $n=1$ is satisfied
because ${\bf P}^2 = {\bf P}$.
We can confirm that the condition~\eqref{M^(0)_n-coderivation} for $n=1$
is satisfied in the following way:
\begin{equation}
\begin{split}
& \triangle \, {\bf P} \, {\bf Q} \, {\bf P}
= ( \, {\bf P} \otimes' {\bf P} \, ) \, \triangle \, {\bf Q} \, {\bf P}
= ( \, {\bf P} \otimes' {\bf P} \, ) \,
( \, {\bf Q} \otimes' {\mathbb I} +{\mathbb I} \otimes' {\bf Q} \, ) \, 
\triangle \, {\bf P} \\
& = ( \, {\bf P} \otimes' {\bf P} \, ) \,
( \, {\bf Q} \otimes' {\mathbb I} +{\mathbb I} \otimes' {\bf Q} \, ) \, 
( \, {\bf P} \otimes' {\bf P} \, ) \, \triangle
= ( \, {\bf P} \, {\bf Q} \, {\bf P} \otimes' {\bf P}
+{\bf P} \otimes' {\bf P} \, {\bf Q} \, {\bf P} \, ) \, \triangle \,.
\end{split}
\end{equation}
Let us next confirm that the condition~\eqref{M^(0)_n-cyclic} for $n=1$
is satisfied.
It follows from $P \, Q = Q \, P$ that
\begin{equation}
[ \, {\bf Q},\, {\bf P} \, ] = 0 \,,
\end{equation}
so we find
\begin{equation}
\langle \omega | \, \pi_2 \, {\bf P} \, {\bf Q} \, {\bf P}
= \langle \omega | \, \pi_2 \, {\bf Q} \, {\bf P}^2 = 0 \,,
\end{equation}
where we used $\langle \omega | \, \pi_2 \, {\bf Q} = 0$.
The condition~\eqref{M^(0)^2-decomposed} for $n=1$ is as follows:
\begin{equation}
[ \, {\bf P} \, {\bf Q} \, {\bf P}, {\bf P} \, {\bf Q} \, {\bf P} \, ] = 0 \,.
\end{equation}
While it is easy to calculate this commutator directly,
we will encounter similar commutators in what follows,
so let us first prove the identity
\begin{equation}
[ \, {\bf P} \, {\bf Q} \, {\bf P}, {\bf P} \, {\bf A} \, {\bf P} \, ]
= {\bf P} \, [ \, {\bf Q}, {\bf A} \, ] \, {\bf P}
\label{PQP-commutator}
\end{equation}
for any operator ${\bf A}$ on $T \mathcal{H}$.
Using $[ \, {\bf Q},\, {\bf P} \, ] = 0$ and ${\bf P}^2 = {\bf P}$,
this identity can be shown as follows:
\begin{equation}
\begin{split}
& [ \, {\bf P} \, {\bf Q} \, {\bf P}, {\bf P} \, {\bf A} \, {\bf P} \, ]
= {\bf P} \, {\bf Q} \, {\bf P} \, {\bf P} \, {\bf A} \, {\bf P}
-(-1)^{{\rm deg} ({\bf A})} \, 
{\bf P} \, {\bf A} \, {\bf P} \, {\bf P} \, {\bf Q} \, {\bf P} \\
& = {\bf P} \, {\bf Q} \, {\bf A} \, {\bf P}
-(-1)^{{\rm deg} ({\bf A})} \, 
{\bf P} \, {\bf A} \, {\bf Q} \, {\bf P}
= {\bf P} \, [ \, {\bf Q}, {\bf A} \, ] \, {\bf P} \,.
\end{split}
\end{equation}
The condition~\eqref{M^(0)^2-decomposed} for $n=1$ thus follows from
$[ \, {\bf Q},\, {\bf Q} \, ] = 0$:
\begin{equation}
[ \, {\bf P} \, {\bf Q} \, {\bf P}, {\bf P} \, {\bf Q} \, {\bf P} \, ]
= {\bf P} \, [ \, {\bf Q}, {\bf Q} \, ] \, {\bf P} = 0 \,.
\end{equation}

Let us next consider ${\bf M}^{(0)}_2$.
Since $M^{(0)}_2 = P \, m_2 \, ( \, P \otimes P \, )$,
we use ${\bf P}$ and ${\bm m}_2$ to construct ${\bf M}^{(0)}_2$ as follows:
\begin{equation}
{\bf M}^{(0)}_2 = {\bf P} \, {\bm m}_2 \, {\bf P} \,.
\end{equation}
This satisfies the condition~\eqref{pi_1-M^(0)_n} for $n=2$:
\begin{equation}
\pi_1 \, {\bf M}^{(0)}_2
= \pi_1 \, {\bf P} \, {\bm m}_2 \, {\bf P}
= P \, m_2 \, ( \, P \otimes P \, ) \, \pi_2 = M^{(0)}_2 \, \pi_2 \,.
\end{equation}
The condition~\eqref{M^(0)_n-P} for $n=2$ is satisfied
because ${\bf P}^2 = {\bf P}$.
We can confirm that the condition~\eqref{M^(0)_n-coderivation} for $n=2$
is satisfied in the following way:
\begin{equation}
\begin{split}
& \triangle \, {\bf P} \, {\bm m}_2 \, {\bf P}
= ( \, {\bf P} \otimes' {\bf P} \, ) \, \triangle \, {\bm m}_2 \, {\bf P}
= ( \, {\bf P} \otimes' {\bf P} \, ) \,
( \, {\bm m}_2 \otimes' {\bf I} +{\bf I} \otimes' {\bm m}_2 \, ) \, 
\triangle \, {\bf P} \\
& = ( \, {\bf P} \otimes' {\bf P} \, ) \,
( \, {\bm m}_2 \otimes' {\bf I} +{\bf I} \otimes' {\bm m}_2 \, ) \, 
( \, {\bf P} \otimes' {\bf P} \, ) \, \triangle
= ( \, {\bf P} \, {\bm m}_2 \, {\bf P} \otimes' {\bf P}
+{\bf P} \otimes' {\bf P} \, {\bm m}_2 \, {\bf P} \, ) \, \triangle \,.
\end{split}
\end{equation}
Let us also confirm that the condition~\eqref{M^(0)_n-cyclic} for $n=2$ is satisfied.
It follows from $P^2 = P$
and $\langle \omega | \, {\mathbb I} \otimes P = \langle \omega | \, P \otimes {\mathbb I}$
that
\begin{equation}
\langle \omega | \, ( \, P \otimes P \, )
= \langle \omega | \, ( \, {\mathbb I} \otimes P \, ) \, ( \, P \otimes {\mathbb I} \, )
= \langle \omega | \, ( \, P \otimes {\mathbb I} \, ) \, ( \, P \otimes {\mathbb I} \, )
= \langle \omega | \, ( \, P \otimes {\mathbb I} \, ) \,.
\label{omega-P-P-1}
\end{equation}
Similarly, we have
\begin{equation}
\langle \omega | \, ( \, P \otimes P \, )
= \langle \omega | \, ( \, P \otimes {\mathbb I} \, ) \, ( \, {\mathbb I} \otimes P \, )
= \langle \omega | \, ( \, {\mathbb I} \otimes P \, ) \, ( \, {\mathbb I} \otimes P \, )
= \langle \omega | \, ( \, {\mathbb I} \otimes P \, ) \,.
\label{omega-P-P-2}
\end{equation}
Using these relations, 
we can confirm that the condition~\eqref{M^(0)_n-cyclic} for $n=2$ is satisfied
in the following way:
\begin{equation}
\begin{split}
& \langle \omega | \, \pi_2 \, {\bf P} \, {\bm m}_2 \, {\bf P} \\
& = \langle \omega | \, ( \, P \otimes P \, ) \,
( \, m_2 \otimes {\mathbb I} \, ) \,
( \, P \otimes P \otimes P \, ) \, \pi_3
+\langle \omega | \, ( \, P \otimes P \, ) \,
( \, {\mathbb I} \otimes m_2 \, ) \,
( \, P \otimes P \otimes P \, ) \, \pi_3 \\
& = \langle \omega | \, ( \, {\mathbb I} \otimes P \, ) \,
( \, m_2 \otimes {\mathbb I} \, ) \,
( \, P \otimes P \otimes P \, ) \, \pi_3
+\langle \omega | \, ( \, P \otimes {\mathbb I} \, ) \,
( \, {\mathbb I} \otimes m_2 \, ) \,
( \, P \otimes P \otimes P \, ) \, \pi_3 \\
& = \langle \omega | \, 
( \, m_2 \otimes {\mathbb I} +{\mathbb I} \otimes m_2 \, ) \,
( \, P \otimes P \otimes P \, ) \, \pi_3 = 0 \,,
\end{split}
\end{equation}
where in the last step we used the cyclic property of $m_2$ in~\eqref{m_2-w_0-cyclic-properties}.
The condition~\eqref{M^(0)^2-decomposed} for $n=2$ is as follows:
\begin{equation}
[ \, {\bf P} \, {\bf Q} \, {\bf P}, {\bf P} \, {\bm m}_2 \, {\bf P} \, ] = 0 \,.
\end{equation}
Using the identity~\eqref{PQP-commutator},
this follows from
$[ \, {\bf Q},\, {\bm m}_2 \, ] = 0$:
\begin{equation}
[ \, {\bf P} \, {\bf Q} \, {\bf P}, {\bf P} \, {\bm m}_2 \, {\bf P} \, ]
= {\bf P} \, [ \, {\bf Q}, {\bm m}_2 \, ] \, {\bf P} = 0 \,.
\end{equation}

The three-string operator $M^{(0)}_3$ contains $h$ defined by
\begin{equation}
h = \frac{b_0}{L_0} \, (1-P) \,.
\end{equation}
Regarding $h$, we have so far used the relation
\begin{equation}
Q \, h +h \, Q = 1-P
\end{equation}
and the fact that $h$ is BPZ even:
\begin{equation}
\langle \omega | \, {\mathbb I} \otimes h
= \langle \omega | \, h \otimes {\mathbb I} \,.
\end{equation}
In what follows we also use the following properties of $h$:
\begin{equation}
h \, P = 0 \,, \qquad
P \, h = 0 \,, \qquad
h^2 = 0 \,.
\end{equation}
In constructing ${\bf M}^{(0)}_3$,
the crucial ingredient is the operator ${\bm h}$ introduced in~\cite{Berglund}.\footnote{
We thank Hiroaki Matsunaga for helpful discussions on the operator ${\bm h}$
used in \cite{Matsunaga:2019fnc}.
}
It is a linear operator on $T \mathcal{H}$ of degree odd
and is defined as follows.
The action on the sector $\mathcal{H}^{\otimes 0}$ vanishes:
\begin{equation}
{\bm h} \, \pi_0 = 0 \,.
\end{equation}
For the sectors $\mathcal{H}$ and $\mathcal{H}^{\otimes 2}$, its actions are as follows:
\begin{align}
{\bm h} \, \pi_1 & = h \, \pi_1 \,, \\
{\bm h} \, \pi_2
& = ( \, h \otimes P +{\mathbb I} \otimes h \, ) \, \pi_2 \,.
\end{align}
The action on the sector $\mathcal{H}^{\otimes n}$
for $n > 2$ is given by
\begin{equation}
{\bm h} \, \pi_n = \Bigl( \, h \otimes P^{\otimes (n-1)}
+\sum_{m=1}^{n-2} \, {\mathbb I}^{\otimes m} \otimes h \otimes P^{\otimes (n-m-1)}
+{\mathbb I}^{\otimes (n-1)} \otimes h \, \Bigr) \, \pi_n
\quad \text{for} \quad n > 2 \,.
\end{equation}
Note that the operator $P$ appears in the definition of ${\bm h}$.
Some examples for the action of ${\bm h}$ are
\begin{equation}
\begin{split}
{\bm h} \, {\bf 1} & = 0 \,, \\
{\bm h} \, A_1 & = h \, A_1 \,, \\
{\bm h} \, ( A_1 \otimes A_2 )
& = h \, A_1  \otimes P A_2 +(-1)^{{\rm deg} (A_1)} A_1 \otimes h \, A_2 \,, \\
{\bm h} \, ( A_1 \otimes A_2 \otimes A_3 )
& = h \, A_1 \otimes P A_2 \otimes P A_3
+(-1)^{{\rm deg} (A_1)} A_1 \otimes h \, A_2 \otimes P A_3 \\
& \quad~ +(-1)^{{\rm deg} (A_1) +{\rm deg} (A_2)} A_1 \otimes A_2 \otimes h \, A_3 \,, \\
{\bm h} \, ( A_1 \otimes A_2 \otimes A_3 \otimes A_4 )
& = h \, A_1 \otimes P A_2 \otimes P A_3 \otimes P A_4
+(-1)^{{\rm deg} (A_1)} A_1 \otimes h \, A_2 \otimes P A_3 \otimes P A_4 \\
& \quad~ +(-1)^{{\rm deg} (A_1) +{\rm deg} (A_2)}
A_1 \otimes A_2 \otimes h \, A_3 \otimes P A_4 \\
& \quad~ +(-1)^{{\rm deg} (A_1) +{\rm deg} (A_2) +{\rm deg} (A_3)}
A_1 \otimes A_2 \otimes A_3 \otimes h \, A_4 \,, \\
& \quad \vdots
\end{split}
\end{equation}
The operator ${\bm h}$ is characterized by
\begin{equation}
\pi_1 \, {\bm h} = h \, \pi_1
\end{equation}
and
\begin{equation}
\triangle \, {\bm h}
= ( \, {\bm h} \otimes' {\bf P} +{\bf I} \otimes' {\bm h} \, ) \, \triangle \,,
\end{equation}
and it has the following properties:
\begin{equation}
[ \, {\bf Q} \,, {\bm h} \, ] = {\bf I} -{\bf P} \,, \qquad
{\bm h} \, {\bf P} = 0 \,, \qquad
{\bf P} \, {\bm h} = 0 \,, \qquad
{\bm h}^2 = 0 \,.
\end{equation}
We then claim that ${\bf M}^{(0)}_3$ is given by
\begin{equation}
{\bf M}^{(0)}_3 = {}-{\bf P} \, {\bm m}_2 \, {\bm h} \, {\bm m}_2 \, {\bf P} \,.
\end{equation}
Let us first calculate $\pi_1 \, {\bf M}^{(0)}_3$
in order to confirm that the condition~\eqref{pi_1-M^(0)_n} for $n=3$ is satisfied.
We find
\begin{equation}
\begin{split}
\pi_1 \, {\bf M}^{(0)}_3
& = {}-P \, m_2 \, ( \, h \otimes P +{\mathbb I} \otimes h \, ) \,
( \, m_2 \otimes {\mathbb I} +{\mathbb I} \otimes m_2 \, ) \,
( \, P \otimes P \otimes P \, ) \, \pi_3 \\
& = {}-P \, m_2 \, ( \, h \, m_2 \otimes P +{\mathbb I} \otimes h \, m_2 \, ) \,
( \, P \otimes P \otimes P \, ) \, \pi_3 \\
& = {}-P \, m_2 \, ( \, h \, m_2 \otimes P +P \otimes h \, m_2 \, ) \,
( \, P \otimes P \otimes P \, ) \, \pi_3
= M^{(0)}_3 \, \pi_3 \,,
\end{split}
\end{equation}
where we used $h \, P =0$ and $P^2 = P$.
The condition~\eqref{M^(0)_n-P} for $n=3$ is satisfied
because ${\bf P}^2 = {\bf P}$.
The condition~\eqref{M^(0)^2-decomposed} for $n=3$ is written as
\begin{equation}
[ \, {\bf M}^{(0)}_1,\, {\bf M}^{(0)}_3 \, ] = {}-{\bf M}^{(0)}_2 \, {\bf M}^{(0)}_2 \,.
\end{equation}
Since
\begin{equation}
\begin{split}
[ \, {\bf M}^{(0)}_1,\, {\bf M}^{(0)}_3 \, ]
& = {}-[ \, {\bf P} \, {\bf Q} \, {\bf P},\,
{\bf P} \, {\bm m}_2 \, {\bm h} \, {\bm m}_2 \, {\bf P} \, ]
= {}-{\bf P} \, [ \, {\bf Q},\, {\bm m}_2 \, {\bm h} \, {\bm m}_2 \, ] \, {\bf P} \\
& = {\bf P} \, {\bm m}_2 \, [ \, {\bf Q},\, {\bm h} \, ] \, {\bm m}_2 \, {\bf P}
= {\bf P} \, {\bm m}_2 \, ( \, {\bf I} -{\bf P} \, ) \, {\bm m}_2 \, {\bf P} \\
& = {\bf P} \, {\bm m}_2 \, {\bm m}_2 \, {\bf P}
-{\bf P} \, {\bm m}_2 \, {\bf P} \, {\bm m}_2 \, {\bf P}
= {}-( \, {\bf P} \, {\bm m}_2 \, {\bf P} \, ) \,
( \, {\bf P} \, {\bm m}_2 \, {\bf P} \, ) \,,
\end{split}
\end{equation}
where we used
$[ \, {\bf Q},\, {\bm m}_2 \, ] = 0$,
$[ \, {\bf Q},\, {\bm h} \, ] = {\bf I} -{\bf P}$,
and ${\bm m}_2^2 = 0$,
the condition~\eqref{M^(0)^2-decomposed} for $n=3$ is satisfied.
In fact, it is easy to generalize this to construct ${\bf M}^{(0)}_n$
such that~\eqref{M^(0)^2-decomposed} is satisfied.
We define ${\bf M}^{(0)}_n$ for $n > 3$ by
\begin{equation}
{\bf M}^{(0)}_n
= (-1)^n \, {\bf P} \, {\bm m}_2 \, ( \, {\bm h} \, {\bm m}_2 \, )^{n-2} \, {\bf P} \,.
\end{equation}
The condition~\eqref{M^(0)^2-decomposed} for $n > 2$ is written as
\begin{equation}
[ \, {\bf M}^{(0)}_1,\, {\bf M}^{(0)}_n \, ]
= {}-\sum_{m=2}^{n-1} {\bf M}^{(0)}_m \, {\bf M}^{(0)}_{n-m+1} \,.
\end{equation}
Since
\begin{equation}
\begin{split}
& [ \, {\bf M}^{(0)}_1,\, {\bf M}^{(0)}_n \, ]
= (-1)^n \, [ \, {\bf P} \, {\bf Q} \, {\bf P},\,
{\bf P} \, {\bm m}_2 \, ( \, {\bm h} \, {\bm m}_2 \, )^{n-2} \, {\bf P} \, ]
= (-1)^n \, {\bf P} \, [ \, {\bf Q},\,
{\bm m}_2 \, ( \, {\bm h} \, {\bm m}_2 \, )^{n-2} \, ] \, {\bf P} \\
& = {}-(-1)^n \, \sum_{m=0}^{n-3} \,
{\bf P} \, {\bm m}_2 \, ( \, {\bm h} \, {\bm m}_2 \, )^m \,
( \, {\bf I} -{\bf P} \, ) \, {\bm m}_2 \,
( \, {\bm h} \, {\bm m}_2 \, )^{n-m-3} \, {\bf P} \\
& = {}-\sum_{m=0}^{n-3} \,
(-1)^{m+2} \,
( \, {\bf P} \, {\bm m}_2 \,
( \, {\bm h} \, {\bm m}_2 \, )^m \, {\bf P} \, ) \,
(-1)^{n-m-1}
( \, {\bf P} \, {\bm m}_2 \,
( \, {\bm h} \, {\bm m}_2 \, )^{n-m-3} \, {\bf P} \, ) \,,
\end{split}
\end{equation}
the condition~\eqref{M^(0)^2-decomposed} for $n > 2$ is satisfied.
While the condition~\eqref{M^(0)_n-P} is also satisfied,
it is not obvious
whether the conditions~\eqref{M^(0)_n-coderivation}
and~\eqref{M^(0)_n-cyclic} are satisfied for $n >2$.
We will show that they are satisfied
in~\S\ref{coderivation-subsubsection}
and~\S\ref{cyclic-subsubsection}.

Before we move on to consider these conditions,
let us prove~\eqref{M^(0)^2=0}
without decomposing ${\bf M}^{(0)}$.
After summing $M^{(0)}_n$ over $n$, ${\bf M}^{(0)}$ is given by
\begin{equation}
{\bf M}^{(0)} = {\bf P} \, {\bf Q} \, {\bf P}
+{\bf P} \, {\bm m}_2 \, \frac{1}{{\bf I}+{\bm h} \, {\bm m}_2}  \, {\bf P} \,,
\end{equation}
where
\begin{equation}
\frac{1}{{\bf I}+{\bm h} \, {\bm m}_2}
\equiv {\bf I} +\sum_{n=1}^\infty \, (-1)^n \, ( \, {\bm h} \, {\bm m}_2 \, )^n \,.
\end{equation}
It follows from this definition that
\begin{equation}
( \, {\bf I}+{\bm h} \, {\bm m}_2 \, ) \,
\frac{1}{{\bf I}+{\bm h} \, {\bm m}_2} = {\bf I} \,.
\label{inverse^(0)}
\end{equation}
The condition~\eqref{M^(0)-P} is satisfied
because ${\bf P}^2 = {\bf P}$.
Let us calculate $[ \, {\bf M}^{(0)} ,\, {\bf M}^{(0)} \, ] \,$.
We find
\begin{equation}
\begin{split}
& [ \, {\bf M}^{(0)},\, {\bf M}^{(0)} \, ] \\
& = [ \, {\bf P} \, {\bf Q} \, {\bf P} ,\, {\bf P} \, {\bf Q} \, {\bf P} \, ]
+2 \, \biggl[ \, {\bf P} \, {\bf Q} \, {\bf P} ,\,
{\bf P} \, {\bm m}_2 \, \frac{1}{{\bf I}+{\bm h} \, {\bm m}_2}  \, {\bf P} \, \biggr]
+2 \, \biggl( \, {\bf P} \, {\bm m}_2 \,
\frac{1}{{\bf I}+{\bm h} \, {\bm m}_2}  \, {\bf P} \, \biggr)^2 \\
& = 2 \, {\bf P} \, \biggl[ \, {\bf Q} ,\,
{\bm m}_2 \, \frac{1}{{\bf I}+{\bm h} \, {\bm m}_2}  \, \biggr] \, {\bf P} \, 
+2 \, \biggl( \, {\bf P} \, {\bm m}_2 \,
\frac{1}{{\bf I}+{\bm h} \, {\bm m}_2}  \, {\bf P} \, \biggr)^2 \\
& = {}-2 \, {\bf P} \, {\bm m}_2 \, \biggl[ \, {\bf Q} ,\,
\frac{1}{{\bf I}+{\bm h} \, {\bm m}_2}  \, \biggr] \, {\bf P} \, 
+2 \, \biggl( \, {\bf P} \, {\bm m}_2 \,
\frac{1}{{\bf I}+{\bm h} \, {\bm m}_2}  \, {\bf P} \, \biggr)^2 \,.
\end{split}
\end{equation}
Since
\begin{equation}
\begin{split}
& {\bm m}_2 \, \biggl[ \, {\bf Q} ,\,
\frac{1}{{\bf I}+{\bm h} \, {\bm m}_2}  \, \biggr]
= {}-{\bm m}_2 \, \frac{1}{{\bf I}+{\bm h} \, {\bm m}_2} \,
[ \, {\bf Q} \,,\, {\bf I}+{\bm h} \, {\bm m}_2 \, ] \,
\frac{1}{{\bf I}+{\bm h} \, {\bm m}_2} \\
& = {}-{\bm m}_2 \, \frac{1}{{\bf I}+{\bm h} \, {\bm m}_2} \,
( \, {\bf I} -{\bf P} \, ) \, {\bm m}_2 \,
\frac{1}{{\bf I}+{\bm h} \, {\bm m}_2}
= {\bm m}_2 \, \frac{1}{{\bf I}+{\bm h} \, {\bm m}_2} \,
{\bf P} \, {\bm m}_2 \,
\frac{1}{{\bf I}+{\bm h} \, {\bm m}_2} \,,
\end{split}
\end{equation}
where we used ${\bm m}_2^2 = 0 \,$, we conclude that
\begin{equation}
[ \, {\bf M}^{(0)},\, {\bf M}^{(0)} \, ] = 0 \,.
\end{equation}

\subsubsection{Coderivation}
\label{coderivation-subsubsection}

Let us show that ${\bf M}^{(0)}_n$ for $n > 2$ given by
\begin{equation}
{\bf M}^{(0)}_n
= (-1)^n \, {\bf P} \, {\bm m}_2 \, ( \, {\bm h} \, {\bm m}_2 \, )^{n-2} \, {\bf P}
\end{equation}
is a coderivation on the projected space:
\begin{equation}
\triangle \, {\bf M}^{(0)}_n
= ( \, {\bf M}^{(0)}_n \otimes' {\bf P} +{\bf P} \otimes' {\bf M}^{(0)}_n \, ) \,
\triangle \,.
\end{equation}
Since
\begin{align}
\triangle \, {\bf P} & = ( \, {\bf P} \otimes' {\bf P} \, ) \, \triangle \,, \\
\triangle \, {\bm m}_2
& = ( \, {\bm m}_2 \otimes' {\bf I} +{\bf I} \otimes' {\bm m}_2 \, ) \,
\triangle \,, \\
\triangle \, {\bm h}
& = ( \, {\bm h} \otimes' {\bf P} +{\bf I} \otimes' {\bm h} \, ) \,
\triangle \,, 
\end{align}
we have
\begin{align}
\triangle \, {\bm h} \, {\bm m}_2 \, {\bf P}
& = ( \, {\bm h} \, {\bm m}_2 \, {\bf P} \otimes' {\bf P}
+{\bf P} \otimes' {\bm h} \, {\bm m}_2 \, {\bf P} \, ) \, \triangle \,, \\
\triangle \, ( \, {\bm h} \, {\bm m}_2 \, )^2 \, {\bf P}
& = ( \, ( \, {\bm h} \, {\bm m}_2 \, )^2 \, {\bf P} \otimes' {\bf P}
+{\bm h} \, {\bm m}_2 \, {\bf P} \otimes' {\bm h} \, {\bm m}_2 \, {\bf P}
+{\bf P} \otimes' ( \, {\bm h} \, {\bm m}_2 \, )^2 \, {\bf P} \, ) \, \triangle \,,
\end{align}
where we used ${\bm h} \, {\bf P} = 0$,
${\bf P} \, {\bm h} = 0$,
and ${\bm h}^2 = 0$.
Suppose that
\begin{equation}
\triangle \, ( \, {\bm h} \, {\bm m}_2 \, )^n \, {\bf P}
= \sum_{m=0}^n ( \, ( \, {\bm h} \, {\bm m}_2 \, )^{n-m} \, {\bf P}
\otimes' ( \, {\bm h} \, {\bm m}_2 \, )^m \, {\bf P} \, ) \, \triangle \,.
\label{triangle-hm}
\end{equation}
We can then show that
\begin{align}
\triangle \, {\bm h} \, {\bm m}_2 \, ( \, {\bm h} \, {\bm m}_2 \, )^n \, {\bf P}
& = ( \, ( \, {\bm h} \, {\bm m}_2 \, )^{n+1} \, {\bf P} \otimes' {\bf P} \, ) \, \triangle
+\sum_{m=0}^n ( \, ( \, {\bm h} \, {\bm m}_2 \, )^{n-m} \, {\bf P}
\otimes' ( \, {\bm h} \, {\bm m}_2 \, )^{m+1} \, {\bf P} \, ) \, \triangle \nonumber \\
& = \sum_{m=0}^{n+1} ( \, ( \, {\bm h} \, {\bm m}_2 \, )^{n+1-m} \, {\bf P}
\otimes' ( \, {\bm h} \, {\bm m}_2 \, )^m \, {\bf P} \, ) \, \triangle \,.
\end{align}
As the assumption~\eqref{triangle-hm} holds for $n=0$, $1$, and $2$,
this proves by induction
that the relation~\eqref{triangle-hm} holds for any non-negative integer $n$.
We then find that
\begin{equation}
\begin{split}
\triangle \, {\bf P} \, {\bm m}_2 \, ( \, {\bm h} \, {\bm m}_2 \, )^n \, {\bf P}
& = \sum_{m=0}^n
( \, {\bf P} \, {\bm m}_2 \, ( \, {\bm h} \, {\bm m}_2 \, )^{n-m} \, {\bf P}
\otimes' {\bf P} \, ( \, {\bm h} \, {\bm m}_2 \, )^m \, {\bf P} \, ) \, \triangle \\
& \quad~
+\sum_{m=0}^n ( \, {\bf P} \, ( \, {\bm h} \, {\bm m}_2 \, )^{n-m} \, {\bf P}
\otimes' {\bf P} \, {\bm m}_2 \, ( \, {\bm h} \, {\bm m}_2 \, )^m \, {\bf P} \, ) \, \triangle \\
& = ( \, {\bf P} \, {\bm m}_2 \, ( \, {\bm h} \, {\bm m}_2 \, )^n \, {\bf P}
\otimes' {\bf P}
+{\bf P} \otimes' {\bf P} \, {\bm m}_2 \, ( \, {\bm h} \, {\bm m}_2 \, )^n \, {\bf P} \, ) \, \triangle \,.
\end{split}
\end{equation}
This completes the proof
that ${\bf M}^{(0)}_n$ for $n >1$ is a coderivation on the projected space.

It is useful to prove that the condition~\eqref{M^(0)-coderivation} is satisfied
without decomposing ${\bf M}^{(0)}$.
We define ${\bm f}^{(0)}$ by
\begin{equation}
{\bm f}^{(0)} \equiv \frac{1}{{\bf I}+{\bm h} \, {\bm m}_2} \, {\bf P} \,.
\end{equation}
It follows from~\eqref{triangle-hm} that
\begin{equation}
\triangle \, {\bm f}^{(0)}
= ( \, {\bm f}^{(0)} \otimes' {\bm f}^{(0)} \, ) \, \triangle \,.
\end{equation}
We also find that
\begin{equation}
{\bf P} \, {\bm f}^{(0)} = {\bf P} \,,
\end{equation}
which follows from ${\bf P} \, {\bm h} = 0$ and ${\bf P}^2 = {\bf P}$.
Then we can write ${\bf M}^{(0)}$ as
\begin{equation}
{\bf M}^{(0)}
= {\bf P} \, {\bf Q} \, {\bf P} +{\bf P} \, {\bm m}_2 \, {\bm f}^{(0)} \,,
\end{equation}
and we find
\begin{equation}
\begin{split}
\triangle \, {\bf M}^{(0)}
& = ( \, {\bf P} \, {\bf Q} \, {\bf P} \otimes' {\bf P}
+{\bf P} \otimes' {\bf P} \, {\bf Q} \, {\bf P} \, ) \, \triangle \\
& \quad~
+( \, {\bf P} \otimes' {\bf P} \, ) \,
( \, {\bm m}_2 \otimes' {\bf I} +{\bf I} \otimes' {\bm m}_2 \, ) \, 
( \, {\bm f}^{(0)} \otimes' {\bm f}^{(0)} \, ) \, \triangle \\
& = ( \, {\bf P} \, {\bf Q} \, {\bf P} \otimes' {\bf P}
+{\bf P} \otimes' {\bf P} \, {\bf Q} \, {\bf P} \, ) \, \triangle \\
& \quad~
+( \, {\bf P} \, {\bm m}_2 \, {\bm f}^{(0)} \otimes' {\bf P}
+{\bf P} \otimes' {\bf P} \, {\bm m}_2 \, {\bm f}^{(0)} \, ) \, \triangle \,,
\end{split}
\end{equation}
where we used~${\bf P} \, {\bm f}^{(0)} = {\bf P}$.
We have thus shown that the condition~\eqref{M^(0)-coderivation} is satisfied.

\subsubsection{Cyclic property}
\label{cyclic-subsubsection}

Let us next prove that the condition~\eqref{M^(0)-cyclic} is satisfied.
For this purpose, it is convenient to introduce a linear operation
which maps $T \mathcal{H} \otimes'T \mathcal{H}$ to $T \mathcal{H}$.
We denote this operation by $\bigtriangledown$,
and it acts by simply replacing $\otimes'$ with $\otimes$.
For example, its action on
$A_1 \otimes A_2 \otimes' A_3 \otimes A_4 \otimes A_5$ is given by
\begin{equation}
\bigtriangledown \,
( \, A_1 \otimes A_2 \otimes' A_3 \otimes A_4 \otimes A_5 \, )
= A_1 \otimes A_2 \otimes A_3 \otimes A_4 \otimes A_5 \,.
\end{equation}

The main identity we will use in what follows is
\begin{equation}
\pi_{m+n} = \bigtriangledown \, ( \, \pi_m \otimes' \pi_n \, ) \, \triangle \,.
\label{main-identity}
\end{equation}
It is easy to see how this identity works.
Let us consider
$\bigtriangledown \, ( \, \pi_2 \otimes' \pi_1 \, ) \, \triangle$
as an example.
Its action on the sector $\mathcal{H}^{\otimes 2}$ vanishes:
\begin{equation}
\bigtriangledown \, ( \, \pi_2 \otimes' \pi_1 \, ) \, \triangle \,
( \, A_1 \otimes A_2 \, )
= \bigtriangledown \, ( \, \pi_2 \otimes' \pi_1 \, ) \,
( \, {\bf 1} \otimes' A_1 \otimes A_2 +A_1 \otimes' A_2
+A_1 \otimes A_2 \otimes' {\bf 1} \, ) = 0 \,,
\end{equation}
as no terms in $\triangle \, ( \, A_1 \otimes A_2 \, )$
survive the projection $\pi_2 \otimes' \pi_1$.
It is clear that the action of
$\bigtriangledown \, ( \, \pi_2 \otimes' \pi_1 \, ) \, \triangle$
on $\mathcal{H}^{\otimes n}$ vanishes unless $n=3$.
Its action on the sector $\mathcal{H}^{\otimes 3}$ is given by
\begin{align}
& \bigtriangledown \, ( \, \pi_2 \otimes' \pi_1 \, ) \, \triangle \,
( \, A_1 \otimes A_2 \otimes A_3 \, ) \nonumber \\
& = \bigtriangledown \, ( \, \pi_2 \otimes' \pi_1 \, ) \,
( \, {\bf 1} \otimes' A_1 \otimes A_2  \otimes A_3
+A_1 \otimes' A_2  \otimes A_3
+A_1 \otimes A_2  \otimes' A_3
+A_1 \otimes A_2  \otimes A_3 \otimes' {\bf 1} \, ) \nonumber \\
& = \bigtriangledown \, ( \, A_1 \otimes A_2  \otimes' A_3 \, )
= A_1 \otimes A_2  \otimes A_3 \,.
\end{align}
We see that only one term in
$\triangle \, ( \, A_1 \otimes A_2 \otimes A_3 \, )$
survives the projection $\pi_2 \otimes' \pi_1$,
and we convince ourselves that
$\bigtriangledown \, ( \, \pi_2 \otimes' \pi_1 \, ) \, \triangle$
is equivalent to $\pi_3$.

The other identity we will use is
\begin{equation}
\bigtriangledown \, ( \, a_m \, \pi_m \otimes' b_n \, \pi_n \, )
= ( \, a_m \otimes b_n \, ) \,
\bigtriangledown \, ( \, \pi_m \otimes' \pi_n \, ) \,,
\end{equation}
where $a_m$ is an $m$-string operator
and $b_n$ is an $n$-string operator.
It is again easy to see how this identity works.
Consider the action of
$\bigtriangledown \, ( \, m_2 \, \pi_2 \otimes' P \, \pi_1 \, )$
on $A_1 \otimes A_2 \otimes' A_3$ as an example.
We find
\begin{equation}
\bigtriangledown \, ( \, m_2 \, \pi_2 \otimes' P \, \pi_1 \, ) \,
( \, A_1 \otimes A_2 \otimes' A_3 \, )
= \bigtriangledown \, ( \, m_2 \, ( \, A_1 \otimes A_2 \, ) \otimes' P \, A_3 \, )
= m_2 \, ( \, A_1 \otimes A_2 \, ) \otimes P \, A_3 \,.
\end{equation}
On the other hand, the action of
$( \, m_2 \otimes P \, ) \, \bigtriangledown \, ( \, \pi_2 \otimes' \pi_1 \, )$
on $A_1 \otimes A_2 \otimes' A_3$ is given by
\begin{equation}
\begin{split}
& ( \, m_2 \otimes P \, ) \, \bigtriangledown \, ( \, \pi_2 \otimes' \pi_1 \, ) \,
( \, A_1 \otimes A_2 \otimes' A_3 \, )
= ( \, m_2 \otimes P \, ) \, \bigtriangledown \, ( \, A_1 \otimes A_2 \otimes' A_3 \, ) \\
& = ( \, m_2 \otimes P \, ) \, ( \, A_1 \otimes A_2 \otimes A_3 \, )
= m_2 \, ( \, A_1 \otimes A_2 \, ) \otimes P \, A_3 \,.
\end{split}
\end{equation}
We thus conclude that
\begin{equation}
\bigtriangledown \, ( \, m_2 \, \pi_2 \otimes' P \, \pi_1 \, )
= ( \, m_2 \otimes P \, ) \, \bigtriangledown \, ( \, \pi_2 \otimes' \pi_1 \, ) \,.
\end{equation}
The generalization to different multi-string operators is straightforward.

Having finished necessary preparations,
let us prove that the condition~\eqref{M^(0)-cyclic} is satisfied.
As we have seen that $\langle \omega | \, \pi_2 \, {\bf P} \, {\bf Q} \, {\bf P}$
vanishes, we have
\begin{equation}
\langle \omega | \, \pi_2 \, {\bf M}^{(0)}
= \langle \omega | \, \pi_2 \, {\bf P} \, {\bm m}_2 \, {\bm f}^{(0)} \,.
\end{equation}
Using the identity
\begin{equation}
\pi_2 = \bigtriangledown \, ( \, \pi_1 \otimes' \pi_1 \, ) \, \triangle \,,
\end{equation}
we find
\begin{equation}
\begin{split}
\langle \omega | \, \pi_2 \, {\bf P} \, {\bm m}_2 \, {\bm f}^{(0)}
& = \langle \omega |
\bigtriangledown \, ( \, \pi_1 \otimes' \pi_1 \, ) \, \triangle \,
{\bf P} \, {\bm m}_2 \, {\bm f}^{(0)} \\
& = \langle \omega |
\bigtriangledown \, ( \, \pi_1 \otimes' \pi_1 \, ) \,
( \, {\bf P} \otimes' {\bf P} \, ) \,
( \, {\bm m}_2 \otimes' {\bf I} \, ) \,
( \, {\bm f}^{(0)} \otimes' {\bm f}^{(0)} \, ) \, \triangle \\
& \quad~ +\langle \omega |
\bigtriangledown \, ( \, \pi_1 \otimes' \pi_1 \, ) \,
( \, {\bf P} \otimes' {\bf P} \, ) \,
( \, {\bf I} \otimes' {\bm m}_2 \, ) \,
( \, {\bm f}^{(0)} \otimes' {\bm f}^{(0)} \, ) \, \triangle \,.
\end{split}
\end{equation}
We can write
$\langle \omega | \bigtriangledown \, ( \, \pi_1 \otimes' \pi_1 \, ) \,
( \, {\bf P} \otimes' {\bf P} \, )$
using~\eqref{omega-P-P-2} as
\begin{equation}
\begin{split}
& \langle \omega | \bigtriangledown \, ( \, \pi_1 \otimes' \pi_1 \, ) \,
( \, {\bf P} \otimes' {\bf P} \, )
= \langle \omega | \bigtriangledown \, ( \, P \, \pi_1 \otimes' P \, \pi_1 \, )
= \langle \omega | \, ( \, P \otimes P \, ) \, \bigtriangledown \,
( \, \pi_1 \otimes' \pi_1 \, ) \\
& = \langle \omega | \, ( \, {\mathbb I} \otimes P \, ) \, \bigtriangledown \,
( \, \pi_1 \otimes' \pi_1 \, )
= \langle \omega | \bigtriangledown \, ( \, \pi_1 \otimes' P \, \pi_1 \, )
= \langle \omega | \bigtriangledown \, ( \, \pi_1 \otimes' \pi_1 \, ) \,
( \, {\bf I} \otimes' {\bf P} \, )
\end{split}
\end{equation}
or using~\eqref{omega-P-P-1} as
\begin{equation}
\begin{split}
& \langle \omega | \bigtriangledown \, ( \, \pi_1 \otimes' \pi_1 \, ) \,
( \, {\bf P} \otimes' {\bf P} \, )
= \langle \omega | \bigtriangledown \, ( \, P \, \pi_1 \otimes' P \, \pi_1 \, )
= \langle \omega | \, ( \, P \otimes P \, ) \, \bigtriangledown \,
( \, \pi_1 \otimes' \pi_1 \, ) \\
& = \langle \omega | \, ( \, P \otimes {\mathbb I} \, ) \, \bigtriangledown \,
( \, \pi_1 \otimes' \pi_1 \, )
= \langle \omega | \bigtriangledown \, ( \, P \, \pi_1 \otimes' \pi_1 \, )
= \langle \omega | \bigtriangledown \, ( \, \pi_1 \otimes' \pi_1 \, ) \,
( \, {\bf P} \otimes' {\bf I} \, )
\end{split}
\end{equation}
to find
\begin{equation}
\begin{split}
& \quad~ \langle \omega |
\bigtriangledown \, ( \, \pi_1 \otimes' \pi_1 \, ) \,
( \, {\bf P} \otimes' {\bf P} \, ) \,
( \, {\bm m}_2 \otimes' {\bf I} \, ) \,
( \, {\bm f}^{(0)} \otimes' {\bm f}^{(0)} \, ) \, \triangle \\
& \quad~ +\langle \omega |
\bigtriangledown \, ( \, \pi_1 \otimes' \pi_1 \, ) \,
( \, {\bf P} \otimes' {\bf P} \, ) \,
( \, {\bf I} \otimes' {\bm m}_2 \, ) \,
( \, {\bm f}^{(0)} \otimes' {\bm f}^{(0)} \, ) \, \triangle \\
& = \langle \omega |
\bigtriangledown \, ( \, \pi_1 \otimes' \pi_1 \, ) \,
( \, {\bf I} \otimes' {\bf P} \, ) \,
( \, {\bm m}_2 \otimes' {\bf I} \, ) \,
( \, {\bm f}^{(0)} \otimes' {\bm f}^{(0)} \, ) \, \triangle \\
& \quad~ +\langle \omega |
\bigtriangledown \, ( \, \pi_1 \otimes' \pi_1 \, ) \,
( \, {\bf P} \otimes' {\bf I} \, ) \,
( \, {\bf I} \otimes' {\bm m}_2 \, ) \,
( \, {\bm f}^{(0)} \otimes' {\bm f}^{(0)} \, ) \, \triangle \\
& = \langle \omega |
\bigtriangledown \, ( \, \pi_1 \, {\bm m}_2 \, {\bm f}^{(0)}
\otimes' \pi_1 \, {\bf P} \, ) \, \triangle
+\langle \omega |
\bigtriangledown \, ( \, \pi_1 \, {\bf P}
\otimes' \pi_1 \, {\bm m}_2 \, {\bm f}^{(0)} \, ) \, \triangle \,,
\end{split}
\end{equation}
where we used ${\bf P} \, {\bm f}^{(0)} = {\bf P}$.

So far we have reduced $\langle \omega | \, \pi_2 \, {\bf M}^{(0)}$ as follows:
\begin{equation}
\langle \omega | \, \pi_2 \, {\bf M}^{(0)}
= \langle \omega |
\bigtriangledown \, ( \, \pi_1 \, {\bm m}_2 \, {\bm f}^{(0)}
\otimes' \pi_1 \, {\bf P} \, ) \, \triangle
+\langle \omega |
\bigtriangledown \, ( \, \pi_1 \, {\bf P}
\otimes' \pi_1 \, {\bm m}_2 \, {\bm f}^{(0)} \, ) \, \triangle \,.
\end{equation}
It follows from~\eqref{inverse^(0)} that
\begin{equation}
( \, {\bf I}+{\bm h} \, {\bm m}_2 \, ) \, {\bm f}^{(0)} = {\bf P} \,,
\end{equation}
and we write ${\bf P}$ as
\begin{equation}
{\bf P} = {\bm f}^{(0)} +{\bm h} \, {\bm m}_2 \, {\bm f}^{(0)}
\end{equation}
to find
\begin{equation}
\begin{split}
& \langle \omega |
\bigtriangledown \, ( \, \pi_1 \, {\bm m}_2 \, {\bm f}^{(0)}
\otimes' \pi_1 \, {\bf P} \, ) \, \triangle \\
& = \langle \omega |
\bigtriangledown \, ( \, \pi_1 \, {\bm m}_2 \, {\bm f}^{(0)}
\otimes' \pi_1 \, {\bm f}^{(0)} \, ) \, \triangle
+\langle \omega |
\bigtriangledown \, ( \, \pi_1 \, {\bm m}_2 \, {\bm f}^{(0)}
\otimes' \pi_1 \, {\bm h} \, {\bm m}_2 \, {\bm f}^{(0)} \, ) \, \triangle \\
& = \langle \omega |
\bigtriangledown \, ( \, \pi_1 \, {\bm m}_2 \otimes' \pi_1 \, ) \,
\triangle \, {\bm f}^{(0)}
+\langle \omega |
\bigtriangledown \, ( \, \pi_1 \, {\bm m}_2
\otimes' \pi_1 \, {\bm h} \, {\bm m}_2 \, ) \, \triangle \, {\bm f}^{(0)} \,.
\end{split}
\end{equation}
The first term on the right-hand side can be transformed as follows:
\begin{equation}
\begin{split}
& \langle \omega |
\bigtriangledown \, ( \, \pi_1 \, {\bm m}_2 \otimes' \pi_1 \, ) \,
\triangle \, {\bm f}^{(0)}
= \langle \omega |
\bigtriangledown \, ( \, m_2 \, \pi_2 \otimes' \pi_1 \, ) \,
\triangle \, {\bm f}^{(0)} \\
& = \langle \omega | \, ( \, m_2 \otimes {\mathbb I} \, ) \,
\bigtriangledown \, ( \, \pi_2 \otimes' \pi_1 \, ) \,
\triangle \, {\bm f}^{(0)}
= \langle \omega | \, ( \, m_2 \otimes {\mathbb I} \, ) \, \pi_3 \, {\bm f}^{(0)} \,.
\end{split}
\end{equation}
The second term on the right-hand side can be transformed as follows:
\begin{equation}
\begin{split}
\langle \omega |
\bigtriangledown \, ( \, \pi_1 \, {\bm m}_2
\otimes' \pi_1 \, {\bm h} \, {\bm m}_2 \, ) \, \triangle \, {\bm f}^{(0)}
& = {}-\langle \omega |
\bigtriangledown \, ( \, \pi_1 \otimes' h \, \pi_1 \, ) \,
( \, {\bm m}_2 \otimes' {\bm m}_2 \, ) \, \triangle \, {\bm f}^{(0)} \\
& = {}-\langle \omega | \, ( \, {\mathbb I} \otimes h \, )
\bigtriangledown \, ( \, \pi_1 \otimes' \pi_1 \, ) \,
( \, {\bm m}_2 \otimes' {\bm m}_2 \, ) \, \triangle \, {\bm f}^{(0)} \,.
\end{split}
\end{equation}
Similarly, we have
\begin{equation}
\begin{split}
& \langle \omega |
\bigtriangledown \, ( \, \pi_1 \, {\bf P}
\otimes' \pi_1 \, {\bm m}_2 \, {\bm f}^{(0)} \, ) \, \triangle \\
& = \langle \omega | \, ( \, {\mathbb I} \otimes m_2 \, ) \, \pi_3 \, {\bm f}^{(0)}
+\langle \omega | \, ( \, h \otimes {\mathbb I} \, )
\bigtriangledown \, ( \, \pi_1 \otimes' \pi_1 \, ) \,
( \, {\bm m}_2 \otimes' {\bm m}_2 \, ) \, \triangle \, {\bm f}^{(0)} \,.
\end{split}
\end{equation}
We therefore find
\begin{equation}
\begin{split}
\langle \omega | \, \pi_2 \, {\bf M}^{(0)}
& = \langle \omega |
\bigtriangledown \, ( \, \pi_1 \, {\bm m}_2 \, {\bm f}^{(0)}
\otimes' \pi_1 \, {\bf P} \, ) \, \triangle
+\langle \omega |
\bigtriangledown \, ( \, \pi_1 \, {\bf P}
\otimes' \pi_1 \, {\bm m}_2 \, {\bm f}^{(0)} \, ) \, \triangle \\
& = \langle \omega | \,
( \, m_2 \otimes {\mathbb I} +{\mathbb I} \otimes m_2 \, ) \, \pi_3 \, {\bm f}^{(0)} \\
& \quad~ {}-\langle \omega | \, ( \, {\mathbb I} \otimes h -h \otimes {\mathbb I} \, )
\bigtriangledown \, ( \, \pi_1 \otimes' \pi_1 \, ) \,
( \, {\bm m}_2 \otimes' {\bm m}_2 \, ) \, \triangle \, {\bm f}^{(0)} \\
& = 0 \,,
\end{split}
\end{equation}
where we used
\begin{equation}
\begin{split}
& \langle \omega | \, ( \, m_2 \otimes {\mathbb I}
+{\mathbb I} \otimes m_2 \, ) = 0 \,, \\
& \langle \omega | \, {\mathbb I} \otimes h
= \langle \omega | \, h \otimes {\mathbb I} \,.
\end{split}
\end{equation}
This completes the proof that the condition~\eqref{M^(0)-cyclic} is satisfied.

\subsection{Open bosonic string field theory with the source term
in the low-energy limit}
\label{with-source-term-subsection}

We have explained the construction of ${\bf M}^{(0)}$
for the effective action of open bosonic string field theory
without the source term.
In this subsection we incorporate the source term
into the construction.

\subsubsection{Multi-string products}
\label{multi-string-products-subsubsection}

Let us first summarize the results in section~\ref{section-6}.
The zero-string product $M_0$ is related to $V_0$ as
\begin{equation}
M_0 = V_0 \,.
\end{equation}
We expand $M_0$ in $\kappa$ as follows:
\begin{equation}
M_0 = \sum_{n=1}^\infty \kappa^n M_0^{(n)} \,,
\end{equation}
where $M_0^{(1)}$, $M_0^{(2)}$, and $M_0^{(3)}$ are 
\begin{align}
M_0^{(1)} & = P \, w_0 \,, \\
M_0^{(2)}
& = P \, m_2 ( \, h \, w_0, \, h \, w_0 \, ) \,, \\
M_0^{(3)}
& = P \, m_2 ( \, h \,
m_2 ( h \, w_0, h \, w_0 ), \, h \, w_0 )
+P \, m_2 ( h \, w_0, \, h \, m_2 ( \, h \, w_0, h \, w_0 ) ) \,.
\end{align}
The one-string product $M_1 ( A_1 )$ is related to $V_1 ( A_1 )$ as
\begin{equation}
M_1 ( A_1 ) = V_1 ( A_1 ) \,.
\end{equation}
We expand $M_1 ( A_1 )$ in $\kappa$ as follows:
\begin{equation}
M_1 ( A_1 ) = \sum_{n=0}^\infty \kappa^n M_1^{(n)} ( A_1 ) \,,
\end{equation}
where $M_1^{(0)} ( A_1 )$, $M_1^{(1)} ( A_1 )$, and $M_1^{(2)} ( A_1 )$ are 
\begin{align}
M_1^{(0)} ( A_1 ) & = Q A_1 \,, \\
M_1^{(1)} ( A_1 )
& = {}-P \, m_2 ( \, h \, w_0,\, A_1 )
{}-P \, m_2 ( A_1, h \, w_0 \, ) \,, \\
M_1^{(2)} ( A_1 ) & =
{}-P \, m_2 ( \, h \, m_2 ( \, A_1,
h \, w_0 \, ), h \, w_0 \, )
{}-P \, m_2 ( \, h \, m_2 ( \, h \, w_0, A_1 \, ) ,\, h \, w_0 \, ) \nonumber \\
& \quad~
{}-P \, m_2 ( \, h \, m_2 ( \, h \, w_0,
h \, w_0 \, ), A_1 \, )
{}-P \, m_2 ( \, A_1, h \, m_2 ( \, h \, w_0, 
h \, w_0 \, ) \, ) \nonumber \\
& \quad~
{}-P \, m_2 ( \, h \, w_0, h \, m_2 ( \, A_1, h \, w_0 \, ) \, )
{}-P \, m_2 ( \, h \, w_0, h \, m_2 ( \, h \, w_0, A_1 \, ) \, ) \,.
\end{align}
The two-string product $M_2 ( A_1, A_2 )$ is related to $V_2 ( A_1, A_2 )$ as
\begin{equation}
M_2 ( A_1, A_2 ) = (-1)^{\rm deg (A_1)} \, V_2 ( A_1, A_2 ) \,.
\end{equation}
We expand $M_2 ( A_1, A_2 )$ in $\kappa$ as follows:
\begin{equation}
M_2 ( A_1, A_2 ) = \sum_{n=0}^\infty \kappa^n M_2^{(n)} ( A_1, A_2 ) \,,
\end{equation}
where $M_2^{(0)} ( A_1, A_2 )$ and $M_2^{(1)} ( A_1, A_2 )$ are 
\begin{align}
M_2^{(0)} ( A_1, A_2 ) & = P \, m_2 ( A_1, A_2 ) \,, \\
M_2^{(1)} ( A_1, A_2 )
& = P \, m_2 ( \, h \, m_2 ( h \, w_0, A_1 ),\, A_2 )
+P \, m_2 ( \, h \, m_2 ( A_1, h \, w_0 ),\, A_2 )
\nonumber \\
& \quad~ +P \, m_2 ( \, h \, m_2 ( A_1, A_2 ),\, h \, w_0 )
+P \, m_2 ( h \, w_0,\, h \, m_2 ( A_1, A_2 ) )
\nonumber \\
& \quad~ +P \, m_2 ( A_1,\, h \, m_2 ( h \, w_0, A_2 ) )
+P \, m_2 ( A_1,\, h \, m_2 ( A_2, h \, w_0 ) ) \,.
\end{align}
The three-string product $M_3 ( A_1, A_2, A_3 )$ is related to $V_3 ( A_1, A_2, A_3 )$ as
\begin{equation}
M_3 ( A_1, A_2, A_3 ) = (-1)^{\rm deg (A_2)} \, V_3 ( A_1, A_2, A_3 ) \,.
\end{equation}
We expand $M_3 ( A_1, A_2, A_3 )$ in $\kappa$ as follows:
\begin{equation}
M_3 ( A_1, A_2, A_3 ) = \sum_{n=0}^\infty \kappa^n M_3^{(n)} ( A_1, A_2, A_3 ) \,,
\end{equation}
where $M_3^{(0)} ( A_1, A_2, A_3 )$ is
\begin{equation}
M_3^{(0)} ( A_1, A_2, A_3 )
= {}-P \, m_2 ( \, h \, m_2 ( A_1, A_2 ),\, A_3 )
-P \, m_2 ( A_1,\, h \, m_2 ( A_2, A_3 ) ) \,.
\end{equation}

\subsubsection{Expression to all orders}

Our goal is to construct
a linear operator ${\bf M}$ on $T \mathcal{H}$
of degree odd which satisfies
\begin{align}
& {\bf P} \, {\bf M} \, {\bf P} = {\bf M} \,,
\label{M-P} \\
& \triangle \, {\bf M}
= ( \, {\bf M} \otimes' {\bf P} +{\bf P} \otimes' {\bf M} \, ) \,
\triangle \,,
\label{M-coderivation} \\
& [ \, {\bf M}, {\bf M} \, ] = 0 \,,
\label{M-A_infinity} \\
& \langle \omega | \, \pi_2 \, {\bf M} = 0
\label{M-cyclic}
\end{align}
and reduces to ${\bf M}^{(0)}$ when $\kappa = 0$,
\begin{equation}
{\bf M} \, \Bigr|_{\kappa=0} = {\bf M}^{(0)} \,,
\end{equation}
with $M_0$ given by
\begin{equation}
M_0 = \kappa \, P \, w_0 +O(\kappa^2) \,,
\end{equation}
where the $n$-string operator $M_n$ is defined in terms of the decomposition
\begin{equation}
\pi_1 \, {\bf M} = \sum_{n=0}^\infty M_n \, \pi_n \,.
\end{equation} 
Actually, the construction of ${\bf M}^{(0)}$ immediately generalizes
to that of ${\bf M}$. We define ${\bf M}$ by
\begin{equation}
{\bf M} = {\bf P} \, {\bf Q} \, {\bf P}
+{\bf P} \, ( \, {\bm m}_2 +\kappa \, {\bm w}_0 \, ) \,
\frac{1}{{\bf I} +{\bm h} \,
( \, {\bm m}_2 +\kappa \, {\bm w}_0 \, )} \, {\bf P} \,,
\end{equation}
where
\begin{equation}
\frac{1}{{\bf I}+{\bm h} \, ( \, {\bm m}_2 +\kappa \, {\bm w}_0 \, )}
\equiv {\bf I} +\sum_{n=1}^\infty \,
(-1)^n \, ( \, {\bm h} \, ( \, {\bm m}_2 +\kappa \, {\bm w}_0 \, ) \, )^n \,.
\end{equation}
It follows from this definition that
\begin{equation}
( \, {\bf I}+{\bm h} \, ( \, {\bm m}_2 +\kappa \, {\bm w}_0 \, ) \, ) \,
\frac{1}{{\bf I}+{\bm h} \, ( \, {\bm m}_2 +\kappa \, {\bm w}_0 \, )} = {\bf I} \,.
\label{inverse}
\end{equation}
The condition~\eqref{M-P} is satisfied because ${\bf P}^2 = {\bf P}$.
It is also obvious that ${\bf M}$ reduces to ${\bf M}^{(0)}$ at $\kappa = 0$.

Let us next prove that the condition~\eqref{M-A_infinity} is satisfied.
We define ${\bm n}$ by
\begin{equation}
{\bm n} \equiv {\bm m}_2 +\kappa \, {\bm w}_0 \,.
\end{equation}
Since ${\bm n}$ is a sum of two coderivations of degree odd,
it is also a coderivation of degree odd.
It has the following properties:
\begin{equation}
[ \, {\bf Q}, {\bm n} \, ] = 0 \,, \qquad
[ \, {\bm n}, {\bm n} \, ] = 0 \,, \qquad
\end{equation}
which follow from
$[ \, {\bf Q}, {\bm m}_2 \, ] = 0$, 
$[ \, {\bf Q}, {\bm w}_0 \, ] = 0$,
$[ \, {\bm m}_2, {\bm m}_2 \, ] = 0$,
$[ \, {\bm m}_2, {\bm w}_0 \, ] = 0$,
and $[ \, {\bm w}_0, {\bm m}_2 \, ] = 0$.
If we compare ${\bf M}$ and ${\bf M}^{(0)}$,
we see that ${\bf M}$ is obtained from ${\bf M}^{(0)}$
by replacing ${\bm m}_2$ with ${\bm n}$.
In the proof that $[ \, {\bf M}^{(0)},\, {\bf M}^{(0)} \, ] = 0$,
the properties we used regarding the coderivation ${\bm m}_2$ of degree odd
were $[ \, {\bf Q}, {\bm m}_2 \, ] = 0$ and
$[ \, {\bm m}_2, {\bm m}_2 \, ] = 0$.
Since the coderivation ${\bm n}$ of degree odd has the properties
$[ \, {\bf Q}, {\bm n} \, ] = 0$ and
$[ \, {\bm n}, {\bm n} \, ] = 0$,
we can replace ${\bm m}_2$ with ${\bm n}$
in the proof that $[ \, {\bf M}^{(0)},\, {\bf M}^{(0)} \, ] = 0$
to conclude that the condition $[ \, {\bf M}, {\bf M} \, ] = 0$ is satisfied.

\subsubsection{Coderivation}

We now move on to the proof that the condition~\eqref{M-coderivation} is satisfied.
We define ${\bm f}$ by
\begin{equation}
{\bm f}
\equiv \frac{1}{{\bf I} +{\bm h} \, ( \, {\bm m}_2 +\kappa \, {\bm w}_0 \, )} \,
{\bf P}
= \frac{1}{{\bf I} +{\bm h} \, {\bm n}} \, {\bf P} \,,
\end{equation}
and we write ${\bf M}$ as
\begin{equation}
{\bf M} = {\bf P} \, {\bf Q} \, {\bf P}
+{\bf P} \, {\bm n} \, {\bm f} \,.
\end{equation}
When we proved in~\S\ref{without-source-term-subsection}
that ${\bm f}^{(0)}$ given by
\begin{equation}
{\bm f}^{(0)} = \frac{1}{{\bf I}+{\bm h} \, {\bm m}_2} \, {\bf P}
\end{equation}
satisfies the relation
\begin{equation}
\triangle \, {\bm f}^{(0)}
= ( \, {\bm f}^{(0)} \otimes' {\bm f}^{(0)} \, ) \, \triangle \,,
\end{equation}
the only property we used regarding ${\bm m}_2$ in ${\bm f}^{(0)}$
is that it is a coderivation of degree odd.
Since ${\bm f}$ is obtained from ${\bm f}^{(0)}$
by replacing ${\bm m}_2$ with ${\bm n}$
and ${\bm n}$ is a coderivation of degree odd,
we conclude that
\begin{equation}
\triangle \, {\bm f}
= ( \, {\bm f} \otimes' {\bm f} \, ) \, \triangle \,.
\end{equation}
We then find that
\begin{equation}
\triangle \, {\bf M}
= ( \, {\bf P} \, {\bf Q} \, {\bf P} \otimes' {\bf P}
+{\bf P} \otimes' {\bf P} \, {\bf Q} \, {\bf P} \, ) \,
\triangle
+( \, {\bf P} \, {\bm n} \, {\bm f}
\otimes' {\bf P} \, {\bm f}
+{\bf P} \, {\bm f}
\otimes' {\bf P} \, {\bm n} \, {\bm f} \, ) \, \triangle \,.
\end{equation}
It follows from ${\bf P} \, {\bm h} = 0$ and ${\bf P}^2 = {\bf P}$ that
\begin{equation}
{\bf P} \, {\bm f} = {\bf P} \,,
\end{equation}
and thus we conclude that the condition~\eqref{M-coderivation} is satisfied.

\subsubsection{Cyclic property}

Finally, let us prove that the condition~\eqref{M-cyclic} is satisfied.
When we reduced $\langle \omega | \, \pi_2 \, {\bf M}^{(0)}$ as
\begin{equation}
\langle \omega | \, \pi_2 \, {\bf M}^{(0)}
= \langle \omega |
\bigtriangledown \, ( \, \pi_1 \, {\bm m}_2 \, {\bm f}^{(0)}
\otimes' \pi_1 \, {\bf P} \, ) \, \triangle
+\langle \omega |
\bigtriangledown \, ( \, \pi_1 \, {\bf P}
\otimes' \pi_1 \, {\bm m}_2 \, {\bm f}^{(0)} \, ) \, \triangle
\end{equation}
in~\S\ref{without-source-term-subsection},
the only property we used regarding ${\bm m}_2$ in ${\bf M}^{(0)}$
is that it is a coderivation of degree odd.
Since ${\bf M}$ is obtained from ${\bf M}^{(0)}$
by replacing ${\bm m}_2$ with ${\bm n}$
and ${\bm n}$ is a coderivation of degree odd,
we conclude that $\langle \omega | \, \pi_2 \, {\bf M}$ can be reduced as
\begin{equation}
\langle \omega | \, \pi_2 \, {\bf M}
= \langle \omega |
\bigtriangledown \, ( \, \pi_1 \, {\bm n} \, {\bm f}
\otimes' \pi_1 \, {\bf P} \, ) \, \triangle
+\langle \omega |
\bigtriangledown \, ( \, \pi_1 \, {\bf P}
\otimes' \pi_1 \, {\bm n} \, {\bm f} \, ) \, \triangle \,,
\end{equation}
where ${\bm f}$ on the right-hand side is obtained from ${\bm f}^{(0)}$
by replacing ${\bm m}_2$ with ${\bm n}$.
It follows from~\eqref{inverse} that
\begin{equation}
( \, {\bf I}+{\bm h} \, {\bm n} \, ) \, {\bm f} = {\bf P} \,,
\label{f-relation}
\end{equation}
and we write ${\bf P}$ as
\begin{equation}
{\bf P} = {\bm f} +{\bm h} \, {\bm n} \, {\bm f}
\end{equation}
to find
\begin{equation}
\begin{split}
& \langle \omega |
\bigtriangledown \, ( \, \pi_1 \, {\bm n} \, {\bm f}
\otimes' \pi_1 \, {\bf P} \, ) \, \triangle \\
& = \langle \omega |
\bigtriangledown \, ( \, \pi_1 \, {\bm n} \, {\bm f}
\otimes' \pi_1 \, {\bm f} \, ) \, \triangle
+\langle \omega |
\bigtriangledown \, ( \, \pi_1 \, {\bm n} \, {\bm f}
\otimes' \pi_1 \, {\bm h} \, {\bm n} \, {\bm f} \, ) \, \triangle \\
& = \langle \omega |
\bigtriangledown \, ( \, \pi_1 \, {\bm n} \otimes' \pi_1 \, ) \,
\triangle \, {\bm f}
+\langle \omega |
\bigtriangledown \, ( \, \pi_1 \, {\bm n}
\otimes' \pi_1 \, {\bm h} \, {\bm n} \, ) \, \triangle \, {\bm f} \,.
\end{split}
\end{equation}
The first term on the right-hand side can be transformed as follows:
\begin{equation}
\begin{split}
& \langle \omega |
\bigtriangledown \, ( \, \pi_1 \, {\bm n} \otimes' \pi_1 \, ) \,
\triangle \, {\bm f}
= \langle \omega |
\bigtriangledown \, ( \, m_2 \, \pi_2 \otimes' \pi_1 \, ) \,
\triangle \, {\bm f}
+\kappa \, \langle \omega |
\bigtriangledown \, ( \, w_0 \, \pi_0 \otimes' \pi_1 \, ) \,
\triangle \, {\bm f} \\
& = \langle \omega | \, ( \, m_2 \otimes {\mathbb I} \, ) \,
\bigtriangledown \, ( \, \pi_2 \otimes' \pi_1 \, ) \,
\triangle \, {\bm f}
+\kappa \, \langle \omega | \, ( \, w_0 \otimes {\mathbb I} \, ) \,
\bigtriangledown \, ( \, \pi_0 \otimes' \pi_1 \, ) \,
\triangle \, {\bm f} \\
& = \langle \omega | \, ( \, m_2 \otimes {\mathbb I} \, ) \, \pi_3 \, {\bm f}
+\kappa \, \langle \omega | \, ( \, w_0 \otimes {\mathbb I} \, ) \, \pi_1 \, {\bm f} \,.
\end{split}
\end{equation}
The second term on the right-hand side can be transformed as follows:
\begin{equation}
\begin{split}
\langle \omega |
\bigtriangledown \, ( \, \pi_1 \, {\bm n}
\otimes' \pi_1 \, {\bm h} \, {\bm n} \, ) \, \triangle \, {\bm f}
& = {}-\langle \omega |
\bigtriangledown \, ( \, \pi_1 \otimes' h \, \pi_1 \, ) \,
( \, {\bm n} \otimes' {\bm n} \, ) \, \triangle \, {\bm f} \\
& = {}-\langle \omega | \, ( \, {\mathbb I} \otimes h \, )
\bigtriangledown \, ( \, \pi_1 \otimes' \pi_1 \, ) \,
( \, {\bm n} \otimes' {\bm n} \, ) \, \triangle \, {\bm f} \,.
\end{split}
\end{equation}
Similarly, we have
\begin{equation}
\begin{split}
\langle \omega |
\bigtriangledown \, ( \, \pi_1 \, {\bf P}
\otimes' \pi_1 \, {\bm n} \, {\bm f} \, ) \, \triangle
& = \langle \omega | \, ( \, {\mathbb I} \otimes m_2 \, ) \, \pi_3 \, {\bm f}
+\kappa \, \langle \omega | \, ( \, {\mathbb I} \otimes w_0 \, ) \, \pi_1 \, {\bm f} \\
& \quad~ +\langle \omega | \, ( \, h \otimes {\mathbb I} \, )
\bigtriangledown \, ( \, \pi_1 \otimes' \pi_1 \, ) \,
( \, {\bm n} \otimes' {\bm n} \, ) \, \triangle \, {\bm f} \,.
\end{split}
\end{equation}
We therefore find
\begin{equation}
\begin{split}
\langle \omega | \, \pi_2 \, {\bf M}
& = \langle \omega |
\bigtriangledown \, ( \, \pi_1 \, {\bm n} \, {\bm f}
\otimes' \pi_1 \, {\bf P} \, ) \, \triangle
+\langle \omega |
\bigtriangledown \, ( \, \pi_1 \, {\bf P}
\otimes' \pi_1 \, {\bm n} \, {\bm f} \, ) \, \triangle \\
& = \langle \omega | \,
( \, m_2 \otimes {\mathbb I} +{\mathbb I} \otimes m_2 \, ) \, \pi_3 \, {\bm f}
+\kappa \, \langle \omega | \,
( \, w_0 \otimes {\mathbb I} +{\mathbb I} \otimes w_0 \, ) \, \pi_1 \, {\bm f} \\
& \quad~ {}-\langle \omega | \, ( \, {\mathbb I} \otimes h -h \otimes {\mathbb I} \, )
\bigtriangledown \, ( \, \pi_1 \otimes' \pi_1 \, ) \,
( \, {\bm n} \otimes' {\bm n} \, ) \, \triangle \, {\bm f} \\
& = 0 \,,
\end{split}
\end{equation}
where we used
\begin{equation}
\begin{split}
& \langle \omega | \, ( \, m_2 \otimes {\mathbb I}
+{\mathbb I} \otimes m_2 \, ) = 0 \,, \\
& \langle \omega | \, ( \, w_0 \otimes {\mathbb I}
+{\mathbb I} \otimes w_0 \, ) = 0 \,, \\
& \langle \omega | \, {\mathbb I} \otimes h
= \langle \omega | \, h \otimes {\mathbb I} \,.
\end{split}
\end{equation}
This completes the proof that the condition~\eqref{M-cyclic} is satisfied.

\subsubsection{Reproducing multi-string products}

We have constructed ${\bf M}$ which satisfies the conditions~\eqref{M-P},
\eqref{M-coderivation}, \eqref{M-A_infinity}, and~\eqref{M-cyclic}
and reduces to ${\bf M}^{(0)}$ when $\kappa =0$.
Let us extract $M_n$ from ${\bf M}$ to confirm
that the expressions for the multi-string products in~\S\ref{multi-string-products-subsubsection}
are reproduced.

The $n$-string operator $M_n$ can be obtained from the decomposition
of $\pi_1 \, {\bf M}$.
We begin the decomposition as follows:
\begin{equation}
\begin{split}
\pi_1 \, {\bf M}
& = \pi_1 \, {\bf P} \, {\bf Q} \, {\bf P}
+\pi_1 \, {\bf P} \, {\bm n} \, {\bm f}
= P \, Q \, P \, \pi_1
+P \, m_2 \, \pi_2 \, {\bm f}
+\kappa \, P \, w_0 \, \pi_0 \, {\bm f} \\
& = \kappa \, P \, w_0 \, \pi_0
+P \, Q \, P \, \pi_1
+P \, m_2 \, \pi_2 \, {\bm f} \,,
\end{split}
\end{equation}
where we used
\begin{equation}
\pi_0 \, {\bm f} = \pi_0 \,.
\end{equation}
The calculation of $\pi_2 \, {\bm f}$ can be reduced
to that of $\pi_1 \, {\bm f}$ as follows:
\begin{equation}
\pi_2 \, {\bm f}
= \bigtriangledown \, ( \, \pi_1 \otimes' \pi_1 \, ) \, \triangle \, {\bm f}
= \bigtriangledown \, ( \, \pi_1 \, {\bm f} \otimes' \pi_1 \, {\bm f} \, ) \, \triangle \,,
\end{equation}
where we used the identity~\eqref{main-identity} with $n=1$ and $m=1$.
We decompose $\pi_1 \, {\bm f}$ as
\begin{equation}
\pi_1 \, {\bm f} = \sum_{n=0}^\infty f_n \, \pi_n \,,
\end{equation}
where $f_n$ is an $n$-string operator of degree odd,
and $\pi_2 \, {\bm f}$ can be expressed in terms of $f_n$ as
\begin{equation}
\begin{split}
\pi_2 \, {\bm f}
& = \sum_{k=0}^\infty \sum_{\ell=0}^\infty
\bigtriangledown \, ( \, f_k \, \pi_k \otimes' f_\ell \, \pi_\ell \, ) \, \triangle
= \sum_{k=0}^\infty \sum_{\ell=0}^\infty \,
( \, f_k \otimes f_\ell \, )
\bigtriangledown \, ( \, \pi_k \otimes' \pi_\ell \, ) \, \triangle \\
& = \sum_{k=0}^\infty \sum_{\ell=0}^\infty \,
( \, f_k \otimes f_\ell \, ) \, \pi_{k+\ell}
= \sum_{n=0}^\infty \sum_{m=0}^n \,
( \, f_m \otimes f_{n-m} \, ) \, \pi_n \,.
\end{split}
\end{equation}
Since
\begin{equation}
\pi_1 \, {\bf M}
=\kappa \, P \, w_0 \, \pi_0
+P \, Q \, P \, \pi_1
+\sum_{n=0}^\infty \sum_{m=0}^n \,
P \, m_2 \, ( \, f_m \otimes f_{n-m} \, ) \, \pi_n \,,
\end{equation}
the $n$-string operator $M_n$ is given by
\begin{align}
M_0 & = \kappa \, P \, w_0 +P \, m_2 \, ( \, f_0 \otimes f_0 \, ) \,, \\
M_1 & = P \, Q \, P
+P \, m_2 \, ( \, f_0 \otimes f_1 \, )
+P \, m_2 \, ( \, f_1 \otimes f_0 \, ) \,, \\
M_n & = \sum_{m=0}^n \,
P \, m_2 \, ( \, f_m \otimes f_{n-m} \, ) \quad \text{for} \quad n > 1 \,.
\end{align}
The calculation of $M_n$ has been reduced to that of $f_n$.
It follows from~\eqref{f-relation} that
\begin{equation}
\pi_1 \, {\bm f} +\pi_1 \, {\bm h} \, {\bm m}_2 \, {\bm f}
+\kappa \, \pi_1 \, {\bm h} \, {\bm w}_0 \, {\bm f} -\pi_1 \, {\bf P} = 0\,.
\end{equation}
Since
\begin{equation}
\begin{split}
& \pi_1 \, {\bm f} +\pi_1 \, {\bm h} \, {\bm m}_2 \, {\bm f}
+\kappa \, \pi_1 \, {\bm h} \, {\bm w}_0 \, {\bm f} -\pi_1 \, {\bf P} \\
& = \sum_{n=0}^\infty f_n \, \pi_n
+\sum_{n=0}^\infty \sum_{m=0}^n h \, m_2 \, ( \, f_m \otimes f_{n-m} \, ) \, \pi_n
+\kappa \,h \, w_0 \, \pi_0
-P \, \pi_1 \\
& = \Bigl[ \, f_0 +h \, m_2 \, ( \, f_0 \otimes f_0 \, )
+\kappa \, h \, w_0 \, \Bigr] \, \pi_0 \\
& \quad~
+\Bigl[ \, f_1 +h \, m_2 \, ( \, f_0 \otimes f_1 \, )
+h \, m_2 \, ( \, f_1 \otimes f_0 \, )
-P \, \Bigr] \, \pi_1 \\
& \quad~
+\sum_{n=2}^\infty \, \Bigl[ \, f_n
+\sum_{m=0}^n h \, m_2 \, ( \, f_m \otimes f_{n-m} \, ) \, \Bigr] \, \pi_n \,,
\end{split}
\end{equation}
we obtain the following relations:
\begin{align}
& f_0 +h \, m_2 \, ( \, f_0 \otimes f_0 \, ) +\kappa \, h \, w_0 = 0 \,,
\label{f_0-relation} \\
& f_1 +h \, m_2 \, ( \, f_0 \otimes f_1 \, )
+h \, m_2 \, ( \, f_1 \otimes f_0 \, ) -P = 0 \,,
\label{f_1-relation} \\
& f_n +\sum_{m=0}^n h \, m_2 \, ( \, f_m \otimes f_{n-m} \, ) = 0 \quad
\text{for} \quad n > 1 \,.
\label{f_n-relation}
\end{align}
We expand $M_n$ and $f_n$ in $\kappa$ as
\begin{equation}
M_n = \sum_{m=0}^\infty \kappa^m \, M_n^{(m)} \,, \qquad
f_n = \sum_{m=0}^\infty \kappa^m \, f_n^{(m)} \,,
\end{equation}
and let us explain how we recursively determine $M_n^{(m)}$ and $f_n^{(m)}$.

The zero-string product $f_0$ vanishes at $\kappa = 0$. We thus have
\begin{equation}
f_0^{(0)} = 0 \,.
\end{equation}
Then the relation~\eqref{f_0-relation} recursively determines $f_0^{(m)}$.
We find
\begin{align}
f_0^{(1)} & = {}-h \, w_0 \,, \\
f_0^{(2)} & = {}-h \, m_2 \, ( \, h \, w_0 \otimes h \, w_0 \, ) \,.
\end{align}
Then $M_0^{(0)}$, $M_0^{(1)}$, $M_0^{(2)}$, and $M_0^{(3)}$ are given by
\begin{align}
M_0^{(0)} & = 0 \,, \\
M_0^{(1)} & = P \, w_0 \,, \\
M_0^{(2)} & = P \, m_2 \, ( \, h \, w_0 \otimes h \, w_0 \, ) \,, \\
M_0^{(3)}
& = P \, m_2 \, ( \, h \, m_2 \, ( \, h \, w_0 \otimes h \, w_0 \, ) \otimes h \, w_0 \, )
+P \, m_2 \, ( \, h \, w_0 \otimes h \, m_2 \, ( \, h \, w_0 \otimes h \, w_0 \, ) \, ) \,.
\end{align}
We have confirmed that $M_0^{(1)}$, $M_0^{(2)}$, and $M_0^{(3)}$
in~\S\ref{multi-string-products-subsubsection} are reproduced.

The relation~\eqref{f_1-relation} recursively determines $f_1^{(m)}$.
We find
\begin{align}
f_1^{(0)} & = P \,, \\
f_1^{(1)} & = h \, m_2 \, ( \, h \, w_0 \otimes P \, )
+h \, m_2 \, ( \, P \otimes h \, w_0 \, ) \,.
\end{align}
Then $M_1^{(0)}$, $M_1^{(1)}$, and $M_0^{(2)}$ are given by
\begin{align}
M_1^{(0)} & = P \, Q \, P \,, \\
M_1^{(1)} & = {}-P \, m_2 \, ( \, h \, w_0 \otimes P \, )
-P \, m_2 \, ( \, P \otimes h \, w_0 \, ) \,, \\
M_1^{(2)}
& = {}-P \, m_2 \, ( \, h \, m_2 \, ( \, h \, w_0 \otimes h \, w_0 \, ) \otimes P \, )
-P \, m_2 \, ( \, P \otimes h \, m_2 \, ( \, h \, w_0 \otimes h \, w_0 \, ) \, ) \nonumber \\
 & \quad~
{}-P \, m_2 \, ( \, h \, w_0 \otimes h \, m_2 \, ( \, h \, w_0 \otimes P \, ) \, )
-P \, m_2 \, ( \, h \, w_0 \otimes h \, m_2 \, ( \, P \otimes h \, w_0 \, ) \, ) \\
& \quad~
{}-P \, m_2 \, ( \, h \, m_2 \, ( \, h \, w_0 \otimes P \, ) \otimes h \, w_0 \, )
-P \, m_2 \, ( \, h \, m_2 \, ( \, P \otimes h \, w_0 \, )  \otimes h \, w_0\, ) \,. \nonumber
\end{align}
We have confirmed that $M_1^{(1)}$ and $M_1^{(2)}$
in~\S\ref{multi-string-products-subsubsection} are reproduced.

The relation~\eqref{f_n-relation} for $n=2$ recursively determines $f_2^{(m)}$.
We find
\begin{equation}
f_2^{(0)} = {}-h \, m_2 \, ( \, P \otimes P \, ) \,.
\end{equation}
Then $M_2^{(0)}$ and $M_2^{(1)}$ are given by
\begin{align}
M_2^{(0)} & = P \, m_2 \, ( \, P \otimes P \, ) \,, \\
M_2^{(1)} & = P \, m_2 \, ( \, h \, w_0 \otimes h \, m_2 \, ( \, P \otimes P \, ) \, )
+P \, m_2 \, ( \, h \, m_2 \, ( \, P \otimes P \, ) \otimes h \, w_0 \, ) \nonumber \\
& \quad~
+P \, m_2 \, ( \, P \otimes h \, m_2 \, ( \, h \, w_0 \otimes P \, ) \, )
+P \, m_2 \, ( \, P \otimes h \, m_2 \, ( \, P \otimes h \, w_0 \, ) \, ) \\
& \quad~
+P \, m_2 \, ( \, h \, m_2 \, ( \, h \, w_0 \otimes P \, ) \otimes P \, )
+P \, m_2 \, ( \, h \, m_2 \, ( \, P \otimes h \, w_0 \, ) \otimes P \, ) \,, \nonumber
\end{align}
and $M_3^{(0)}$ is given by
\begin{equation}
M_3^{(0)} = {}-P \, m_2 \, ( \, P \otimes h \, m_2 \, ( \, P \otimes P \, ) \, )
-P \, m_2 \, ( \, h \, m_2 \, ( \, P \otimes P \, ) \otimes P \, ) \,.
\end{equation}
We have confirmed that $M_2^{(1)}$
in~\S\ref{multi-string-products-subsubsection} is reproduced.

\section{Conclusions and discussion}
\label{section-9}

In this paper we discussed gauge-invariant operators of open bosonic string field theory
in the low energy.
We added source terms for gauge-invariant operators to the action
and derived the effective action for massless fields
obtained by classically integrating out massive and tachyonic fields.
While the gauge-invariant operators depend linearly on the open string field
and do not resemble the corresponding operators such as the energy-momentum tensor
in the low-energy limit,
we find that nonlinear dependence is generated
in the process of integrating out massive and tachyonic fields.
We also find that the gauge transformation is modified
in such a way that the effective action and the modified gauge transformation
can be written in terms of the same set of multi-string products
satisfying weak $A_\infty$ relations,
and we presented explicit expressions for the multi-string products.
The effective action is in general highly complicated,
but relatively compact expressions are possible
for open bosonic string field theory based on the star product
compared to closed string field theory,
and the technology associated with the weak $A_\infty$ structure
provides us with analytic control over our all-order expressions
for the multi-string products.

Our discussion is motivated by the low-energy limit
in the context of the AdS/CFT correspondence,
and we are interested in the low-energy region
compared to the scale determined by $\alpha'$
of the effective action for massless fields.
After taking this low-energy limit,
the theory will be invariant under the ordinary gauge transformation.
Invariance under the ordinary gauge transformation
requires familiar constraints,
and, for example, the $\alpha'$ expansion
of the effective action for the gauge field of the open string
takes the form of a linear combination of gauge-invariant terms.
However, invariance under the ordinary gauge transformation
does not constrain the coefficients in front of such gauge-invariant terms.
On the other hand, the effective action with an $A_\infty$ structure
does not take the form of a linear combination of gauge-invariant terms,
and constraints from invariance under the gauge transformation
associated with the $A_\infty$ structure have a more dynamical flavor.
Furthermore, the insight we obtained from the analysis of this paper
is that correlation functions of gauge-invariant operators
are similarly constrained from a weak $A_\infty$ structure
before strictly taking the low-energy limit.
Since the weak $A_\infty$ structure is closely related
to the world-sheet picture, we hope that the dynamics
of gauge-invariant operators
dictated by the weak $A_\infty$ structure will help us
reveal quantum gravity from the open string sector.

\bigskip
\noindent
{\normalfont \bfseries \large Acknowledgments}

\medskip
We would like to thank Hiroshige Kajiura and Hiroaki Matsunaga for useful discussions.
The results in this paper were presented
at the online conference ``{\it 2020 Workshop on String Field Theory and Related Aspects}''
organized by ICTP-SAIFR/IFT-UNESP in S\~{a}o Paulo
and the International Institute of Physics in Natal,
which greatly benefited us.
The work of Y.O.\ was supported in part
by Grant-in-Aid for Scientific Research~(C) 17K05408
from the Japan Society for the Promotion of Science (JSPS).

\end{document}